\documentclass[12pt]{article}

\usepackage{graphicx} %
\usepackage{natbib}
\usepackage{multibib}
\newcites{app}{Appendix References}

\makeatletter
\def\app@toctype{toc}
\newcommand{\beginappendixtoc}{%
  \@starttoc{atoc}%
  \let\ORIG@addcontentsline\addcontentsline
  \renewcommand{\addcontentsline}[3]{%
    \ORIG@addcontentsline{##1}{##2}{##3}%
    \def\@tempappc{##1}%
    \ifx\@tempappc\app@toctype \ORIG@addcontentsline{atoc}{##2}{##3}\fi}%
}
\makeatother
\usepackage{amsfonts}
\usepackage{amssymb}
\usepackage{amsmath}
\usepackage{amsthm}
\usepackage{hyperref}
\usepackage[margin=3.5cm]{geometry}
\usepackage{setspace}
\usepackage{xcolor}
\usepackage{tikz}
\usepackage{pgfplots}
\usepackage{array}
\usepackage{booktabs}
\usepackage{colortbl}
\usepackage{ragged2e}
\usepackage{caption}
\usepackage{float}
\usepackage{threeparttable}
\usepackage{multirow}
\usepackage{longtable}
\usepackage[T1]{fontenc}
\usepackage{comment}
\usepackage{makecell}
\usepackage{pdflscape}
\usepackage{forest}
\usepackage[normalem]{ulem}
\usepackage{bbm}

\hypersetup{
    colorlinks=true,
    linkcolor=black,
    filecolor=black,     
    citecolor=black,
    urlcolor=cyan,
    pdftitle={Overleaf Example},
    pdfpagemode=FullScreen,
    }

\newcommand{\madeup}[1]{\textbf{TBD}}

\DeclareMathOperator*{\argmin}{argmin\;}
\DeclareMathOperator*{\diag}{diag\;}

\title{Literature Review and Evidence Aggregation: a Toolkit for Applied Micro\footnote{We thank Anna Dreber Almenberg, Martha Bailey, Jonathan Cohen, Emma Duchini, Alex Frankel, Ryan Kellogg, Jack Mountjoy, Virginia Minni, Matt Notowidigo, Mikkel Plagborg-Møller, Charlie Rafkin, Shanthi Ramanth, Lutz Sager, and Witold Więcek for helpful conversations. We thank Giuseppe Cognata, Shravan Haribalaraman, Ari Jacob, Sedona Jolly, and Min Seo Kim for their excellent research assistance. Ganong thanks Pascal Noel for asking questions which were a source of inspiration to start working on this topic.}}

\author{Peter Ganong\footnote{University of Chicago Harris School of Public Policy and National Bureau of Economic Research. \href{mailto:ganong@uchicago.edu}{ganong@uchicago.edu}.} \and Avik Garg\footnote{University of Chicago. \href{mailto:avikg@uchicago.edu}{avikg@uchicago.edu}.} \and Maximilian Kasy\footnote{Department of Economics, University of Oxford. \href{mailto:maximilian.kasy@economics.ox.ac.uk}{maximilian.kasy@economics.ox.ac.uk}.}}

\begin{document}

\maketitle

\begin{abstract}
  
Consider an analyst interested in predicting the size of an effect. She has identified a set of prior published studies of similar effects. 
We provide a toolkit for (i) summarizing the prior literature, (ii) making predictions of effects in new contexts, and (iii) correcting for the bias from selectivity in the prior literature. We illustrate these methods with empirical examples from labor, public, behavioral, environmental, and development economics. Some of the tools are relevant even when only three prior studies are available. We show how it is possible to use covariates to transparently make predictions for a new context by reweighting prior estimates. The mean effect---after correcting for selectivity---is between 12\% and 21\% of the simple mean in our empirical examples. 
We conclude with a cookbook for practitioners producing meta-analyses.

\end{abstract}

\newpage
\tableofcontents

\newpage
\singlespacing
\pagenumbering{arabic}
\section{Introduction}

Consider an analyst interested in estimating or predicting the size of an effect -- the estimand $\theta_0$. 
She has a dataset with estimates from several prior studies of similar causal effects with standard errors.
We provide a toolkit for three goals that she might have:
\begin{enumerate}
    \item She wants to compare a new estimate $\hat \theta_0$ to the prior literature.
    \item She observes covariates $X_i$ for both prior studies where estimates are available, plus a new context where no estimate is available, and wants to predict $\theta_0$ for this new context.
    \item She is worried that published estimates are ``missing''  estimates due to specification searching, p-hacking or publication bias, and wants to estimate how much selectivity is present. If it is present, she wants to estimate the distribution of effects adjusting for the bias from selectivity.
\end{enumerate}

To develop this toolkit, we alternate between theoretical discussions of methods and practical case-studies. 
We rely on an empirical Bayes approach to provide a unified theoretical framework. This framework assumes that the estimands  $\theta_i$ are drawn from some distribution $\mu$, while the observed estimates are conditionally normal, $\hat \theta_i \sim N(\theta_i, \sigma_i^2)$. If there is selectivity, then observability of study $i$ might depend on the $Z$-statistic $Z_i = \hat \theta_i / \sigma_i$. The distribution $\mu$ might be of independent interest, but more importantly it serves as a device to construct predictions for $\theta_0$ for some new context.

To illustrate these methods, we present both small case studies with three to five prior estimates (drawing on \citealt{sager2025}, \citealt{blundell2025}, \citealt{bailey2025}), and larger case studies with dozens of estimates. The larger case studies are drawn from several fields of applied economics, studying the effect of active labor market policy \citep{card2018works}, unemployment insurance \citep{cohen2025disemployment}, nudges \citep{dellavigna2022rctstoscale}, and unconditional cash transfers \citep{crosta2024unconditional}.

\paragraph{Question 1: Comparing estimates to the prior literature}
This question appears in every applied micro paper that estimates a quantity also studied in prior work. The analyst begins by asking whether the estimates in the prior literature differ primarily because of sampling variation in $\hat\theta_i$   or because of between-study variation in underlying effects $\theta_i$.  
Between-study variation can be measured by comparing the dispersion of estimates $\hat{\theta}_i$ to the standard errors $\sigma_i$. Then, she can compute the precision-weighted mean of the prior estimates.

An estimate of the mean and distribution of the underlying effects is useful in several scenarios. Suppose an analyst has a new estimate which addresses a bias in the existing literature; she can test whether her new estimate could have emerged from the distribution of effects in the existing literature. Alternatively, suppose that the analyst has a new better-powered identification strategy and she wants to quantify the improvement in precision. Finally, suppose the prior literature contains meaningful between-study variation and the analyst has a new estimate to add to this literature. She can point to meaningful between-study variation as a motivation to produce estimates for additional contexts and to search for mechanisms that predict heterogeneity.

We illustrate that each of these three scenarios actually arises using papers published in \textit{American Economic Journal: Economic Policy} (AEJ Policy) in 2025. 
In fact, these scenarios appear to emerge frequently. Of the 61 papers published in 2025 in AEJ Policy, 27 papers discuss prior estimates of similar targets from the literature. A crude lower bound is therefore that meta-analysis can be used---without the author finding any additional related studies---in nearly half of recently-published research in one prominent journal in applied micro. 

In settings with dozens of estimates, an analyst can go further. If the underlying distribution of effects is normal, then the empirical Bayes leads to linear shrinkage estimators of effects.
However, she can also use richer models which do not assume that the underlying effects are distributed normally. Using maximum likelihood estimation, they can estimate a $t$ distribution to account for fat tails, or even use a nonparametric distribution to account for skewness or multi-modality. Empirically, we find that we can reject the normal model in favor of a richer model in three out of our four larger case studies.

\paragraph{Question 2: Predicting effects for new contexts}
The second question---what is the predicted effect $\theta_0$ in a new context---can be addressed by meta-regression or a generalization using Gaussian Process priors. The classic linear meta-regression approach to this question involves simply regressing estimates on covariates and using the coefficients from the regression for a new set of covariate values. Meta-regression's prediction for the new context can be rewritten as a weighted average of the prior estimates. Reweighting is attractive because it gives the analyst a transparent understanding of how the existing studies contribute to a prediction for the new $\theta_0$. 

Gaussian Process priors provide a way to predict $\theta_0$ which maintains transparency through weighted averages, but does not require the restrictive linear additivity assumption of meta-regression. Instead, the researcher chooses a hyperparameter which corresponds to how much they are willing to import insights from existing studies in distinct contexts. When a large number of prior estimates are available for the same context, meta-regressions and Gaussian Process priors make similar predictions. However, when a researcher wants to extrapolate to a new context with no prior studies or only a few prior studies, the distinction between the methods becomes meaningful.\\ 

We illustrate the predictions from these methods using a dataset on the employment effects of active labor market policies from \cite{card2018works}. Consider an analyst who seeks to predict the long-term effect of training for the long-term unemployed (LTU). This prediction exercise necessarily involves extrapolation, because the review does not contain any studies of this exact horizon by program by population cell. Fortunately, there are related studies of: (i) the long-term effects of training for a different population (\textit{unemployment benefit recipients}) and (ii) the \textit{short-term} effects of training for LTUs. 

The analyst may rely on these studies interpreted through the lens of a meta-regression or a Gaussian Process prior. Meta-regression or a Gaussian Process prior with a long length scale---meaning that the analyst is willing to draw heavily on studies from contexts that are further away---deliver a precise posterior for the program's forecasted effect. The standard deviation for the posterior mean is 1 p.p. or less, which is quite small relative to the a posterior mean of roughly 10 p.p. In contrast, if the analyst uses a Gaussian Process prior with a short length scale, the distribution for the prediction is far more uncertain, with a posterior standard deviation of 7 p.p. We discuss how the researcher's preferred \textit{economic} model of employment determination can be used to inform the \textit{statistical} choice of the length scale hyperparameter.

\paragraph{Question 3: Detecting and correcting selectivity}
Turning to the third question, we finally examine selectivity, where reporting and publication decisions can depend on an estimate's $Z$-statistic. We review two types of approaches to selectivity. We first discuss methods for testing for the presence of selectivity based on the distribution of $p$-values across studies. 
We then turn to methods which leverage the richer information contained in the joint distribution of estimates $\hat \theta_i$ and standard errors $\sigma_i$. Leveraging this richer information allows us to not only test for selectivity, but also to estimate its extent, and to obtain effect distributions that correct for selectivity. We provide  graphical intuition for identification in this setting.

We find that selectivity is ubiquitous and that accounting for selectivity can lead to dramatically different estimates of parameters of interest in the context of our four larger case studies. First, insignificant findings are much less likely to be published: the probability of publishing an insignificant estimate is between 5\% and 28\% as compared to a positive significant estimate. Second, failing to correct for selectivity leads researchers to overstate the mean effect. The mean effect---after correcting for selectivity---is between 12\% and 21\% of the simple mean across the four applications we study. Ignoring selectivity leads to severely biased conclusions about the underlying effects.

\paragraph{Roadmap}
The remainder of the article is structured as follows: Section \ref{sec:setup} introduces the formal setup, Section \ref{sec:aggregation_no_x} discusses aggregation, Section \ref{sec:aggregation_with_x} discusses aggregation with covariates, Section \ref{sec:selectivity} discusses selectivity,  Section \ref{sec:unified model} discusses how to combine selectivity and covariates, and Section 7 provides a cookbook for practitioners. Before introducing our formal setup, the remainder of this section provides a brief review of some of the relevant methodological literature.

\subsection{Literature}

\paragraph{Key theoretical references for our review}
In the following, we draw on several literatures in statistical theory that provide a foundation for the more specific questions encountered in evidence aggregation.
The \textit{empirical Bayes} (or random-effects) framework that organizes our discussion throughout originates with \citet{robbins1956empirical}, \citet{stein1981estimation}, and \citet{Morris1983}; 
We rely on both parametric and non-parametric implementations: kernel deconvolution \citep{meister2009}, convex programming for non-parametric maximum likelihood of the prior \citep{koenker2014convex}, Gaussian process priors for flexible extrapolation across covariates \citep{williams2006gaussian}, and Tweedie-type shrinkage of individual estimates \citep{efron2011tweedie}. 
An approach closely related to empirical Bayes imposes priors on hyper-parameters, yielding \textit{hierarchical Bayes} estimators, cf. \citet[chapter~5]{gelman2014bayesian}; we discuss this approach in \autoref{sec:empiricalbayes_with_x}.

A central focus of our review is \textit{selective reporting and publication} of empirical findings. Concerns about selectivity in published economics research date back at least to \citet{stanley1989metaregression}. Bias tests via meta-regression were introduced by \citet{egger1997bias} and refined by \citet{stanley2017finding}. \citet{brodeur2016star} document bunching of reported test statistics just above conventional significance thresholds, an approach extended to a broader cross-section of identification strategies by \citet{brodeur2020methods}; \citet{ioannidis2017power} highlight the prevalence of severely underpowered findings and propose to focus on highly-powered studies. \citet{elliott2022detecting} provide non-parametric tests for selectivity based on violations of monotonicity of the p-curve. Likelihood-based methods that estimate selection from the joint distribution of estimates and standard errors are developed in \citet{publicationbias2019}; we build on these in Section~\ref{sec:selectivity}.  \citet{christensen2018transparency} survey the broader research-transparency movement concerned with selective reporting.

\paragraph{Meta-analysis in economics}
Quantitative research synthesis has a long tradition in economics, with early applications such as \citet{CardKrueger1995}'s synthesis of the minimum-wage literature. Methodological tools for meta-regression and the detection of publication selection are developed by \citet{Stanley2008} and \citet{stanley2014meta}, and surveyed in the practitioner's guide of \citet{irsova2023meta}. A more recent wave exploits standardized designs and larger samples of studies, including the comparison of academic and at-scale nudge trials in \citet{dellavigna2022rctstoscale}, the replication exercise of \citet{camerer2016evaluating}, and the generalizability analysis of \citet{vivalt2019much}.

\paragraph{Approaches beyond the scope of the present review}
There are a number of exciting and insightful developments in the recent literature on evidence aggregation which are beyond the scope of the present review: 

An increasing set of studies, in particular in development and experimental economics, make micro-data available.
In settings where this is the case, meta-studies can directly analyze the \textit{pooled micro-data}, yielding insights beyond those available from aggregate statistics; see, e.g., \citep{bandiera2021social, meager2022aggregating, kremer2023cleanwater, lund2024mentalhealth, mullins2025welfare}.

Meta-regressions, as discussed in \autoref{sec:empiricalbayes_with_x}, are often used to explore possible drivers of effect heterogeneity. Parts of the literature use priors for regression coefficients putting a point-mass at 0. This leads to \textit{Bayesian model averaging} approaches \citep{irsova2023meta}. In our review (and in line with the recommendations of \citealt{gelman2014bayesian}), we instead focus on continuous priors that do not assume true sparsity in a correctly specified linear model.

A central focus of our review is prediction of effects $\theta_i$ for specific instances. A decision-theoretic foundation for the proposed approaches can be found in the theory of compound decision problems \cite{zhang2003compound}. An active literature in theoretical econometrics explores the broader \textit{decision-theoretic foundations} and ramifications of evidence aggregation, e.g.,\ \cite{manski2020toward}, \cite{christensen2026optimal}, \cite{ishihara2024evidence}, and \cite{montielolea2023decision}.

\section{Formal Setup}
\label{sec:setup}

Throughout, our review is based on the following framework. This framework allows us to provide a unified discussion of the issues involved in meta-studies, including information aggregation and extrapolation, as well as p-hacking and publication bias.

Suppose we are interested in a set of empirical studies indexed by $i \in 1,\ldots,n$. Corresponding to each of these studies there is an (unknown) estimand $\theta_i$. This might for example be the true average treatment effect, if $i$ is an experimental study, or it might be the elasticity of some behavioral relationship.  Each study reports an estimate $\hat \theta_i$ of $\theta_{i}$. 
Each study is furthermore characterized by study-level covariates $X_i$, which might include features such as sampling frame and site of the study, study design and identifying variation, treatment or policy characteristics, etc. We will use $\theta$ to denote $(\theta)_{i\in1, \ldots, n}$ and use $\hat\theta, X$ analogously.

Not all studies are available to us when we conduct a meta-analysis. Some studies are never published, either because of researcher decisions (possibly reflecting issues such as p-hacking or specification searching), or because of reviewer and editor decisions (which might lead to publication bias). Some studies might also simply be excluded from the sampling frame of a meta-study, for instance based on the journal or year they were published in. We indicate observability and inclusion of a study in a meta-analysis by $D_i \in \{0,1\}$, where $D_{i}=1$ for studies that are included.

We assume that the sampling distribution of $\hat{\theta}_{i}$ is given by
$$\hat \theta_i |\theta_i, \sigma_i^2,X_{i} \sim N(\theta_i, \sigma_i^2),$$
where the standard error $\sigma_{i}$ is known to us whenever a study is published. This assumption implies that $\hat{\theta}_{i}$ is unbiased for $\theta_{i}$; alternatively, we can think of this assumption as simply \emph{defining} the estimand $\theta_{i}$ as the expectation of $\hat{\theta}_{i}$. This assumption also implies that $\hat{\theta}_{i}$ is normally distributed. Normality is typically justified by asymptotic arguments based on the central limit theorem. The same asymptotic arguments also imply that we can treat $\sigma_i^2$ as known.

Under normality, it is useful to define the $Z$-statistic $Z_{i}$, the corresponding normalized parameter $\omega_{i}$, and the p-value $P_{i}$ for a one-sided test of the null hypothesis that $\theta_{i}=0$, via
$$
\begin{aligned}
Z_{i} &= \hat{\theta}_{i} / \sigma_i, &
\omega_{i} &= \theta_{i} / \sigma_{i}, &\text{ and }
P_{i} &= \Phi(-Z_{i}),
\end{aligned}
$$
where $\Phi$ is the cumulative distribution function of the standard normal distribution.

We furthermore assume that the estimands $\theta_{i}$ are themselves sampled from some unknown distribution $\mu$ across studies,
$$
\theta_{i}\sim^{iid} \mu.
$$
The distribution $\mu$ is the population counterpart to the sample distribution $\{\theta_{1}, \dots, \theta_{n}\}$. The sample distribution is well-defined, regardless of the substantive relationship between the estimands of different studies $i$. We discuss the interpretation of the distribution $\mu$  in more detail below in Section~\ref{sec:empiricalbayes}.

We will at times rely on the stronger assumption $\theta_{i} |\sigma_i \sim^{iid} \mu$, so that the estimands vary independently of estimator precision across studies. For the most part, this assumption only serves as a convenient simplification for the construction of estimators. The one point where this assumption is truly required is identification of selection models using meta-studies, discussed in \autoref{sec:selectivity}.

The observability $D_{i}$ of studies, which reflects reporting and publication decisions, can in principle depend on $\hat{\theta}_{i}, \sigma_{i}$, and $X_{i}$ in complicated ways. We will however assume that the probability of being observable only depends on the $Z$-statistic $Z_{i}$,
$$P(D_i =1 | \hat \theta_i, \sigma_i) = \bar{d}(Z_{i}).$$
This is a restrictive assumption, which is maintained in much of the literature on p-hacking and publication bias (cf. Section~\ref{sec:selectivity}).

\section{Evidence aggregation}
\label{sec:aggregation_no_x}
\subsection{Theory -- Empirical Bayes}
\label{sec:empiricalbayes}

We start our methodological review by considering methods for evidence aggregation. For clarity of exposition, throughout this section we assume that there is no selective reporting, so that $\bar{d} \equiv 1$. The natural framework for discussing evidence aggregation is the empirical Bayes framework, which was first introduced by \cite{robbins1956empirical}. 
The approach, as applied to our setting, can be summarized as follows:

\begin{enumerate}
	\item Specify a model for the sampling distribution of $\hat \theta_i |\theta_i, \sigma_i^2,X_{i}$. In our context, we have
$\hat \theta_i |\theta_i, \sigma_i^2,X_{i} \sim N(\theta_i, \sigma_i^2)$.
	\item Specify a model for $\mu$ (the ``latent distribution'') for the distribution of estimands $\theta_{i}$ across $i$, possibly conditional on observed covariates $X_{i}$. This model might be parametric or non-parametric (unrestricted); we will consider both cases.
	\item Based on the observed distribution of $(\hat{\theta}_{i}, \sigma_{i},X_{i})$, estimate the distribution $\mu$ of $\theta_{i}$ (possibly conditional on $X_{i}$ and $\sigma_i$), by maximizing the marginal likelihood
$$
	f(\hat \theta_i | \sigma_i^2,X_{i}) = \int \tfrac{1}{\sigma_{i}} \varphi\left( \tfrac{\hat{\theta}_{i}-\theta_{i}}{\sigma_{i}} \right) d\mu(\theta_{i} |X_{i}, \sigma_i),
$$
where $\varphi$ is the standard normal density. Alternatively, one might use method of moments estimators that are based on this marginal distribution.
	\item Lastly, form study-level estimates of $\theta_{i}$ using the posterior mean of $\theta_{i}$, $E[\theta_{i}|\hat{\theta}_i, \sigma_i^2,X_{i}]$. This posterior mean is calculated using the ``prior'' $\mu$ estimated in step 3. 
	
	Additionally, in line with the goals listed under items 1 and 2 in our introduction,
	form predictions for a new problem instance $i=0$ given $X_{0}$, $P(\theta_{0}|X_{0})$, which might in turn be used for a downstream decision problem; we will discuss examples later.
\end{enumerate}

Many meta-studies stop at step 3 of this recipe, characterizing properties such as the expectation $E[\theta_{i}]$ (the average effect), the variance $Var(\theta_{i})$ (effect heterogeneity), or the covariance $Cov(\theta_{i}, X_{i})$ (predictors of effect heterogeneity).
Step 4 in this recipe is, however, necessary whenever the goal is to solve a specific policy problem, where policies are chosen for a specific site $i$, rather than across all possible sites.

Viewed through the lens of these four steps, the distribution $\mu$ is best thought of as a conceptual device. It is an input to the construction of empirical Bayes estimators, decision procedures in compound decision problems, and as a device to discuss questions of external validity. However, the justification of these procedures does not necessarily rely on the family of priors. In addition, we note that empirical Bayes has provably good properties for such compound decision problems, \emph{conditional} on the $\theta_{i}$; see for instance \cite{stein1981estimation}.\footnote{Compound decision problems involve repeated decisions across the instances $i$, and loss that is averaged across these instances. A leading example is estimation of the vector $(\theta_1, \dots, \theta_n)$, with mean squared error $\tfrac1n \sum_i (\hat \theta_i-\theta_i)^2$.} For a general review, see \cite{zhang2003compound}. The primary object of interest for both analysts and policymakers will typically be the effect $\theta_0$ for new instances, rather than the distribution $\mu$ itself.

We start by considering the parametric empirical Bayes approach absent covariates, with a normal family of priors (settings with covariates are discussed in \autoref{sec:aggregation_with_x} below). In doing so, we follow \cite{Morris1983}; see also \cite{rubin1981estimation}, and \cite{dersimonian1986meta}. We next generalize by allowing that $\mu$ has heavy tails, considering a family of $t$-distributions. We can finally leave the distribution of $\theta_{i}$ fully unrestricted. 

\subsubsection{Parametric Empirical Bayes (normal latent distribution)}
\label{sec: param eb}

Let us first consider the simplest empirical Bayes setting, following \cite{Morris1983}, where we ignore covariates and assume that the effects $\theta_i$ are normally distributed across studies, independently of $\sigma_i$, so that $\mu$ is given by
$$\theta_i | \sigma_{i} \sim N(\bar{\theta}, \tau^2).$$
Under this assumption, the marginal distribution of $\hat{\theta}_{i}$, integrating out the latent $\theta_{i}$, is given by
$$\hat{\theta}_{i}|\sigma_{i} \sim N(\bar{\theta}, \tau^{2} + \sigma_{i}^{2}).$$
A simple method-of-moments estimator of $\bar{\theta}$ is given by $\hat{\bar{\theta}} = \tfrac{1}{n}\sum_{i} \hat{\theta}_{i}$. 
A simple estimator of $\tau^{2}$ is given by
\begin{equation}
\label{eqn: tau mom}
\hat{\tau}^{2} =\max\left( \tfrac{1}{n}  \sum_{i} \left[ \left(\hat{\theta}_{i} - \hat{\bar{\theta}}\right)^{2} -  \sigma_{i}^{2} \right], 0\right)    
\end{equation}
The efficiency of the estimator of $\bar{\theta}$ could then be further improved by using a precision-weighted average,

\begin{equation}
\label{eqn: prec wtd mean}
\hat{\bar{\theta}}' = \tfrac{1}{\sum_{i} \tfrac{1}{\hat{\tau}^{2} + \sigma_{i}^{2}}}\cdot\sum_{i} \frac{\hat \theta_{i}}{\hat{\tau}^{2} + \sigma_{i}^{2}}. 
\end{equation}

Alternatively, both $\bar{\theta}$ and $\tau^{2}$ can be estimated simultaneously using maximum likelihood.
In the absence of any further information about a new instance $\theta_0$, we can think of $\hat{\bar{\theta}}$ as a predictor for $\theta_0$.

To estimate $\theta_{i}$ for individual instances $i$ where we do have an estimator $\hat{\theta}_{i}$, we can then apply step 4 of the empirical Bayes recipe, where
\begin{equation}
\label{eqn: study level correction}
E[\theta_{i}|\hat{\theta}_{i}, \sigma_{i}] = \bar{\theta}+ \frac{\tau^{2}}{\tau^{2} + \sigma_{i}^{2}}\cdot(\hat{\theta}_i-\bar{\theta}).    
\end{equation}
Substituting estimates for $\bar{\theta}$ and $\tau^{2}$ into this expression yields the empirical Bayes estimator of $\theta_i$. It is interesting to note that in the special case where $\sigma^{2}_{i}$ is constant across $i$, the empirical Bayes approach essentially recovers the James-Stein shrinkage estimator (up to a small degrees of freedom correction).
This shrinkage estimator is known to uniformly dominate $(\hat{\theta}_i)_{i=1}^n$ as estimator of the vector of $\theta_i$'s; see \cite{stein1981estimation} for a proof.

\subsubsection{Heavy tails and non-parametric empirical Bayes}
\label{ssec:nonparametric_eb_summary}

The normality assumption $\theta_i \sim N(\bar{\theta}, \tau^2)$ underlying equation~\eqref{eqn: study level correction} is restrictive: in many applications the latent distribution $\mu$ has heavy tails, is skewed, or is multi-modal. 
We can generalize our baseline model to relax this assumption. 

First, one can take $\mu$ to be a (possibly scaled and shifted) $t$-distribution with $\nu$ degrees of freedom,
\begin{equation*}
\label{eqn:tweedie}
\theta_i \mid \sigma_i \sim t_{df}(\bar{\theta}, \tau^2)
\end{equation*}
which allows for heavier tails than the normal; smaller $\nu$ implies heavier tails. 

Second, one can leave $\mu$ entirely unrestricted. In this case, $\mu$ can be estimated by nonparametric maximum likelihood \citep{koenker2014convex} or by deconvolution \citep{meister2009}, and the posterior mean of $\theta_i$ admits a model-free expression known as \emph{Tweedie's formula} \citep{efron2011tweedie}:
\begin{equation}
   \label{eqn:tweedie}
E\left[\theta_{i}\,\big|\, \hat{\theta}_{i}, \sigma_{i}\right] = \hat{\theta}_{i} + \sigma_{i}^{2} \cdot \frac{ \partial }{ \partial \hat{\theta_i} } \log f(\hat{\theta}_{i} \,|\, \sigma_{i}),
\end{equation}
where $f(\hat{\theta}_i\,|\,\sigma_i)$ is the marginal density of $\hat{\theta}_i$ given $\sigma_i$. Intuitively, the posterior mean adjusts the raw estimate $\hat{\theta}_i$ in the direction of higher marginal density: the magnitude and direction of shrinkage are determined by the local shape of $f$ rather than by a global normality assumption. We illustrate the empirical relevance of this generalization, which allows us to appropriately adjust for heavy tails, skewness, and multi-modality, in Section~\ref{sssec:heavy_tails_npmle}. The derivation of Tweedie's formula and its connection to deconvolution are given in Appendix~\ref{app:heavy_tails}.

\subsection{Empirical examples}

\subsubsection{Average and dispersion -- ``small'' meta-analyses}

\label{subsec: small_meta_analyses}
The primary focus of many applied micro papers is to credibly estimate a quantity of interest and then to interpret the new estimate in part by comparing it to prior work. To illustrate how meta-analysis might be able to add additional insights to this type of exercise, we start from the studies which a paper \textit{already cites} as estimating a similar object and ask what additional insights can be extracted based on applying the equations from Section \ref{sec:empiricalbayes}. 

We distinguish between three uses of meta-analysis. In the first example, an analyst has what they believe to be a superior methodology and wants to know if their estimate is meaningfully different from the prior literature. In the second example, an analyst has a new estimate and they want to construct the most likely estimate for the true effect, taking into account the distribution of effects in the prior literature. In the third example, an analyst has a new estimate and they want to quantify the improvement in precision.

\paragraph{Case Study \#1: Correcting Bias} \citet{sager2025} explore different strategies to estimate the effect of pollution on housing prices, with an eye towards correcting bias in the methods used in the prior literature. They describe their paper as contributing three insights to the literature on pollution damages. After describing the first two insights, they write
\begin{quote}
Our third insight is that pollution damages might be even larger than previously thought... The implied elasticity [of house prices] with respect to [pollutants] $PM_{2.5}$ of around -1.4 is around twice that found for $PM_{10}$ \citep{bento2015} and up to four times the elasticity for Total Suspended Particles (TSP or $PM_{100}$) \citep{chay2005daq}.
\end{quote}

\begin{table}[h!]
\centering
\caption{Elasticity of Housing Prices with Respect to Air Pollution}
\label{tab:pollution_elasticities}
\begin{minipage}{0.9\linewidth}
\centering
\begin{tabular}{lrr}
\toprule\toprule
Study & Estimate & SE \\
\midrule
\textit{Panel A: Prior estimates (n=3)} & & \\
Chay and Greenstone (2005) & -0.27 & (0.10) \\
Bento et al. (2015) & -0.60 & (0.23) \\
Grainger (2012) & -0.37 & (0.15) \\
\midrule
\textit{Panel B: Mean of prior estimates} & & \\
Simple mean & -0.41 & (0.10) \\
Precision-weighted mean $\hat{\bar{\theta}}'$ (equation 2) & -0.34 & (0.08) \\
\midrule
\textit{Panel C: Dispersion of prior estimates} & & \\
Variance of estimates & 0.019 & $-$ \\
Average squared standard error & 0.028 & $-$ \\
Between-study std.\ dev.\ $\hat{\tau}$ (equation 1) & 0.00 & $-$ \\
\midrule
\textit{Panel D: Summary of prior literature} & & \\
\multicolumn{3}{c}{$\theta_i \sim N(-0.34, 0.00)$} \\
\midrule
\textit{Panel E: New estimate} & & \\
Sager and Singer (2025) & -1.44 & (0.35) \\
$p$-value that new estimate drawn from $N(-0.34, 0.00)$ & $<$0.01 & $-$ \\
\bottomrule\bottomrule
\end{tabular}

\vspace{0.6em}
{\footnotesize\raggedright \textit{Note:} The variance of the estimates is $\tfrac{1}{n}\sum_{i}\left(\hat{\theta}_{i}-\bar{\theta}\right)^{2}$, the average squared standard error is $\tfrac{1}{n}\sum_{i}\sigma_{i}^{2}$, and equation~(\ref{eqn: tau mom}) is the former minus the latter, subject to a floor of 0. In Panel D, $N(\cdot,\cdot)$ reports the mean and standard deviation of the fitted distribution of treatment effects. See appendix~\ref{sec: appendix housing prices} for details on how estimates in panels A and E are calculated.\par}
\end{minipage}

\end{table}

A meta-analytic framework offers additional lessons which complement the discussion in \citet{sager2025}. 
With the estimates from Table \ref{tab:pollution_elasticities}, one could therefore have expanded on the ``third insight'' discussion in that paper, adding
\begin{quote}
A meta-analysis of the prior literature indicates that the existing estimates are consistent with a homogeneous treatment effect of pollution with an elasticity of -0.34.  Our elasticity estimate of -1.4 is significantly larger (more negative) than the prior literature. Endogeneity issues in identification strategies used in the prior literature lead them to understate the importance of pollution for affecting housing prices. However, we note that the standard error for our estimate is larger than much of the prior literature; had our design found an elasticity similar to the prior literature, we would be unable to reject the hypothesis that our estimate is drawn from the distribution of prior estimates.
\end{quote}
\paragraph{Case Study \#2: Best Estimate of True Effect}  \citet{blundell2025} is interested in estimating the effect of pay transparency laws on the gender pay gap. They study a 2018 UK law that requires firms with over 250 employees to publicly disclose ``gender equality indicators.'' When discussing their paper in the context of the broader literature, they mention three other papers that attempt to estimate a similar quantity, writing:
\begin{quote}
    The closest studies to ours are \citet{bennedsen2022}; \citet{baker2023}; and \citet{gulyas2023}...
    Both \citet{bennedsen2022} and \citet{baker2023} find that transparency leads to pay compression by slowing down men’s wage growth. In contrast, \citet{gulyas2023} find no impact on individuals’ wages or the gender pay gap.
\end{quote}

Meta-analysis yields additional insights which help to put this paper's findings in context. Unlike in \cite{sager2025}, \cite{blundell2025} believes the prior literature to be relevant to their context, so this new estimate is the most efficient estimate available. 
With the estimates from Table \ref{tab:disclosure_laws} one could therefore have expanded on the discussion in that paper, adding

\begin{quote}
A meta-analysis of the prior literature indicates that of the four disclosure laws studied (including ours), the mean effect is -11.6\% and and the standard deviation across studies is 7.6\%. This is useful for two purposes. First, the fact that there is substantial variation across studies in whether disclosure laws reduce gender pay gaps motivates an analysis in our paper of the mechanisms which give rise to variation in effectiveness. We focus in particular on a binary covariate:whether the law requires \textit{public} disclosure or simply requires disclosure to employees. Second, combining our new study with what is already known from the prior studies in the literature, we estimate that the effect of the 2018 UK law was -14.9\%. This is lower than our econometric estimate because the Empirical Bayes framework shrinks our estimate of -18.8\% toward the mean estimate from the literature as a whole. 
\end{quote}

\begin{table}[h!]
\centering
\caption{Effect of Pay Transparency Laws on the Gender Pay Gap}
\label{tab:disclosure_laws}
\begin{minipage}{0.9\linewidth}
\centering
\begin{tabular}{lrr}
\toprule\toprule
Study & Estimate & SE \\
\midrule
\textit{Panel A: Prior and new estimates (n=4)} & & \\
Baker et al. (2023) & -27.2 & (8.3) \\
Bennedsen et al. (2022) & -13.0 & (5.2) \\
Gulyas et al. (2023) & 0.0 & (0.6) \\
Blundell et al. (2025) (new estimate) & -18.8 & (8.1) \\
\midrule
\textit{Panel B: Mean of estimates} & & \\
Simple mean & -14.8 & (4.9) \\
Precision-weighted mean $\hat{\bar{\theta}}'$ (equation 2) & -11.6 & (4.7) \\
\midrule
\textit{Panel C: Dispersion of estimates} & & \\
Variance of estimates & 98.0 & $-$ \\
Average squared standard error & 40.5 & $-$ \\
Between-study std.\ dev.\ $\hat{\tau}$ (equation 1) & 7.6 & $-$ \\
$p$-value that $\hat{\tau} = 0$ & $<$0.01 & $-$ \\
\midrule
\textit{Panel D: Summary of literature} & & \\
\multicolumn{3}{c}{$\theta_i \sim N(-11.6, 7.6)$} \\
\midrule
\textit{Panel E: Forecast for new estimate} & & \\
Forecast $E[\theta_0 \mid \hat{\theta}_0, \sigma_0]$ (equation 3) & -14.9 & (5.5) \\
\bottomrule\bottomrule
\end{tabular}

\vspace{0.6em}
{\footnotesize\raggedright \textit{Note:}  See appendix~\ref{sec: appendix pay transparency} for details on how estimates in panel A are calculated.\par}
\end{minipage}

\end{table}

The three examples in this section are drawn from articles published in \textit{American Economic Journal: Economic Policy} in 2025. To assess the generalizability of the examples listed here, we reviewed all 61 articles published in the journal in 2025. In essentially all of these articles, the primary object of interest is a causal effect; however, many of the articles do not directly compare their estimates to prior papers or only compare their estimates to a single prior paper. Table~\ref{tab:aej_candidates} lists 27 papers which compare their new estimates to at least two prior estimates from the literature.  A crude lower bound is therefore that meta-analysis can be used---without identifying any additional related studies---in almost half of high-quality research in one journal in applied micro. 

\paragraph{Case Study \#3: Improving Precision}

\citet{bailey2025} is interested in the effect of paid family leave on mothers' earnings and employment. They use a regression discontinuity design in California to estimate the impact of this policy. Describing the prior literature, they write 

\begin{quote}
For the United States, empirical evidence regarding the employment [and earnings] effects of these policies has been inconclusive owing to small sample sizes and incomplete data. While some studies suggest that paid leave improves women’s short-term career outcomes (Rossin-Slater, Ruhm, and Waldfogel, 2013; Campbell, Chin, Chyn, and Hastings, 2018), estimates tend to be imprecise... $\left[ \text{description of data and research design...} \right]$ 
These microdata are around 200 times larger than publicly available surveys, which increases the precision of the estimates...
\end{quote}

\begin{table}[h!]
\centering
\caption{Effect of Paid Family Leave on Mothers' Short-Run Earnings}
\label{tab: paid leave}
\begin{minipage}{0.9\linewidth}
\centering
\begin{tabular}{lrr}
\toprule\toprule
Study & Estimate & SE \\
\midrule
\textit{Panel A: Prior estimates (n=4)} & & \\
Rossin-Slater, Ruhm, and Waldfogel (2013) & 7.5 & (4.0) \\
Campbell, Chin, Chyn, and Hastings (2018) & 10.2 & (10.7) \\
Baum and Ruhm (2016) & 3.6 & (3.4) \\
Timpe (2024) & -5.6 & (2.4) \\
\midrule
\textit{Panel B: Mean of prior estimates} & & \\
Simple mean & 3.9 & (3.0) \\
Precision-weighted mean $\hat{\bar{\theta}}'$ (equation 2) & -0.3 & (1.7) \\
\midrule
\textit{Panel C: Dispersion of prior estimates} & & \\
Variance of estimates & 35.9 & $-$ \\
Average squared standard error & 37.2 & $-$ \\
Between-study std.\ dev.\ $\hat{\tau}$ (equation 1) & 0.0 & $-$ \\
\midrule
\textit{Panel D: Summary of prior literature} & & \\
\multicolumn{3}{c}{$\theta_i \sim N(-0.3, 0.0)$} \\
\midrule
\textit{Panel E: New estimate and updated literature (n=5)} & & \\
Bailey, Byker, Patel, and Ramnath (2025) (new estimate) & -0.8 & (0.8) \\
Precision-weighted mean incl.\ new $\hat{\bar{\theta}}'$ (equation 2) & -0.3 & (1.3) \\
\bottomrule\bottomrule
\end{tabular}

\vspace{0.6em}
{\footnotesize\raggedright \textit{Note:} Estimates are for the effect of paid leave availability. See appendix~\ref{sec: appendix paid leave} for details on how estimates in panel A are calculated.\par}
\end{minipage}

\end{table}

Table \ref{tab: paid leave} conducts a meta-analysis for the effect of offering paid family leave on earnings. With such a meta-analysis in hand, the discussion might have continued

\begin{quote}
Our estimates reduce statistical uncertainty about the effects of paid family leave because our data and research design enable a substantial improvement in precision. The standard error for our research design is 0.8\%, while the standard errors for the prior studies ranges from 2.4\% to 10.7\%. As a result, the standard error for the precision-weighted mean including our design is 1.3\%, whereas it would be 1.7\% without our study. Because our estimate of -0.8\% is so similar to the mean of the existing estimates, adding our study to the pool leaves the mean estimate unchanged (up to one decimal point).
\end{quote}

The simple meta-analyses done in this section are likely to generalize to other research in applied micro. All three examples are drawn from articles published in \textit{American Economic Journal: Economic Policy} in 2025. To assess the generalizability of the examples listed here, we reviewed all 61 articles published in the journal in 2025. In essentially all of these articles, the primary object of interest is a causal effect; however, many of the articles do not directly compare their estimates to prior papers or only compare their estimates to a single prior paper.

Appendix  \ref{app:aej_review} lists 27 papers which compare their new estimates to at least two prior estimates from the literature.  A crude lower bound is therefore that meta-analysis can be used---without identifying any additional related studies---in almost half of high-quality research in one journal in applied micro. 

\subsubsection{Average and dispersion -- ``large'' meta-analyses}
\label{sec: empirical examples}

\paragraph{Applications} 
In the remainder of the paper, we re-analyze datasets from four existing meta-analyses: \cite{card2018works}, \cite{cohen2025disemployment}, \cite{crosta2024unconditional}, and \cite{dellavigna2022rctstoscale}.  
These meta-analyses cover four different subfields of economics --  public, development, behavioral, and labor -- which allows us to emphasize considerations specific to different context. Unlike the ``small'' meta-analyses in the previous section, each of these meta-analyses is based on a relatively large number of studies and uses covariates to predict treatment effects. The larger $n$ and the covariates allow us to use richer methodologies to extract additional insights.

\cite{cohen2025disemployment} analyzes the effect of unemployment benefit generosity on unemployment duration using 93 independent estimates from 57 studies, most of which are based on natural experiments. There are no randomized controlled trial (RCT) experiments in the sample, and a few older studies rely on a selection-on-observables assumption rather than a natural experiment. The paper finds that insignificant results are about 10\% as likely to be published as significant results. The paper also finds that the baseline replacement rate for unemployment benefits is a meaningful predictor of the benefit duration elasticity, a point to which we return in Section \ref{sec: unified model empirical}.

\cite{card2018works} analyzes the effect of active labor market policies (ALMPs) on employment. About one-fifth of the studies the paper considers use an RCT design, and more than half of the studies occurred in Europe. The paper has an extensive heterogeneity analysis by covariates, which we revisit in Section 
\ref{sec: ckw regression}. Standard errors are missing from many of the studies in the original meta-analysis. We drop those studies for a final sample of 169 estimates from 45 studies.

\cite{dellavigna2022rctstoscale} analyzes the effect of nudges on a wide range of outcomes. The paper considers the effect of RCTs from two different samples: 74 estimates that were published in academic journals and 241 estimates from studies conducted by two government-run ``nudge units.'' \cite{dellavigna2022rctstoscale} find substantial publication bias in the sample of academic studies, relative to the nudge unit sample. We will combine both the nudge unit and academic journal estimates. In the selectivity section, we will assume no selectivity, $\bar d \equiv 1$, for estimates from the nudge unit subsample.

\cite{crosta2024unconditional} analyzes the effect of unconditional cash transfer programs in low- and middle-income countries on several different outcomes. The paper exclusively considers RCT studies.  \cite{crosta2024unconditional} concludes, based on 75 estimates, that for each additional \$100 of transfer, recipients have monthly consumption that is \$3.30 higher.

\paragraph{Evidence aggregation}
\autoref{tab:intercept_only_models} illustrates four lessons about evidence aggregation. First, for all four applications, there is substantial dispersion in estimated treatment effects beyond what would be expected based only on standard errors. This is reflected in the fact that we estimate $\tau^2 > 0$ both when using a normal distribution or a $t$ distribution for the latent distribution of estimates.

\begin{table}[h!]
\centering
\caption{Parameter Estimates of Distribution of Treatment Effects}
\label{tab:intercept_only_models}
\setlength{\tabcolsep}{3pt}
\small
\makebox[\textwidth][c]{%
  \resizebox{1.1\textwidth}{!}{%
    \begin{tabular}{llccc}
\toprule \toprule
Treatment Effect \& Dataset & Method & $\bar{\theta}$ & $\tau$ & $df$ \\
\midrule
\multirow{4}{*}{\shortstack[l]{Elasticity of Unemp Duration\\w.r.t. Unemp Benefits\\ \cite{cohen2025disemployment}}} & Simple Mean & 0.513 (0.050) & -- & -- \\
 & Prec-wtd. Mean (eqn 2) & 0.448 (0.044) & 0.345 (0.059) & -- \\
 & MLE (normal) & 0.447 (0.050) & 0.338 (0.053) & -- \\
 & MLE (t) & 0.382 (0.046) & 0.240 (0.032) & 4.04 (1.69) \\
\midrule
\multirow{4}{*}{\shortstack[l]{Effect of Active Labor Market\\ Program on Employment\\ \cite{card2018works}}} & Simple Mean & 0.083 (0.008) & -- & -- \\
 & Prec-wtd. Mean (eqn 2) & 0.040 (0.007) & 0.000 (2.615) & -- \\
 & MLE (normal) & 0.074 (0.008) & 0.086 (0.007) & -- \\
 & MLE (t) & 0.057 (0.020) & 0.064 (0.023) & 4.11 (4.13) \\
\midrule
\multirow{4}{*}{\shortstack[l]{Effect of \$1 Transfer\\ on Monthly Consumption\\ Crosta et al. (2024)}} & Simple Mean & 0.127 (0.049) & -- & -- \\
 & Prec-wtd. Mean (eqn 2) & 0.011 (0.004) & 0.000 (2.212) & -- \\
 & MLE (normal) & 0.028 (0.004) & 0.020 (0.003) & -- \\
 & MLE (t) & 0.028 & 0.020 & $\infty$ \\
\midrule
\multirow{4}{*}{\shortstack[l]{Effect of Nudge on\\ Targeted Outcome\\ \cite{dellavigna2022rctstoscale}}} & Simple Mean & 0.031 (0.004) & -- & -- \\
 & Prec-wtd. Mean (eqn 2) & 0.026 (0.003) & 0.070 (0.002) & -- \\
 & MLE (normal) & 0.024 (0.004) & 0.050 (0.008) & -- \\
 & MLE (t) & 0.004 (0.001) & 0.005 (0.001) & 1.00 (0.084) \\
\bottomrule
\end{tabular}
  }%
}
\begin{minipage}{\textwidth}
\footnotesize
Notes: This table shows estimates of the distribution of treatment effects for four empirical applications. $\bar{\theta}$ is the median for all three distributions. Standard errors are shown in parentheses. For Precision Weighted Mean and Maximum Likelihood Estimate (MLE) normal, the variance is $\tau^2$. For MLE $t$, the variance is $\tau^2\frac{df}{df - 2}$ if $df>2$ and $\infty$ otherwise. For Precision Weighted Mean, we use the $\tau^2$ formula from equation \ref{eqn: tau mom}. We do not show standard errors with method MLE $t$ for \cite{crosta2024unconditional} since they are numerically unstable.
\end{minipage}
\end{table}

Second, using the MLE normal (or equation \ref{eqn: prec wtd mean}) provides a principled way to reduce weight on noisy outlier estimates. This is most easily seen in the context of \cite{crosta2024unconditional}, where the simple mean and MLE normal deliver meaningfully different conclusions. The simple mean implies that a \$1 transfer raises consumption by \$0.13 while MLE normal implies that \$1 transfer raises consumption by \$0.03. This pattern arises because \cite{crosta2024unconditional} is a setting where a few estimates have unusually large treatment effects and standard errors. The simple mean weights these outliers equally to the other estimates. Indeed, when the 10 percent of estimates with the largest standard errors are excluded, the method of moments estimator shrinks to \$0.04, in line with the MLE normal estimate. 

Third, MLE estimates of the mean can be much more precise than the method of moments estimates. The precision advantages of the MLE are greatest when there is (i) variation in sampling uncertainty ($\sigma_i^2$) and (ii) between-study variation  is much smaller than sampling uncertainty ($\tau^2<<\sigma_i^2$). This is illustrated in the estimates for \cite{crosta2024unconditional}, which meets both of these conditions. At the opposite extreme, when $\tau^2 >> \sigma_i^2$, as in \cite{dellavigna2022rctstoscale}, the weights on each study will be similar, leading MLE normal to yield a similar standard error to the method of moments. Standard errors for MLE estimators are calculated using a sandwich estimator. 
We include standard errors for $\hat\tau$ in the ``large'' meta-analyses in  \autoref{tab:intercept_only_models} but not in the ``small'' meta-analyses in Tables \ref{tab:pollution_elasticities} and \ref{tab:disclosure_laws} since these standard errors are imprecisely estimated in small samples.

Fourth, a non-normal model often provides a better fit to the data. For three of the applications (\citealt{cohen2025disemployment}, \citealt{card2018works}, and \citealt{dellavigna2022rctstoscale}), we reject that the latent distribution is normal. This is reflected in the finite degrees of freedom parameter $df$. In these three applications, assuming that the latent distribution is $t$ leads to a much lower estimate for the latent median $\bar{\theta}$ (as compared to assuming that the latent distribution is normal). This finding will serve as a justification for why we prefer to use a $t$ distribution in the context of analyzing selectivity in Section \ref{sec: selectivity empirical}. For \cite{crosta2024unconditional}, assuming the latent distribution is $t$ leads to pathological results since the normal model is the best fit.

\subsubsection{Non-parametric empirical Bayes}
\label{sssec:heavy_tails_npmle}

Until now, we have studied models which assume that the latent distribution of effects can be modeled using a normal or a $t$ distribution. When the latent distribution is in fact skewed or multi-modal, these parametric assumptions will lead to biased conclusions. A more flexible alternative---the nonparametric MLE \citep{koenker2014convex}---combined with Tweedie's formula (equation \eqref{eqn:tweedie}) allow the shape of the posterior mean to adapt to the data.
\begin{figure}[h!]
\centering
\caption{Empirical Bayes shrinkage under skewness, fat tails and multimodality}
\label{fig:shrinkage_curves}
\begin{minipage}{\textwidth}
\centering
\includegraphics[width=\textwidth]
{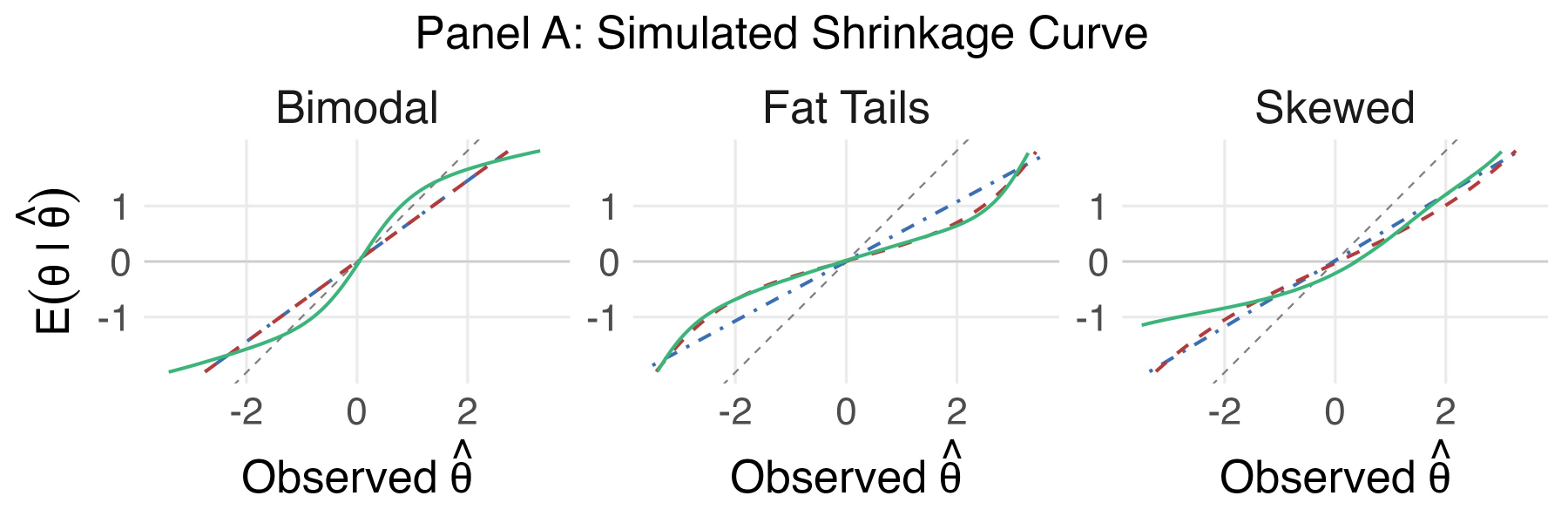}
\includegraphics[width = \textwidth]{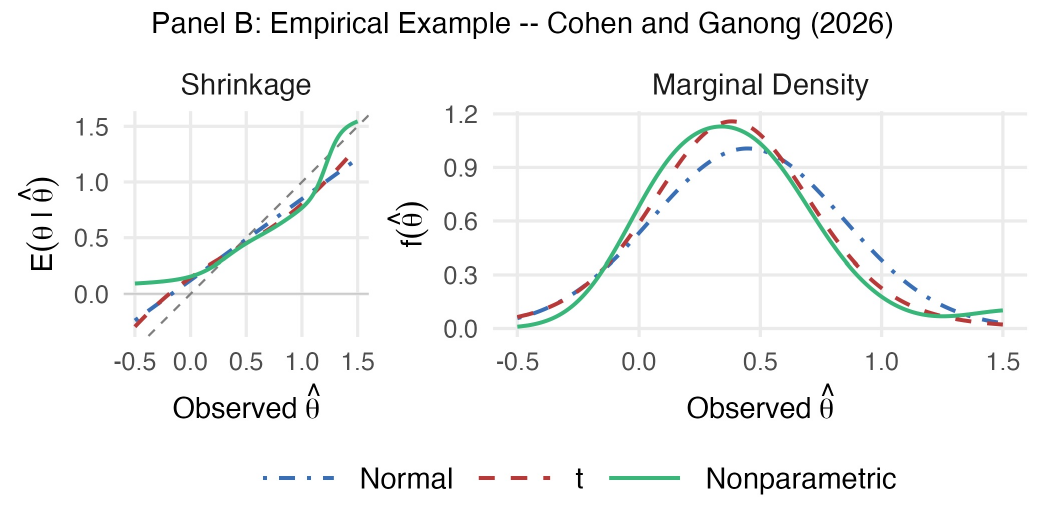}\hfill
\begin{minipage}[b]{0.95\textwidth}\footnotesize
\textit{Notes:} Panel A shows posterior means $E[\theta_{i}\mid\hat{\theta}_{i},\sigma_{i}]$ vs.~$\hat{\theta}_{i}$. The left plot uses draws from a mixture of $N(-1.5, 0.6^2)$ and $N(1.5, 0.6^2)$, the center plot uses draws from $0.5t_{2.5}$, and the right plot uses draws from $0.5(\chi_3^2 - 3).$ For all three, sampling noise $\sigma^2$ is drawn from $\text{Uniform}(0.4, 1.6)$. The reference sampling variance for the new estimate $\hat\theta_i$ is $\sigma_i^2 = 1$ Panel B shows the shrinkage curve (left) and estimated marginal density of $\hat{\theta}_{i}$  (right) for data from \cite{cohen2025disemployment}. The reference sampling variance for the new estimate $\hat\theta_i$ is $\sigma_i^2 = 0.04$, the 75th percentile sampling variance in \cite{cohen2025disemployment}. On all shrinkage plots, the dashed 45-degree line represents no shrinkage.
\end{minipage}
\end{minipage}
\end{figure}

\autoref{fig:shrinkage_curves}, Panel A, shows shrinkage curves under three simulated latent distributions that violate normality: a bimodal mixture of two normal distributions, a heavy-tailed $t_{3}$, and a right-skewed (centered) $\chi^{2}_{3}$. In each case the NPMLE curve departs sharply from the normal benchmark in the tails. In the bimodal case it is strongly non-monotone in slope, pulling observations toward the closer mode rather than a single global average. The $t$-model captures heavy tails but cannot recover skewness or multimodality. In the skewness case, right-tail estimates far away from zero are shrunk less than closer estimates. Panel B shows the analogous comparison for the meta-analysis of unemployment elasticities from \cite{cohen2025disemployment}: the marginal density of $\hat\theta_i$ is visibly right-skewed with a heavier right tail than a matching normal, and the NPMLE shrinkage curve is correspondingly asymmetric---it shrinks negative estimates aggressively but flattens out for large positive $\hat\theta_i$ rather than pulling them back to the global mean.

The choice of latent distribution is most consequential for \emph{study-level} estimates of individual $\theta_{i}$---ranking treatments, identifying ``best'' effects, or cutoff-based decisions. For aggregate quantities like $\bar{\theta}$, the differences across models are typically modest. As a practical guideline: when $n\gtrsim 50$ and the empirical distribution of $\hat \theta_i$ shows visible deviations from normality, the NPMLE (or at least a $t$-model) should be preferred; with small $n$, parametric models might remain more reliable.

\section{Evidence aggregation with covariates}
\label{sec:aggregation_with_x}

Thus far, we have studied what a meta-analysis can discover based on estimates and standard errors alone. Dispersion in the distribution of effects across studies ($\tau^{2}>0$) naturally motivates a search for predictors of study-level heterogeneity. In each of our four applications, readers and researchers will naturally wonder \textit{why}  treatments are more effective in some instances than others. In this section, we consider what more can be learned when study-level covariates are available.

\subsection{Theory}
\label{sec:empiricalbayes_with_x}
The basic empirical Bayes approach can be generalized by allowing for covariates, which enter as linear predictors of $\theta_{i}$; see for example \cite{chetty2014land}. A further generalization, using tools that might be less familiar to readers in economics, allows for a non-linear relationship between the covariates $X_{i}$ and the conditional mean of $\theta_{i}$ given $X_i$, and does not restrict the functional form of this relationship. This can be implemented using Gaussian process priors \citep{williams2006gaussian}, which allow for closed form predictions, despite the unrestricted functional form.  

This approach is particularly useful when we want to think about the external validity of a meta-study, and the extrapolation of estimates to new problem instances. This approach allows for an assessment of (posterior) uncertainty, when extrapolating outside the support of previously encountered instances. By contrast, extrapolations based on (linear) functional forms might give a false sense of certainty. 

\subsubsection{Meta-regressions}

To make predictions of $\theta_0$ for new instances, one needs to learn the relation of effect heterogeneity to contextual covariates $X_i$, to make predictions for new instances characterized by $X_0$. This motivates meta-regressions of the form 
\begin{equation*}
    \hat{\theta}_i = X_i \cdot \beta + \epsilon_i,
\end{equation*} 
as estimated by \cite{stanley1989metaregression} and by many authors since. We emphasize that the variation in $X_i$ is not exogenous and so this regression identifies a \textit{correlation} between the covariate and the treatment effect rather than a \textit{causal} relationship.

To rationalize such meta-regressions, and generalizing the basic parametric empirical Bayes model, assume that
\begin{equation}
    \theta_i |X_i = x \sim N(x \cdot \beta, \tau^2).
    \label{eq:metaregression}
\end{equation}
Here the coefficient vector $\beta$  takes the role previously held by $\bar{\theta}$.
For this model, we then get
$$\hat{\theta}_{i} |X_{i} = x, \sigma_{i} \sim N(x \cdot \beta, \tau^{2} + \sigma_{i}^{2}).$$

Generalizing the method-of-moments approach of Section~\ref{sec: param eb}, $\beta$ can be estimated using either OLS, or a weighted least squares (WLS) regression where the weights are $\frac{1}{\sigma_i^2}$. As a further generalization, we can run a regression of $\hat{\theta}_i$ on $x$, and $\tau^{2}$ can then be estimated as the average of $e_i^{2} - \hat{\sigma}_i^{2}$, where $e_i$ are the estimated residuals of the WLS regression. We could then re-run the regression, weighting each observation by the inverse of its variance, $\frac{1}{\tau^2 + \sigma_i^2}$, which yields a weighted least squares random effects (WLS RE) predictor.

Alternatively we could estimate the hyper-parameters $\tau^2$ and $\beta$ jointly by maximizing the marginal likelihood, as in the empirical Bayes approach. Or we could impose a prior on both $\tau^2$ and $\beta$ to obtain a hierarchical Bayes model, as discussed further below.

\subsubsection{Prediction and Gaussian process priors}
\label{ssec:gpp}

It is common to think of predictions based on regressions as a two-step process: First, the regression coefficients $\beta$ are estimated. Then the effect $\theta_0$ for a new instance is predicted as $X_0 \cdot \widehat{\beta}$.
There is an alternative, numerically equivalent way to obtain the same result: we can leap-frog estimation of $\beta$ and directly predict $\theta_0$.

This alternative perspective, which we will explain next, has two advantages for our purposes. First, it makes explicit that predictions are weighted averages of the available estimates in a meta-study, and clarifies the weights used. Second, it lends itself naturally to generalizations from the linear model to non-linear (but smooth) relationships between $\theta$ and $X$. Using such more general models in turn is helpful because it avoids undue reliance on functional form for extrapolation.

\paragraph{Prediction in the linear model (Ridge)} 

In this spirit, let us now characterize the posterior for $\theta_{0}$ as a function of $ (\hat{\theta}_{1}, \dots, \hat{\theta}_{n})$. In addition to the linear model of Equation \eqref{eq:metaregression}, assume that $\tau^2$ is known (possibly from first-stage maximum likelihood estimation), and suppose that we have a prior for the coefficient vector $\beta$ of the form
$$
\beta \sim N(0, \rho^2 \cdot I),
$$
where $\rho^2$ is known. Then the prior distribution of $(\theta_{0}, \hat{\theta}_{1}, \dots, \hat{\theta}_{n})$ is jointly normal conditional on $\tau^2$ and conditional on the observed covariates $(X_0, X_1, \dots, X_n)$; we will leave the conditioning implicit in our notation.
The prior moments of this joint normal distribution are given as follows, for $i,j=0,1,\dots,n$:
$$
\begin{aligned}
E[\theta_i] &= E[X_i\cdot \beta] = X_i \cdot E[\beta] = 0, \\
C_{i,i} = Var(\theta_i) &= 
Var(X_i \cdot \beta) + Var(\theta_i|\beta)
=
\rho^2 \cdot\|X_i\|^2 + \tau^{2}, \\
C_{i,j} = Cov(\theta_{i}, \theta_{j}) &=  
Cov(X_i \cdot \beta, X_j \cdot \beta) =
\rho^2 \cdot X_{i} X_{j}',\text{ for } i\neq j.
\end{aligned}
$$
where $\rho^{2} + \tau^{2}$ governs the prior variance of $\theta_i$. The ratio  $\rho^{2} / ( \rho^{2} + \tau^{2})$ determines the expected share of variation of $\theta_i$ that is explained by $X_i$ (analogous to an $R^2$ statistic). We collect these terms in the $n$-vector of prior covariances $c$ and the $n\times n$ prior variance matrix of the estimands $C$, 
\begin{align*}
c &=(C_{0,1},\dots,C_{0,n}), 
& C&= \begin{pmatrix}
C_{1,1}&\dots &C_{1,n} \\
\vdots & \ddots & \vdots \\
C_{n,1}&\dots &C_{n,n}
\end{pmatrix},
\end{align*}
to get
\begin{equation}  
    (\theta_{0}, \hat{\theta}_{1}, \dots, \hat{\theta}_{n}) \sim N\left( 
        \begin{pmatrix}
            0 \\
            0
        \end{pmatrix},
        \begin{pmatrix}
            C_{{0,0}} & c, \\
            c' & C + diag(\sigma_{i}^{2})
        \end{pmatrix}
        \right).
        \label{eq:normalprior}
\end{equation}
 
Given joint normality of $(\theta_{0}, \hat{\theta}_{1}, \dots, \hat{\theta}_{n})$, the posterior mean of $\theta_0$ is equal to the posterior best linear predictor,
\begin{equation}
 E[\theta_{0} | (\hat{\theta}_{1}, \dots, \hat{\theta}_{n})]=
\underbrace{ Cov(\theta_{0}, (\hat{\theta}_{1}, \dots, \hat{\theta}_{n})) }_{ c } \cdot \underbrace{ Var((\hat{\theta}_{1}, \dots, \hat{\theta}_{n}))^{-1}  }_{ (C+diag(\sigma^{2}))^{-1}}\cdot (\hat{\theta}_{1}, \dots, \hat{\theta}_{n})'.
\label{eq:GPposterior}   
\end{equation}

For intuition, note that this formula for the best linear predictor is the multivariate counterpart of the familiar formula for the univariate OLS slope, which is equal to the covariance of prediction target and predictor, divided by predictor variance. 
The posterior distribution of $\theta_{0}$ is thus given by

\begin{align}
\theta_{0} | (\hat{\theta}_{1}, \dots, \hat{\theta}_{n}) \sim N\left( 
c \cdot (C + \diag(\sigma_{i}^{2}))^{-1} \cdot (\hat{\theta}_{1}, \dots, \hat{\theta}_{n})',\right.\nonumber\\
\left.C_{0,0} -  c \cdot (C + \diag(\sigma_{i}^{2}))^{-1} \cdot c'
\right).
\end{align}

Note that the posterior mean of $\theta_{0}$ is a weighted average of the $\hat{\theta}_{i}$, with larger weights put on studies $i$ that have covariate values $X_{i}$ closer to $X_{0}$ (so that $C_{0,i}= \rho^2\cdot X_0  X_i'$ is larger), and on studies that have smaller values of $\sigma_{i}$. The covariances $C_{0,i}$ measure how similar (close) instance $0$ is to instance $i$. The coefficients $c \cdot (C + diag(\sigma_{i}^{2}))^{-1}$ then give the optimal weights for a posterior prediction. The variance $C_{0,0} -  c \cdot (C + diag(\sigma_{i}^{2}))^{-1} \cdot c'$ measures the posterior uncertainty for $\theta_{0}$.

\paragraph{Nonlinear generalization (Gaussian Process)}

The linear model that we just discussed specifies the prior co-variance of $\theta_i$ and $\theta_j$ as $C_{ij}=K(X_i, X_j)$ for $i\neq j$ and $C_{ii}=K(X_i, X_i) + \tau^2$, where $K(x,x') = \rho^2\cdot  x x'$ . The function $K(x,x')$ is called a covariance kernel. As we saw in equation \eqref{eq:GPposterior}, the predicted value of $\theta_0$ can be calculated using just this covariance kernel, without any explicit reference to the regression slopes $\beta$.

We have assumed thus far that the prediction function $\bar{\theta}(x) = E[\theta_i | X_i=x]$ is linear; this is reflected in the prior covariances $K(x,x') = Cov(\bar{\theta}(x), \bar{\theta}(x'))$.
This approach can be generalized beyond linear models. If we assume that the $\theta_i$ are jointly normal (conditional on covariates), with prior covariances specified by a general covariance kernel $K(X_i, X_j)$, then we obtain a so-called Gaussian process prior for $\bar{\theta}(x)$ \citep{williams2006gaussian}.
Denoting again  $C_{ij}=K(X_i, X_j)$ for $i\neq j$ and $C_{ii}=K(X_i, X_i) + \tau^2$, the exact same expression as before (Equation \eqref{eq:normalprior}) describes the joint prior distribution of $\theta_0$ and the observed estimates $(\hat{\theta}_{1}, \dots, \hat{\theta}_{n})$. Correspondingly, the same expression (Equation \eqref{eq:GPposterior}) continues to describe the posterior distribution of $\theta_0$ given the observed estimates.

The linear model specifies $K(x,x') = \rho^2\cdot x x'$.
A popular alternative covariance kernel, which does not impose linearity, is the squared exponential kernel,
\begin{equation}
    K(x,x') = \rho^{2} \cdot \exp \left(- \frac{\lVert x-x' \rVert^{2}}{l^{2}}   \right), 
\label{eqn: se kernel}
\end{equation}

The ``length scale'' $l^{2}$ governs the smoothness of the relationship between covariates $X$ and effects $\theta$, where larger $l$ imply smoother functions.
This length scale thus place a role analogous to the bandwidth in kernel regressions.
These hyper-parameters $\rho^{2}, \tau^2$ and $l^2$ could be chosen a priori, for a fully Bayesian specification, or in a data-dependent manner, as in the empirical Bayes approach.

There are many other kernels which have been used in the literature. Spline regression, for example, corresponds to a Gaussian Process prior where $\bar{\theta}(\cdot)$ follows an integrated Brownian motion. While the squared exponential kernel leads to infinitely differentiable predictions, spline regressions are only twice differentiable, in general.

\subsubsection{Hierarchical Bayes}
\label{ssec:hier_bayes}

The empirical Bayes approach estimates hyper-parameters, such as $\tau^2$ and $\beta$ for the linear model, by looking at the data. Alternatively, one might put a prior on hyper-parameters (i.e. $\beta \sim \pi_\beta, \quad \tau \sim \pi_\tau$) and obtain a posterior distribution for hyper-parameters and study-level estimates in one step. The posterior based on the meta-regression model from \autoref{eq:metaregression} can be written as

$$P(\theta, \beta, \tau |\hat\theta, \sigma_i, X) \propto \pi_\beta\pi_{\tau}\prod \frac{1}{\sigma_i}\varphi\left(\frac{\hat\theta_i - \theta_i}{\sigma_i}\right)\frac{1}{\tau}\varphi\left(\frac{\theta_i - X_i\beta}{\tau}\right).$$
This yields hierarchical Bayes estimators (see e.g. \citealt{meager2019understanding}). 

If we put a normal prior $\beta \sim N(0, \rho^2 I)$ on the coefficients and a normal prior with very large variance on $\tau$,
we recover the same estimated $\beta$ as the linear GP model (Ridge) discussed above. In particular
$$
E[\beta | \hat{\theta}_1,\ldots, \hat{\theta}_n, \tau^2] = \argmin_b \sum_i \frac{(\hat{\theta}_i-X_i\cdot b)^{2}}{ \tau^{2}+\sigma_i^{2} } +  \frac{\|b\|^2}{\rho^2} .
$$
We use Hierarchical Bayes in \autoref{sec: ckw regression} and \autoref{sec: unified model empirical}.\footnote{We also use a variant of Hierarchical Bayes in Section \ref{sec: predictive models with x} when we do estimation with Gaussian Process priors. In that case, we choose the hyper-parameters manually. This can be thought of as a prior which is a point mass (rather than a distribution).} One advantage of the hierarchical Bayes approach is that it facilitates uncertainty quantification for predictions, taking into account hyper-parameter uncertainty. This matters especially in the context of smaller meta-studies, where hyper-parameter uncertainty is non-negligible.\footnote{An alternative route to uncertainty quantification, which stays within the empirical Bayes framework rather than imposing priors on hyper-parameters, is provided by the robust empirical Bayes confidence intervals of \citet{armstrong2022robust}.}

\subsubsection{Binary implementation decisions}
A leading downstream use of the apparatus developed above is the choice of binary implementation decisions $A_{i}\in \{0,1\}$ -- for example, whether to scale up a treatment evaluated in study $i$, or whether to implement a treatment in a new instance characterized by covariates $X_{0}$. If $\theta_{i}$ captures all relevant costs and benefits of the treatment, so that welfare is $U(A_{i},\theta_{i}) = A_{i}\cdot \theta_{i}$, a hypothetical decision-maker who observed $\theta_{i}$ might choose $A_{i} = \mathbf 1(\theta_{i}\geq 0)$. In practice we observe only $(\hat\theta_{i}, \sigma_{i}, X_{i})$, and the optimal Bayesian decision becomes
\begin{equation}
    A_{i} \;=\; \mathbf 1\bigl( E[\theta_{i}\mid \hat\theta_{i}, \sigma_{i}, X_{i}] \geq 0\bigr),
\end{equation}
\label{eqn: action}

\noindent
where the posterior mean is computed using the empirical Bayes or Gaussian process expressions above. Note that the sign of this posterior mean may differ from that of $\hat\theta_{i}$ when $\hat\theta_{i}$ is small relative to its standard error or inconsistent with the effects predicted by the covariates. For a new problem instance for which $\hat\theta_0$ is not observed (but covariates $X_{0}$ are observed), the analogous rule is $A_{0} = \mathbf 1\bigl( E[\theta_{0}\mid X_{0}, \hat{\theta}_1,\ldots, \hat{\theta}_n] \geq 0\bigr)$, using the Gaussian process posterior mean for $\theta_{0}$ from Equation~\eqref{eq:GPposterior}. 
We discuss other implementation decisions in more detail in Appendix~\ref{app:binary_decisions}.

\subsection{Empirical examples}

\subsubsection{Regression interpretation: precision and latent heterogeneity}
\label{sec: ckw regression}

\cite{card2018works} studies how the effectiveness of active labor market programs (ALMPs) varies with the program type (training, job search assistance, subsidized private sector employment, subsidized public sector employment), with the gender targeted (mixed-gender, female, male), with the time horizon (short-term, medium-term, long-term), and with the population targeted (UI recipients, long-term unemployed, or disadvantaged). The paper considers models where the first group in each parenthetical constitutes the reference group and asks how changes in program type, gender targeted, time horizon, and population targeted affect the probability of employment. The models used in the paper give each estimate equal weight when assessing the role of covariates.\footnote{This estimation approach is quite common. \cite{card2018works} and \cite{dellavigna2022rctstoscale} do this via Ordinary Least Squares, while \cite{cohen2025disemployment} does this via Bayesian Model Averaging.} We capture several of the research questions examined in their paper by estimating

\begin{equation}
\label{eqn: ckw linear additive}
    \hat{\theta}_i = \alpha + \beta_1\,TimeH_i + \beta_2\,ProgramType_i + \beta_3\,PopTargeted_i + \beta_4\,Gender_i + \epsilon_i.
\end{equation}

\begin{figure}[t]
\centering
\caption{Covariate heterogeneity in effect of active labor market programs \citep{card2018works}}
\label{fig: ckw coefficients}
\includegraphics[width=0.9\textwidth]{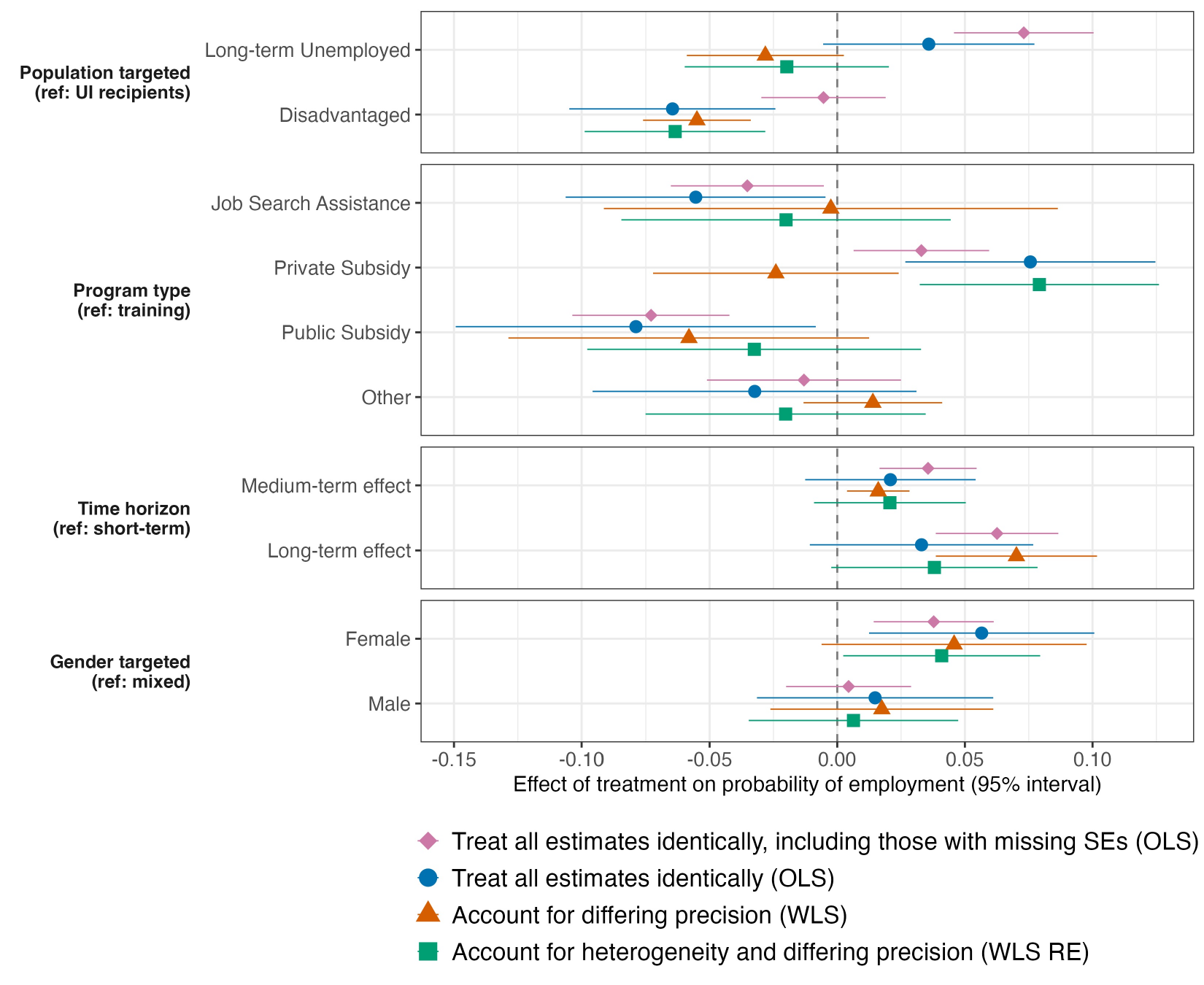}
\end{figure}

Figure~\ref{fig: ckw coefficients} shows how the effectiveness of ALMPs varies with these characteristics, comparing estimates from four types of models. The first regression weights every estimate equally and includes studies with missing standard errors, as in the original \cite{card2018works} analysis.\footnote{In our analysis below, we treat the $\theta_i$ as independent draws. As noted in Section \ref{sec: empirical examples}, in the studies reviewed by \cite{card2018works}, there are often multiple estimates $i$ per study $j$. Generalizing our approach, these might be modeled as $\theta_{ij} = \bar \theta_i + \eta_{ij}$, with each of the $\bar \theta_i$ and $\eta_{ij}$ independent. We do not pursue this here.} Using this model, the paper concludes that programs which target the long-term unemployed are more effective at raising employment.  We replicate this finding (shown in purple diamonds): targeting the long-term unemployed raises employment effects by about 7 percentage points. The second limits the sample to studies with non-missing standard errors; the finding holds in this subsample as well. 

Applying the frameworks from Section~\ref{sec: param eb}, we explore alternative models that account for differences in precision across estimates and allow for a latent distribution of treatment effects. The third (``WLS'') weights each estimate by the inverse of its variance, $w_i = 1/\sigma_i^2$. The fourth (``WLS with random effects'') additionally allows for a latent distribution of treatment effects, weighting each estimate by $w_i = 1/(\tau^2 + \sigma_i^2)$, with $\hat{\tau^2} = 0.074$ obtained by a hierarchical Bayes model as discussed in \autoref{ssec:hier_bayes}. This is reminiscent of the method-of-moments weights in Equation~\eqref{eqn: prec wtd mean}.

\begin{figure}[h!]
\centering
\caption{Study-level treatment effects for programs targeting the long-term unemployed}
\label{fig: ckw ltu}
\includegraphics[width=0.9\textwidth,height=0.7\textheight,keepaspectratio]{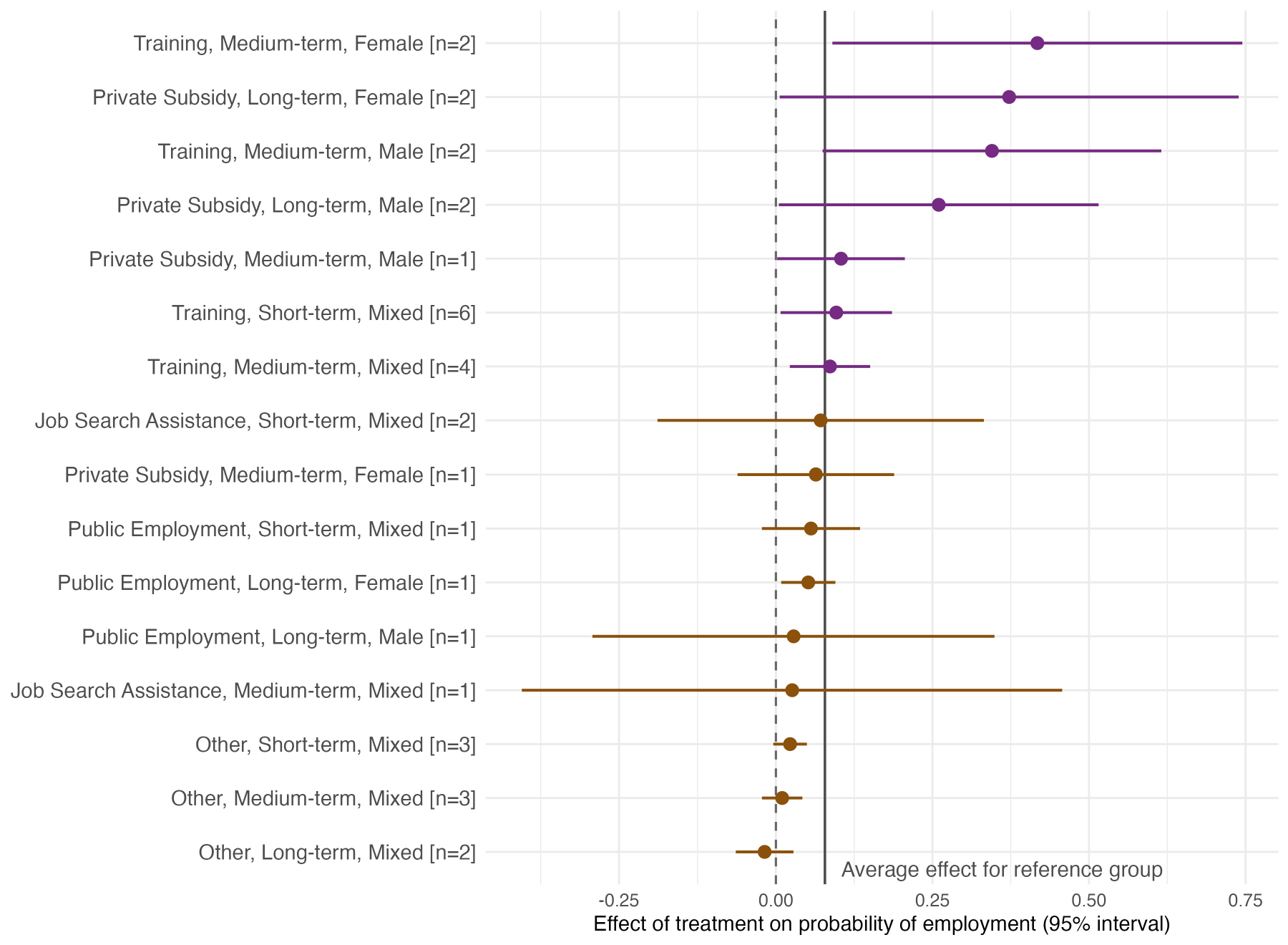}
\begin{minipage}{\textwidth}
\footnotesize
\textit{Notes:} Each point is the mean treatment effect across the estimates from a single study targeting the long-term unemployed, with [n=k] giving the number of such estimates from that study in the \cite{card2018works} data; the whiskers are 95\% intervals based on the average standard error across those estimates. The dark vertical line marks the precision-weighted average treatment effect for the reference group of studies targeting UI recipients (= 0.036). Studies with a mean effect above this average are purple and those below are brown. \citet{vodopivec1998} is excluded because of its extreme standard errors.
\end{minipage}
\end{figure}

Accounting for differences in precision across estimates undoes -- and in fact reverses -- the estimated positive effect of targeting the long-term unemployed. The intuition is most easily seen by looking at the study-level estimates and standard errors for programs targeting this group. Figure~\ref{fig: ckw ltu} shows the estimated treatment effect and 95\% confidence interval for every study targeting the long-term unemployed. A vertical dark line marks the precision-weighted average treatment effect for all studies targeting the reference group (UI recipients). Two studies report treatment effects of at least 30 percentage points -- an enormously large effect relative to much of the literature -- but their estimates are imprecise, with standard errors larger than for many of the other studies. The studies with small or negative effects, by contrast, have small standard errors (or no standard errors at all). A precision-weighted regression gives little weight to the first group and substantial weight to the second; this is why, in Figure~\ref{fig: ckw coefficients}, the coefficient for targeting the long-term unemployed becomes negative under WLS. The same qualitative pattern emerges when we additionally account for latent heterogeneity using $w_i = 1/(\tau^2 + \sigma_i^2)$.

\subsubsection{Predictive models, external validity, and the weights on prior studies}
\label{sec: predictive models with x}

Regressions such as equation (\ref{eqn: ckw linear additive}) should be interpreted as predictive models for the effect $\theta_0$ in a new problem instance characterized by covariates $X_0$, with posterior given by Equation~\eqref{eq:GPposterior}. The predicted effect is a weighted average of the previously reported estimates, with weights determined by the prior covariance kernel $K(\cdot,\cdot)$ and by the standard errors $\sigma_i$. This connects directly to the question of \textit{external validity}: how confident can a policymaker or researcher be that the conclusions of the existing literature carry over to a new setting?

To illustrate, consider two hypothetical new contexts in which a policymaker is contemplating a training program for recipients of mixed gender evaluated at the long-term horizon, again using the \cite{card2018works} data:
\begin{enumerate}
    \item \textbf{Context A (target UI recipients, within support).} A \textit{training} program targeted to UI recipients of mixed gender, . Ten of the 169 estimates in the sample fall in this cell, so the data speak directly to this context.
    \item \textbf{Context B (target long-term unemployed, outside support).} A \textit{training} program targeted to long-term unemployed (LTU) of mixed gender, evaluated at the \textit{long-term} horizon. No estimate in the sample falls in this cell, so prediction is fully extrapolated along the recipient dimension.
\end{enumerate}
We focus on the long-term unemployed for continuity with the meta-regression discussion above, and compare to UI recipients because they are the most commonly observed recipient group in the meta-analysis sample.

For each context, we compute the posterior mean prediction for $\theta_0$, the posterior standard deviation, and the implicit weights $w_i$ on each prior estimate, under three models that differ only in the prior covariance kernel:
(i)~weighted least squares with random effects;
(ii)~a squared-exponential GPP with a long length scale $\ell$, smoothing more aggressively across the covariate space; and
(iii)~a squared-exponential GPP with a short length scale $\ell$, putting most weight on very close neighbours.

\paragraph{Predictions} Prior hyperparameters are set to sensible defaults based on the data. Covariates are pre-normalised (mean 0, variance 1) before kernels are evaluated. The sample variance of $\hat\theta_i$ is $\widehat{\mathrm{Var}}(\hat\theta) \approx 0.012$ and the median sampling variance is $\widetilde{\sigma}^2 \approx 0.001$, giving a signal variance $V \equiv \widehat{\mathrm{Var}}(\hat\theta) - \widetilde{\sigma}^2 \approx 0.011$. 
For the linear model, we set $\tau^2 = \rho^2 = V/(q+2)$, where $q=\dim(X_i)$.\footnote{The number of covariates $q$ is 10, because the equation \ref{eqn: ckw linear additive} is estimated using the ten dummy variables shown in Figure \ref{fig: ckw coefficients}.} 
For the squared exponential kernel, we set $\tau^2 = \rho^2 = V/2$. These choices ensure that the unconditional prior variance of $\theta_i$ matches the observed variance.
The squared-exponential length scales are $\ell = \sqrt{q}/2 \approx 1.58$ (short) and $\ell = 2\sqrt{q} \approx 6.32$ (long).

\begin{landscape}

\begin{figure}[h]
\vspace{-5em}
\hspace{-5em}
\centering
\includegraphics[width=1.1\linewidth]{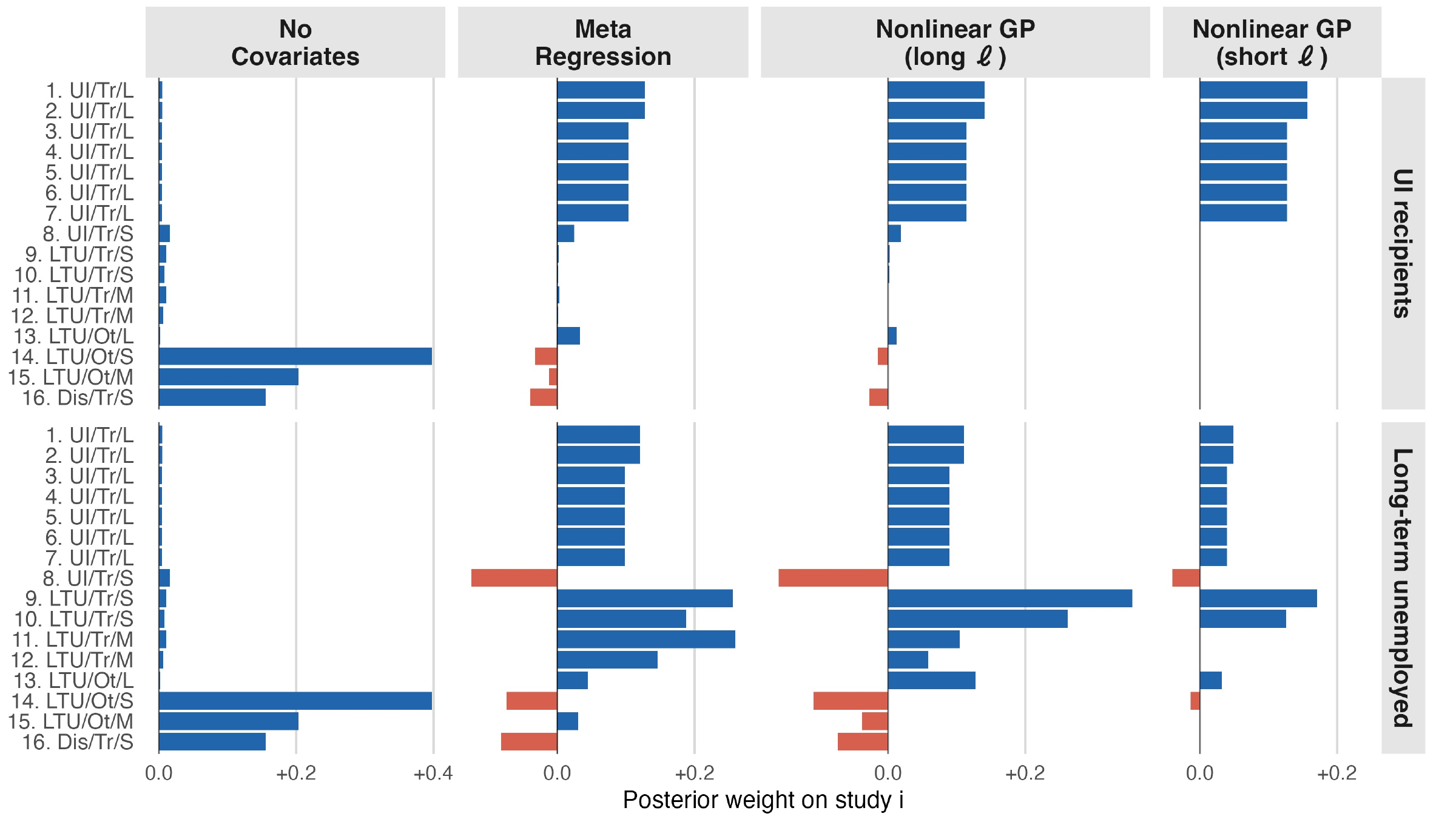}
\caption{Predictive weights on prior estimates for two hypothetical CKW contexts}
\label{fig: ckw gp weights}
\begin{minipage}{1.1\linewidth}
    \footnotesize
\textit{Notes:} Each column is a prediction method (the leftmost, ``No Covariates,'' is a precision-only baseline that ignores covariates); the top row of panels is Context~A (UI recipients, in support) and the bottom row is Context~B (long-term unemployed, out of support), both for mixed-gender Training programs at the long-term horizon. Bars show the posterior weight each study receives in the predictor of $\theta_0$, coloured by sign (blue positive, red negative). The y-axis lists the sixteen studies that appear among the ten highest-weight estimates in either context, labelled by target/program/horizon: target --- UI = registered unemployed, LTU = long-term unemployed, Dis = disadvantaged; program --- Tr = training, Ot = other; horizon --- S = short, M = medium, L = long. Hyperparameters and per-study estimates, standard errors, and kernel distances are reported in Table~\ref{tab:ckw_tr_long_mixed_ui_vs_ltu_panel_b}.
\end{minipage}
\end{figure}
\end{landscape}

Table~\ref{tab: ckw_tr_long_mixed_ui_vs_ltu panel A} reports posterior means and standard deviations under each method, and Figure~\ref{fig: ckw gp weights} displays the weights graphically for the most heavily-weighted estimates (positive or negative). The per-study estimates, standard errors, distances to each test point, citations, and sample sizes for every row displayed in Figure~\ref{fig: ckw gp weights} are reported in Appendix~\ref{app:covariates}. Across the three specifications, the posterior means for the employment effects range from about 12 to 15 percentage points for Context~A and from about 6 to 10 percentage points for Context~B. Context~A exceeding Context~B is a faint echo of the negative LTU coefficient seen in the regression analysis above. 

\begin{table}[htbp]
\centering
\caption{Posterior distribution of employment effects of training}
\label{tab: ckw_tr_long_mixed_ui_vs_ltu panel A}
\scriptsize
\renewcommand{\arraystretch}{0.9}
\begin{tabular}{llccc}
\toprule
 & & WLS & GPP long $l$ & GPP short $l$ \\
\midrule
UI recipients & Posterior mean & 0.122 & 0.136 & 0.149 \\
 & Posterior SD & 0.003 & 0.003 & 0.004 \\
\midrule
Long-term unemployed & Posterior mean & 0.094 & 0.096 & 0.058 \\
 & Posterior SD & 0.005 & 0.010 & 0.067 \\
\bottomrule
\end{tabular}
\end{table}

\paragraph{Policy Recommendations} From predicted effects, it is then a short step to policy recommendations using the results from equation \ref{eqn: action}. Suppose that the decisionmaker is willing to pay \$10,000 for each additional employed person and the program costs \$1,000 to administer per-participant, regardless of who is served. The estimates then imply that they would fund the program for UI recipients for whom the expected value is positive ($\$10,000\times0.122-\$1,000>0$) but not for the long term unemployed for whom the expected value is negative ($\$10,000\times0.094-\$1,000<0$) . After receiving such a recommendation, it would be natural for the policymaker (or the researcher generating the recommendations) to want to understand which estimates are most influential for generating the policy recommendation. 

\paragraph{Mechanisms} A detailed review of Figure~\ref{fig: ckw gp weights} clarifies the mechanisms underlying the predictions. The left-most column of Figure~\ref{fig: ckw gp weights} shows a reference case that ignores covariates entirely: a fixed-effect inverse-variance-weighted meta-analysis ($\tau^2 = 0$) where each estimate's weight is proportional to $1/\sigma_i^2$. Estimates~\#14, ~\#15, and ~\#16 receive the most weight because their reported standard errors are roughly an order of magnitude smaller than every other estimate in the sample. Assigning these estimates such a high weight is of course nonsense because they describe programs which are very different from the context of interest.

Focusing first on context A (in sample prediction for UI recipients), Figure~\ref{fig: ckw gp weights} shows that the three models do essentially the same thing, which is to construct an inverse-variance weighted mean of the ten estimates whose covariates exactly match this context. Estimates \#1--\#7 have small standard errors and receive meaningful weight; the other three estimates are sufficiently noisy that they receive little weight.

The differences between the models are most consequential and interesting in context B (out of sample prediction for the long-term unemployed); two patterns are noteworthy. First, meta regression and GP with long $\ell$ place substantial negative weights on a wide set of estimates. With little or no data at the target cell, the linear predictor effectively interpolates from neighbouring cells using contrasts of the form
\begin{equation}
\label{eq:ckw contrast}
    E[\theta_0 \mid \text{LTU, L}] = \underbrace{E[\theta_0 \mid \text{UI, L}]}_{\text{estimates \#1--\#7}} + \underbrace{E[\theta_0 \mid \text{LTU, S}]}_{\text{estimates \#9, \#10}} - \underbrace{E[\theta_0 \mid \text{UI, S}]}_{\text{estimate \#8}}.
\end{equation}
This assumption is analogous to the parallel trends assumption in a difference-in-difference design.

This re-weighting pattern is visible in Figure~\ref{fig: ckw gp weights} for Context~B under meta regression: positive weight on UI training at the long horizon which we used in Context A (Estimates~\#1--\#7), positive weight on LTU training at the short horizon (Estimates~\#9--\#10), and negative weight on UI training at the short horizon (Estimate~\#8). The same mechanism also generates the other negative meta-regression weights visible in the figure.\footnote{For example, negative weight is consistently assigned to Estimate~\#16 (Dis/Tr/S, whose unusually small standard error pins down the disadvantaged-inflow indicator and is then subtracted off because the test point is not disadvantaged), To a lesser extent, there are also negative weights on Estimates~\#14, and \#15. This occurs for the same underlying reason:  because the test point is neither LTU/Other nor Disadvantaged, the model nets out their identifying contributions.} Negative weights aren't a problem in this setting: they are how a linear predictor (or long $\ell$)  extrapolates by subtracting off distant test points.

Second, the posterior standard deviations differ dramatically across methods for the out-of-support Context~B, even though the point predictions agree. For meta regression, the posterior standard deviation in Context~B is essentially the same as in Context~A. For the long length scale it grows roughly threefold. For the short length scale it grows by a factor of about seventeen, reflecting that the data are silent about a recipient $\times$ horizon cell with zero observations under a strictly local kernel.

To understand the mechanism by which the posterior standard deviation is so tight in context A (unconditionally) and in context B (when using a long length scale), but is so large in context B when using a short length scale, it is helpful to review Table~\ref{tab:ckw_tr_long_mixed_ui_vs_ltu_panel_b}, which reports the kernel covariances from equation \ref{eqn: se kernel} between the prior studies and the new context. With long $\ell$, the model can confidently predict behavior in the new context. With short $\ell$, the model assumes little covariance between the existing studies and the new context. Instead, the model's prediction is much closer to the precision of the prior before any data is observed (which is the unconditional prior variance equal discussed above under ``Predictions'').

How much to extrapolate across the covariate space is ultimately a question about the economic model the researcher is willing to entertain, and the choice of length scale makes that model explicit. A researcher who views employment impacts of the training program as an \emph{additive} function of the population served and the time horizon will want to borrow heavily across cells. Even if a researcher doesn't view the relationships as literally additive, so long as they are comfortable with extrapolation from far away test points, they can use long $\ell$ GP. Concretely, in the context of this application, these models are suitable if the researcher believes that impacts for UI and LTU differ by one constant, and likewise that  long- and short-term impacts differ by a second constant.\footnote{This is reminiscent of the literature on surrogates, where short-term impacts are used to predict long-term outcomes \citep{atheyetal2025}.} Under such additivity, the long-term effect of training on the long-term unemployed can be reconstructed by adding a population adjustment to the training effect for UI recipients, and a long length scale is appropriate.

A researcher who instead believes that the long-term unemployed respond to training \emph{differently} --- in either direction --- will prefer a short length scale that declines to make this extrapolation. The structural literature supplies models of both signs. Training may be \emph{more} effective for the long-term unemployed: in the equilibrium search model of \citet{kospentaris2021unobserved}, skill loss and reduced search effort account for much of the decline in job-finding among the long-term unemployed, so that human-capital programs do the most for precisely the workers who have lost the most. Or, training may be \emph{less} effective for them: if employers screen on unemployment duration \citep{kroft2013duration, jarosch2019statistical}, so that callback rates fall sharply as a spell lengthens, then even retrained long-term unemployed workers may be unable to convert their new skills into offers. Either mechanism implies a training-by-population interaction that the additive predictor cannot represent, and a researcher who finds either one credible should shorten the length scale accordingly.

\section{Selectivity}
\label{sec:selectivity}

\subsection{Identification}
\label{sec:selectivity theory}

In Section \ref{sec:aggregation_with_x}, we focused on evidence aggregation and extrapolation of findings to new contexts, in the absence of selectivity in the publication process. In practice, however, selectivity is very much a problem for inference based on published evidence: the probability $\bar{d}$ that an empirical finding is published might be a function of that finding itself.

In this chapter, we discuss identification in the presence of selectivity. Such selectivity might be due to decisions by various parties, including researchers (who might engage in p-hacking or specification searching), reviewers and editors (who selectively accept papers, leading to publication bias), as well as due to those conducting a meta-study (who always face the choice of which studies to include). Selectivity does not necessarily reflect improper research methods, but it complicates the interpretation of published findings.\footnote{As discussed formally in \cite{whichfindings2018}, there are conflicting objectives: 
validity of conventional inference requires the \emph{absence of selectivity}.
If published findings are to inform subsequent decision-making, however, then \emph{surprising findings} should be selected, because those are the ones which will have a meaningful impact on decisions. But if there is uncertainty over the validity of some identification strategy or data-source, then surprising findings might be less plausible, and plausibility might require the selection of \emph{unsurprising findings}.}

Different identification methods differ in terms of their intended purpose and ambition. They might be designed to just test for the presence of selectivity ($\bar{d} \not\equiv 1$).
More ambitiously, they might estimate the degree of selectivity and the shape of the selection function $\bar{d}(\cdot)$. They might also provide selection-corrected estimates of average effects $\bar{\theta}$, or more broadly of the distribution $\mu$ of $\theta_i$ across studies. Finally, some methods might provide study-level corrected estimates of the $\theta_i$ themselves.

Different methods for identification furthermore leverage data of varying degrees of richness. First, they might only use the (marginal) distribution of published $Z$-statistics or p-values, and seek to test for jumps or non-monotonicity of the p-curve (density of published p-values). 
Second, using richer information, they might use the joint distribution of estimates and standard errors (or sample sizes), and seek to leverage a possible dependence between estimates and standard errors to test for the presence of selectivity, or --  more ambitiously -- to identify the degree and shape of selectivity.\footnote{The function $\bar d(\cdot)$ might for instance be U-shaped, if there is a preference for significant results, or inverse U-shaped, if there is a preference for insignificant results, or monotonically increasing/decreasing, if there is a preference for positive/negative effects, etc. See also \cite{forking2021}.}

Lastly, methods for identifying selectivity might additionally leverage unselected samples of ground-truth estimates. Such estimates might come from research efforts that were pre-registered, for instance by grant agencies, non-academic research labs, or systematic replication studies. Because such non-selected ground truth samples are only rarely available, we focus our review here on the first two approaches, using either just the p-curve, or alternatively the joint distribution of estimates and standard errors.

The results in this section focus on non-parametric identification, and avoid reliance on functional form assumptions. We then discuss both parametric and non-parametric estimation based on these non-parametric identification results in the context of the empirical examples below.

\subsubsection{Distribution of p-values (\cite{elliott2022detecting})}
\label{subsec: elliott}

\paragraph{Testing for selectivity: }
The derivation in the preceding section assumed that publication was non-selective, so that $D_{i}$ is independent of $Z_{i}$ across studies. In the presence of selectivity, however, we have in general that $\bar{d}(z) = P(D_i=1|Z_i=z)$ is not constant as a function of $z$, and thus, denoting $\tilde{d}(p) = \bar{d}(z)$, 
$$
g(p) \cdot \frac{\tilde{d}(p)}{E[D_i]} \neq g(p).
$$
In words, the density of published p-values deviates from the 
density $g(p)$ of latent p-values by a factor proportional to the publication probability $\tilde{d}(p)$.

The latent density of $Z$-statistics is given by the convolution $f(z) = \int \varphi(z - \omega) d \nu(\omega)$, where $\nu$ is the distribution of $\omega_{i} = \theta_{i} / \sigma_{i}$.
The corresponding density of p-values equals $g(p) = f(z) / \varphi(z)$ for $z= \Phi^{-1}(1-p)$.
\cite{elliott2022detecting} characterize this density, showing that it is smooth and non-increasing.
In Appendix~\ref{app:pcurve_derivation} we provide a short proof of these properties.\\

This suggest the following testable hypotheses, which are implied by the null hypothesis of no selectivity:
\begin{enumerate}
	\item $H_0'$: $f(p)$ is continuous at $p$, for conventional significance levels such as $p= .05$.
	\item $H_0''$: $f(p)$ is non-increasing in $p$, across all values of $p$.
\end{enumerate}

Tests based of the first of these derived hypotheses (i.e. continuity/smoothness) have power if $\tilde{d}(p)$ is discontinuous at conventional levels. Tests based on the second of these derived hypotheses do not have power if $\tilde{d}(p)$ is also non-increasing: In that case $f(p) = g(p) \cdot \tilde{d}(p)$ is non-increasing, since both $g(p)$ and $\tilde{d}(p)$ are. Typically, researchers prefer smaller p-values, and ``significant'' results. Whenever that is the case, then $\tilde{d}(p)$ is indeed non-increasing. 

There are, however, scenarios where the opposite is true. This includes notably cases where researchers may wish to find an insignificant result, for instance when testing for pre-trends in a difference-in-differences context, and in the context placebo tests for the validity of some research design more broadly.

\subsubsection{Meta-studies}
\label{sec: meta studies}
The approaches that we have just discussed are based solely on the distribution of p-values $P_{i}$ or, equivalently, the distribution of $Z$-statistics $Z_{i} = \hat{\theta}_i / \sigma_i$. These approaches only allow us to test for the presence of selectivity, but not to identify its shape or magnitude, nor to correct published estimates.
In meta-studies, however, we typically have not only P-values at our disposal, but instead observe both point estimates $\hat{\theta}_{i}$ and standard errors $\sigma_{i}$. We can therefore leverage their \emph{joint} distribution for identification, if we are willing to impose some additional assumptions. 
This also allows us to be more ambitious, going beyond mere tests of the presence of selectivity, to estimate the magnitude and shape of selectivity, and to implement selection corrections.\\

Approaches that leverage this joint distribution typically rely on two assumptions, both of which are substantively restrictive. The first of these is the assumption that estimands and standard errors vary independently across studies, so that
\begin{equation}
\label{eqn: theta se indep}
\theta_{i} \perp \sigma_{i}.
\end{equation}
The second is the assumption that selection is only based on p-values,
\begin{equation}
\label{eqn: select only on z}
E[D_i =1 | \widehat \theta_i = \theta, \sigma_i = \sigma] = \bar{d}(Z_{i}).
\end{equation}
These two assumption are easily generalized to allow for independence conditional on observed covariates $X_i$; we do so in \autoref{sec:unified model theory} below.
The first of these two assumptions might be violated if sample sizes in experiments are chosen based on power calculations, and if the researchers preparing these experiments have correct priors regarding how the magnitude of treatment effects varies across studies \citep{allcott2015, gechter2024}. The second of these assumptions might be violated if selectivity depends on the magnitude of estimates, in addition to p-values, for instance because only certain magnitudes are considered substantively (economically) meaningful by authors or reviewers, regardless of statistical significance.\\

We will discuss three versions of identification based on the joint distribution of $\hat{\theta}_i$ and $\sigma_i$: (1) meta-regressions \citep{egger1997bias},
(2) the ``weighted average of adequately powered'' estimates \citep{stanley2017finding}, and (3) simultaneous estimation of selection function and distribution of estimands \citep{publicationbias2019}. 

All three approaches are based on the same two assumptions: independence and selection based on p-values.
The theoretical arguments made below to justify the approaches of \cite{egger1997bias} and of \cite{stanley2017finding} do not appear explicitly in the original references, but they rationalize the methods proposed in these papers.

\paragraph{Testing for selectivity: Meta-regressions \citep{egger1997bias}}
Consider the null-hypothesis that there is no selectivity, so that $\bar{d}\equiv 1$. Under this null hypothesis, the independence assumption $\theta_{i} \perp \sigma_{i}|X_{i}$ implies conditional mean independence $E[\hat{\theta}_{i} | \sigma_{i},X_{i}] = E[\theta_{i} | \sigma_{i},X_{i}] = E[\theta_{i} | X_{i}]$. As proposed by \cite{egger1997bias}, this motivates linear regressions of $\hat{\theta}_{i}$ on $\sigma_{i}$ (and possibly $X_i$, which we omit for notational simplicity in the following), 
$$
\hat{\theta}_i = \alpha+\beta\cdot \sigma_i + \epsilon _i,
$$
or equivalently (this is the version originally proposed by \citealt{egger1997bias}), after dividing both sides by $\sigma_i$,
\begin{equation}
\label{eqn: egger}
    Z_i = \beta + \alpha\cdot \tfrac{1}{\sigma_{i}} + \tfrac{\epsilon _i}{\sigma _i}.
\end{equation}
This latter form of the regression can be motivated as a weighted least-squares estimator that down-weights noisier observations. Note that $\alpha$ and $\beta$ trade places as intercept and slope, after this transformation.

Under the null hypothesis, we obtain $\beta=0$: There should be no systematic relation between the magnitude of estimates $\hat{\theta}_i$ and the size of standard errors $\sigma_{i}$. A conventional $t$-test can be used to test this implication, based on either version of the regression. Rejection of the null that $\beta=0$ can then be interpreted as evidence of selectivity.\\

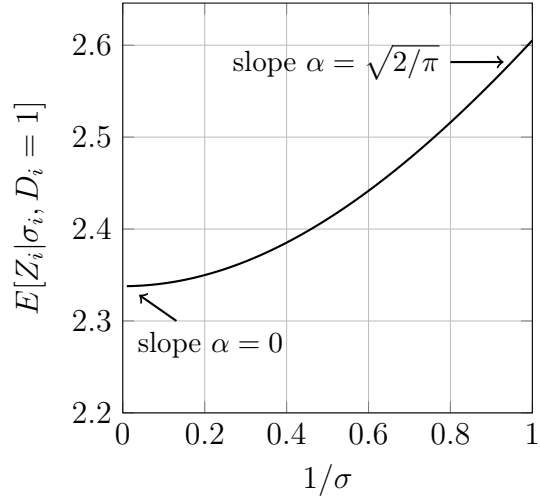
\begin{figure}[H]
	\caption{Nonlinearity of meta-regressions.}
	\label{fig:nonlinear_metaregression}
	\begin{center}
		\begin{tikzpicture}
			\begin{axis}[
				xlabel=$1/\sigma$,
				ylabel={$E[Z_i|\sigma_i, D_{i} = 1]$},
				grid=major,
				major grid style={gray!50},
				xmin=0,
				xmax=1,
				ymin=2.2,
				width=7cm,
				height=7cm,
				samples=100
			]
			
			\addplot[thick] coordinates {
			(0.010, 2.337865)
			(0.020, 2.337956)
			(0.030, 2.338108)
			(0.040, 2.338320)
			(0.050, 2.338592)
			(0.060, 2.338926)
			(0.070, 2.339320)
			(0.080, 2.339774)
			(0.090, 2.340288)
			(0.100, 2.340863)
			(0.110, 2.341497)
			(0.120, 2.342192)
			(0.130, 2.342947)
			(0.140, 2.343761)
			(0.150, 2.344635)
			(0.160, 2.345568)
			(0.170, 2.346561)
			(0.180, 2.347612)
			(0.190, 2.348723)
			(0.200, 2.349892)
			(0.210, 2.351119)
			(0.220, 2.352405)
			(0.230, 2.353749)
			(0.240, 2.355151)
			(0.250, 2.356611)
			(0.260, 2.358127)
			(0.270, 2.359701)
			(0.280, 2.361332)
			(0.290, 2.363019)
			(0.300, 2.364762)
			(0.310, 2.366562)
			(0.320, 2.368417)
			(0.330, 2.370327)
			(0.340, 2.372293)
			(0.350, 2.374313)
			(0.360, 2.376387)
			(0.370, 2.378516)
			(0.380, 2.380699)
			(0.390, 2.382935)
			(0.400, 2.385224)
			(0.410, 2.387565)
			(0.420, 2.389960)
			(0.430, 2.392406)
			(0.440, 2.394904)
			(0.450, 2.397453)
			(0.460, 2.400053)
			(0.470, 2.402704)
			(0.480, 2.405405)
			(0.490, 2.408155)
			(0.500, 2.410955)
			(0.510, 2.413804)
			(0.520, 2.416702)
			(0.530, 2.419648)
			(0.540, 2.422642)
			(0.550, 2.425684)
			(0.560, 2.428772)
			(0.570, 2.431908)
			(0.580, 2.435089)
			(0.590, 2.438317)
			(0.600, 2.441589)
			(0.610, 2.444907)
			(0.620, 2.448270)
			(0.630, 2.451677)
			(0.640, 2.455128)
			(0.650, 2.458623)
			(0.660, 2.462160)
			(0.670, 2.465740)
			(0.680, 2.469363)
			(0.690, 2.473027)
			(0.700, 2.476733)
			(0.710, 2.480480)
			(0.720, 2.484268)
			(0.730, 2.488096)
			(0.740, 2.491964)
			(0.750, 2.495872)
			(0.760, 2.499818)
			(0.770, 2.503804)
			(0.780, 2.507827)
			(0.790, 2.511889)
			(0.800, 2.515988)
			(0.810, 2.520125)
			(0.820, 2.524298)
			(0.830, 2.528508)
			(0.840, 2.532754)
			(0.850, 2.537035)
			(0.860, 2.541352)
			(0.870, 2.545704)
			(0.880, 2.550090)
			(0.890, 2.554511)
			(0.900, 2.558965)
			(0.910, 2.563453)
			(0.920, 2.567974)
			(0.930, 2.572528)
			(0.940, 2.577114)
			(0.950, 2.581733)
			(0.960, 2.586383)
			(0.970, 2.591064)
			(0.980, 2.595777)
			(0.990, 2.600520)
			(1.000, 2.605293)};

			\node[below right] (lower-label) at (axis cs:0.01,2.3) {\small slope $\alpha=0$};
			\draw[->, thick] (lower-label) -- (axis cs:0.040, 2.328320);

			\node[above left] (upper-label) at (axis cs:0.8,2.55) {\small slope $\alpha=\sqrt{2/\pi}$};
			\draw[->, thick] (upper-label) -- (axis cs:0.930, 2.581733);
			
			\end{axis}
			\end{tikzpicture}
	\end{center}
	\footnotesize
	\textit{Notes}:  This figure plots $E[Z_i|\sigma_i, D_{i} = 1]$ for $D_i = \mathbf 1(Z_i \geq \bar{z})$ with $\bar{z} = 1.96$ and $\theta \sim N(0,1)$.
	For this example, $E[Z_i|\sigma_i, D_{i} = 1] = \rho \cdot \frac{\varphi(\bar{z} / \rho)}{1-\Phi(\bar{z} / \rho)}$, where $\rho = \sqrt{ 1+\frac{1}{\sigma^{2}} }$, and correspondingly $E[\hat{\theta}_i|\sigma_i, D_{i} = 1] = \sqrt{ 1+\sigma^{2}} \cdot \frac{\varphi(\bar{z} / \rho)}{1-\Phi(\bar{z} / \rho)}$;
    this is a special case of equation \eqref{eqn:metaregnonlinear}.

    The meta-regression approach, as in \cite{egger1997bias}, fits a linear approximation to the non-linear function $E[Z_i|\sigma_i, D_{i} = 1]$.
\end{figure}

Note that linearity of the relationship between $\hat{\theta}_i$ and $\sigma_i$ is trivially satisfied \textit{under the null}, since they are mean-independent. Linearity is however generically violated \textit{under the alternative}, when the published evidence is selected, so that $\bar{d}\not\equiv 1$: The function
\begin{equation}
\label{eqn:metaregnonlinear}
    E[Z_i|\sigma_i, D_{i} = 1] = \int \int \frac{\bar{d}(z)}{E[\bar{d}(Z)|\sigma_{i}]}\cdot z\cdot\varphi\left( z-\tfrac{\theta}{\sigma_{i}} \right)dz \;d\mu(\theta)
\end{equation}
depends nonlinearly on $\tfrac{1}{\sigma}$.
This is illustrated by Figure~\ref{fig:nonlinear_metaregression}, which plots  $E[Z_i|\sigma_i, D_{i} = 1]$, for $D_i = \mathbf 1(Z_i \geq \bar{z})$ with $\bar{z} = 1.96$ and $\theta \sim N(0,1)$, against $1/\sigma$. 

This observation is important because many papers aim to interpret $\alpha$ in regression \eqref{eqn: egger} as an estimate of $\lim_{ \sigma \to 0 } E[\hat{\theta}_{i} | \sigma_{i} = \sigma, D_{i}=1]$ (which in turn is taken to identify the average effect $E[\theta_{i}]$). Such an interpretation is \textit{not} valid in general.
In the example of Figure~\ref{fig:nonlinear_metaregression}, \textit{any} coefficient $\alpha$ between $0$ and $\sqrt{2/\pi}$ could be obtained, depending on the distribution of $\sigma_i$.
Linear extrapolation will therefore necessarily lead to biased estimates of the average effect $\bar{\theta}$.

\paragraph{Estimating average effects: Highly powered studies \citep{stanley2017finding} }

Meta-regressions allow us to construct a test of the null that there is no selectivity (as was the case for our discussion of the distribution of p-values). An alternative goal, in line with traditional meta-studies, is to obtain an estimate of the expectation $\bar{\theta}$ of $\theta_{i}$ across studies, that is, of the average effect.
Assume that $\bar{d}(z) \to 1$ as $|z| \to \infty$, which means that findings with large $z$-statistics are always published. By definition, and recalling our assumption that selection depends only on the $Z$-statistic,
$$
E[\hat{\theta}_{i} | \sigma_{i} = \sigma, D_{i}=1] = \frac{E[\hat{\theta}_{i} \cdot \bar{d}(\hat{\theta}_{i} / \sigma) | \sigma_{i} = \sigma]}{E[ \bar{d}(\hat{\theta}_{i} / \sigma) | \sigma_{i} = \sigma]}.
$$
By assumption, $\bar{d}({\theta} / \sigma) \to 1$ as $\sigma \to 0$, whenever ${\theta} \neq 0$. By the dominated convergence theorem (assuming that $E[|\theta_i|] < \infty$ and $P(\theta_i=0)=0$), this implies
$$
\lim_{ \sigma \to 0 } E[\hat{\theta}_{i} | \sigma_{i} = \sigma, D_{i}=1] = E[\theta_{i}].
$$
It follows that the average estimate for ``highly powered'' studies, with small $\sigma_{i}$, is approximately equal to the average estimand $E[\theta_{i}]$. Put differently, highly powered studies are not subject to selection, because they always yield large $z$-statistics. 

This argument justifies the proposed estimator of $E[\theta_{i}]$ in \cite{stanley2017finding} (see also \citealt{ioannidis2017power}), who suggest to focus on a weighted average of highly powered studies, in order to to estimate $\bar{\theta} = E[\theta_{i}]$. To be empirically feasible, this approach requires the meta-study to include observations in the vicinity of $\sigma_{i}=0$. Put differently, there need to be studies that are highly powered in the sense that $E[\bar{d}(Z_i) | \sigma_i] \approx 1$.
In practice, it might well be the case that the latent distribution of $\theta_i$ has mass around $0$, so that even studies with small $\sigma_i$ might not be highly powered. In this case, no reliable estimate of $\bar \theta$ can be formed by using the approach of \cite{stanley2017finding}. We return to this point in Section \ref{sec: selectivity empirical}.

Note also that the assumption that $\bar{d}(z) \to 1$ as $|z| \to \infty$ requires that both positive and negative estimates are published, whenever $|z|$ is large enough. This excludes one-sided selection, where only positive significant estimates are published, for instance. In the example in Figure~\ref{fig:nonlinear_metaregression} this is not the case.
For this example, $\lim_{ \sigma \to 0 } E[\hat{\theta}_{i} | \sigma_{i} = \sigma, D_{i}=1] = 2 \cdot \varphi(0) = \sqrt{2/\pi} \neq 0 =E[\theta_i]$.

\paragraph{Estimating selectivity and effect distributions \citep{publicationbias2019}}
\label{sec: ak theory}
Meta-regressions allow us to test for the presence of selectivity. Highly powered studies allow us to estimate average effects. As it turns out, however, we can achieve considerably more under our assumptions. We can not only test for the presence of selectivity, but we can also
identify \textit{how much} more likely significant estimates are to be published. And we can not only identify the average effect $\bar{\theta}$ when there are highly powered studies, but we can non-parametrically identify the distribution $\mu$ of $\theta_{i}$ across studies -- for instance the \textit{variance} of effects across studies, or the presence of \textit{fat tails}.
This was proven in \cite{publicationbias2019}: both the distribution $\mu$ of $\theta$ and the selection function $\bar{d}(z)$ are non-parametrically identified under the assumptions that $\hat \theta_i |\theta_i, \sigma_i^2 \sim N(\theta_i, \sigma_i^2)$, $\theta_{i} \sim \mu$, $\theta_{i} \perp \sigma_{i}$, and $E[D_i =1 | \widehat \theta_i, \sigma_i] = \bar{d}(Z_{i})$.

To provide some intuition for this identification result, we provide both algebra and graphical examples. We first describe how to construct functions of the data which are purged of selectivity, and then describe how to recover the distribution  $\mu$ of $\theta$. Algebra and graphical examples are used to \textit{illustrate} the source of non-parametric identification in the model. Construction of the actual estimators uses maximum likelihood estimation.

Suppose that studies can have only two levels of the standard error: $\sigma_1$ and $\sigma_2$. The distributions of estimates, conditional on either standard error, draw from the same latent distribution of true effects and are affected by the same selection process. They differ only in the fact that $\sigma_2$ is noisier than $\sigma_1$. Assume further (again just for illustration) that the latent distribution  $\mu$  is normal.\footnote{\cite{publicationbias2019} prove identification in a setting where the distributions of true effects and standard errors are left unrestricted, and selectivity can based on an arbitrary function of the $Z$-statistic.}

\begin{figure}[H]
  \centering
    \caption{Distribution of true effects and all estimates -- published and unpublished}
    \label{fig: effects unobserved}
    \includegraphics[width=\textwidth]{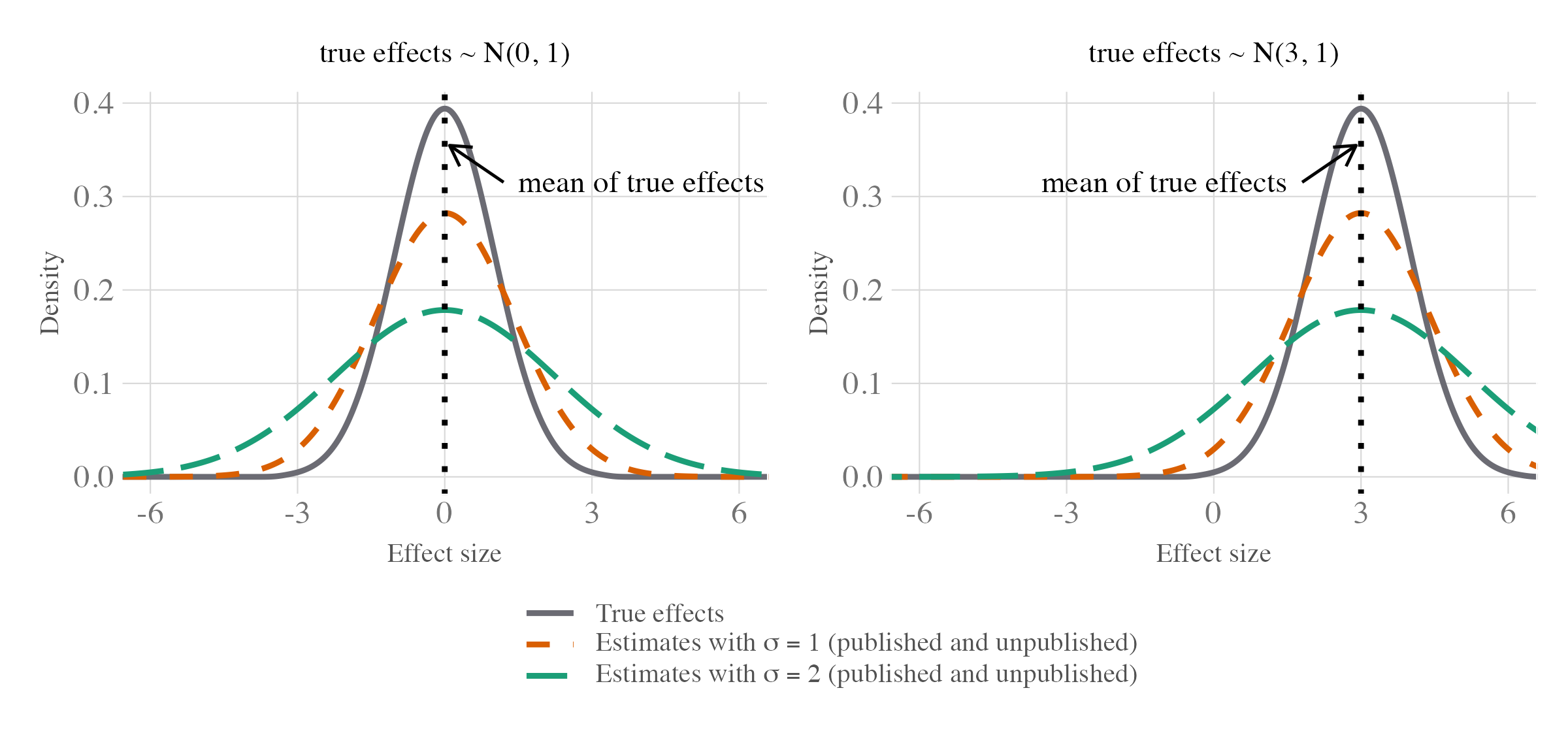}
    \label{fig:panel-a}

\end{figure}

Figure \ref{fig: effects unobserved} illustrates several distributions in this simplified environment. Two latent distributions are shown in black: in the left panel the mean is zero and in the right panel the mean is three. Both latent distributions have variance, $\tau^2 = 1$. In addition, the figure shows the distribution of estimates drawn from the latent distribution with noise $\sigma = 1$ (published and unpublished) and the distribution of estimates with noise $\sigma = 2$ (published and unpublished). 

\captionsetup{font=small}
\captionsetup[sub]{font=small,skip=0.0cm} %
\vspace{-2em}

\begin{figure}[H]
  \centering
    \caption{Distribution of $Z$-statistics -- published and unpublished}
    \label{fig: z unobserved}
    \includegraphics[width=\textwidth]{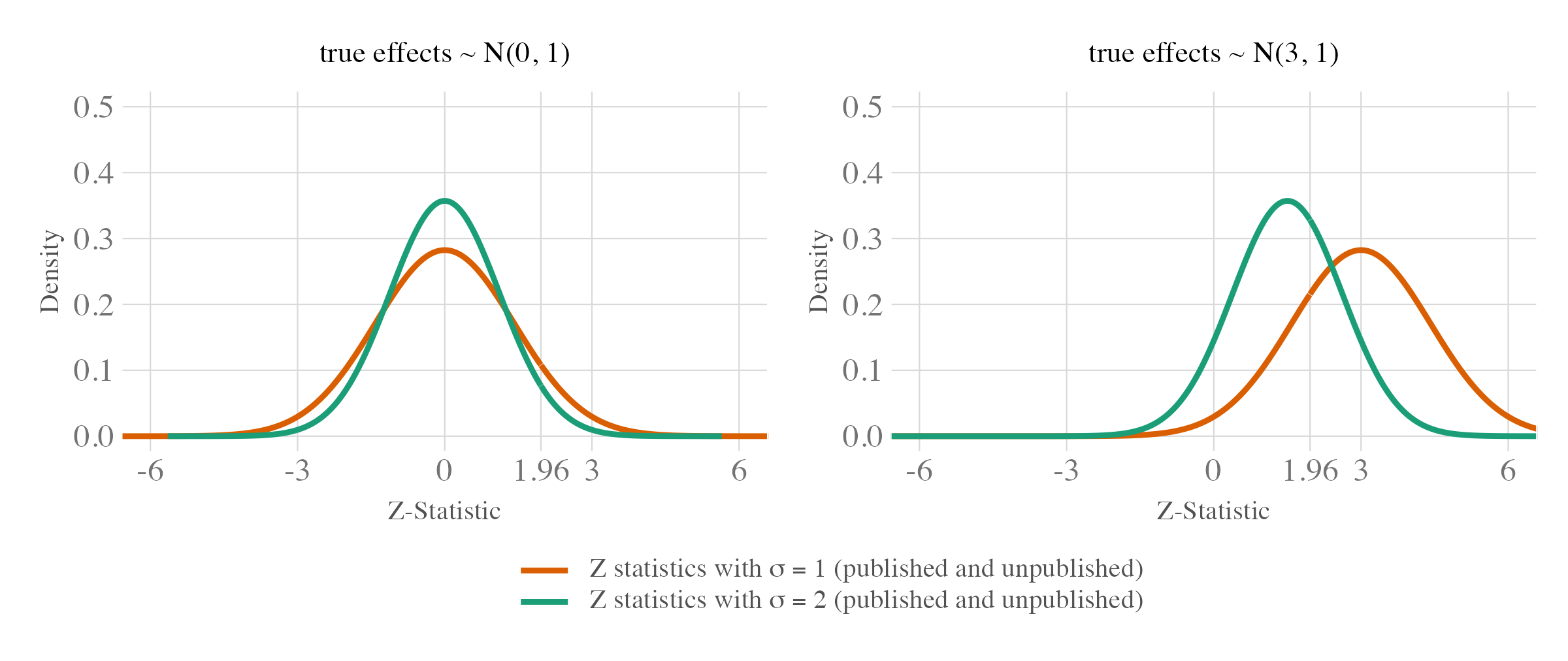}
    \label{fig:panel-b}
\end{figure}

Consider the density of the $Z$-statistic across studies (both published and unpublished), conditional on the standard error:
$$
f(z|\sigma) = \int \varphi\left( z - \tfrac{\theta}{\sigma} \right) d \mu(\theta)
$$
Figure \ref{fig: z unobserved} shows the distributions of the $Z$-statistics for the illustrative example. Larger standard errors are associated with \textit{smaller} $Z$-statistics so the distribution for $\sigma = 1$  is more dispersed than the distribution for $\sigma = 2$. 
\vspace{-2em}

The density of the $Z$-statistic conditional on the standard error and conditional on observability ($D_{i} = 1$) is then given by
$$
f(z|\sigma, D=1) = \frac{\bar{d}(z)}{E[\bar{d}(Z)|\sigma]}\cdot f(z|\sigma).
$$
\begin{figure}[H]

  \centering
    \caption{Distribution of $Z$-statistics -- published only}
    \label{fig: z pub}
    \includegraphics[width=\textwidth]{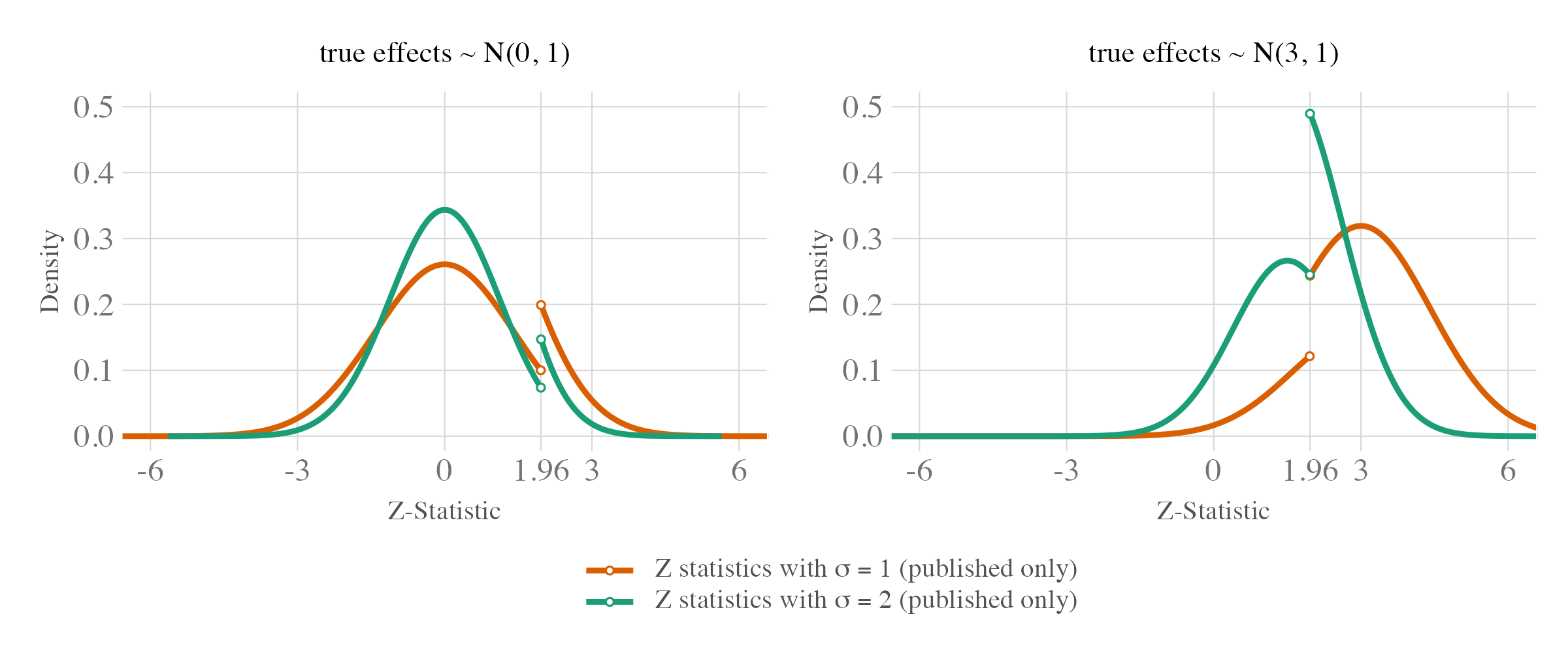}
    \label{fig:panel-c}
\end{figure}
\vspace{-2em}

Figure \ref{fig: z pub} shows the distributions of the $Z$-statistics in the presence of selectivity.  We assume that selectivity can be characterized by two values: the publication rate for significant estimates ($Z > 1.96$), which is normalized to 1, and the publication rate for insignificant estimates ($Z < 1.96$), which is assumed in this example to be 0.5. Both distributions therefore show a discontinuity with a doubling of the density at $Z = 1.96$.\footnote{This plot makes apparent the connection between the meta-studies approach method and the approach based on discontinuities in the distribution of p-values from Section \ref{subsec: elliott}. The latter effectively zooms in on the density close to the discontinuity. 
The meta-studies method uses all of the information in the density but requires the independence assumption; it also yields much richer results.} 

The key identification idea behind the meta-studies model in \cite{publicationbias2019} is that if we take the ratio of the density $f(z|\sigma, D=1)$ across these two values of $\sigma$, then the selection probability $\bar{d}(z)$ appears in both the numerator and the denominator so it drops out:
\begin{align}
\label{eqn: density ratio}
    \frac{f(z|\sigma = 1, D=1)}{f(z|\sigma = 2, D=1)} = & \frac{\frac{\bar{d}(z)}{E[\bar{d}(Z)|\sigma=1]}\cdot f(z|\sigma = 1)}{\frac{\bar{d}(z)}{E[\bar{d}(Z)|\sigma=2]}\cdot f(z|\sigma = 2)} \\
    = & const \cdot \frac{\int \varphi\left( z - \frac{\theta}{\sigma = 1} \right) d \mu(\theta)}{\int \varphi\left( z - \frac{\theta}{\sigma = 2} \right) d \mu(\theta)},
\end{align}
where $const.$ is a constant that does not depend on $z$.

\begin{figure}[H]

  \centering
    \caption{Ratio of observed $\sigma = 1$ distribution to $\sigma = 2$ distribution}
    \label{fig: density ratio}
    \includegraphics[width=\textwidth]{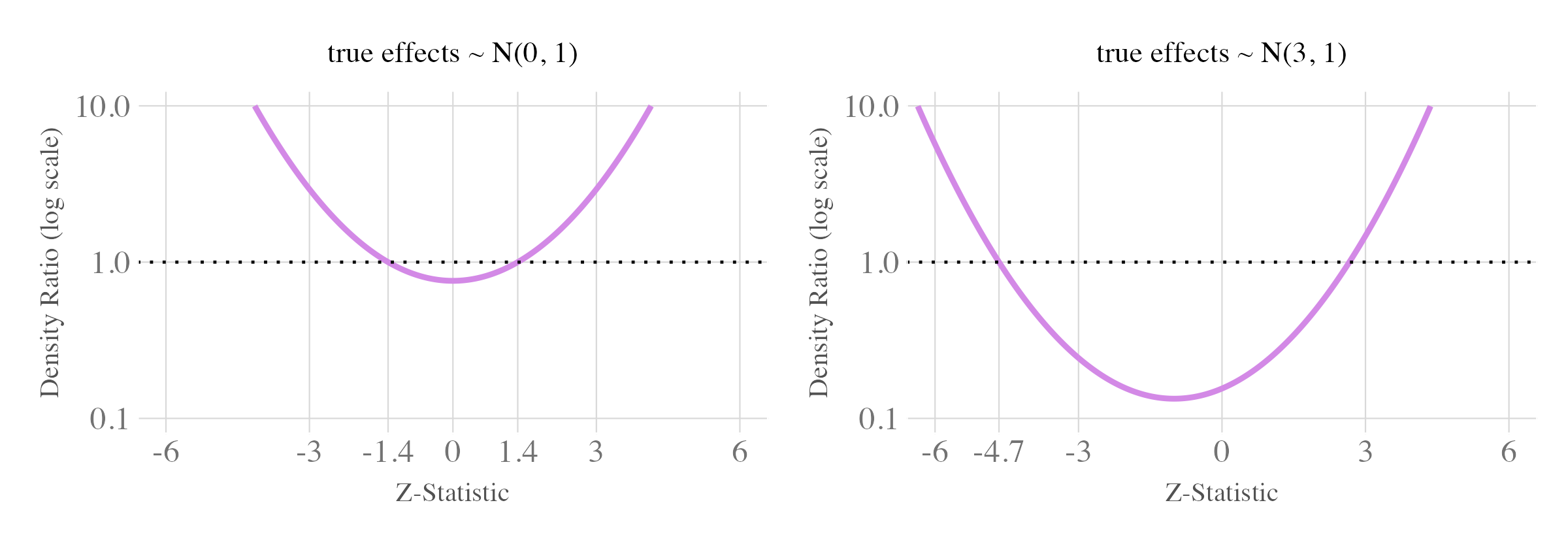}
    \label{fig:panel-d}
\end{figure}
Figure \ref{fig: density ratio} shows the ratio of the two densities. Notice the ratio of densities in Figure \ref{fig: z unobserved} is the same as the ratio of densities in Figure \ref{fig: z pub}. Although the densities in Figure \ref{fig: z pub} are distorted by selectivity, the \textit{ratio} of the densities is not distorted. This is because the density function for $\sigma = 1$ jumps by a factor of $\frac{1}{0.5} = 2$ at the cutoff, and the density function for $\sigma = 2$ also jumps by a factor of two at the cutoff. Having established how to construct functions of the data which are purged of selectivity, we next turn to how parameters of the distribution can be identified using these functions of the data. 

The relative distribution of $Z$-statistics---encapsulated by the density ratio---provides a ``signature'' which can be uses to identify the mean of the latent distribution. In Figure \ref{fig: density ratio}, when the latent mean is zero, the density ratio of the $\sigma=1$ distribution to the $\sigma=2$ distribution is greater than 1 when $z<-1.4$ or $z>1.4$. In contrast, when the latent mean is three, the density ratio of the  $\sigma=1$ distribution to the $\sigma=2$  distribution is less than $1$ for a much larger region (from $z>-4.7$ to $z<2.7$). Put simply, when the latent mean is positive, estimates from the precise distribution will more often be associated with large positive $Z$-statistics than estimates from the imprecise distribution. The larger the region where the density ratio is less than 1, the greater is the latent mean.
This is the ``reduced form'' signature of a positive latent mean in the presence of selectivity. We show this same signature exists in data from \cite{cohen2025disemployment} in \autoref{app: cg density ratio}.

The result of \cite{publicationbias2019} formalizes this intuition, showing the distribution $\mu$ of $\theta$ is uniquely pinned down by the ratio shown in \autoref{eqn: density ratio}. Once $\mu$ is identified, we can recover $\bar{d}$ (up to a multiplicative constant) from
$$
\bar{d}(z) = const. \cdot\frac{f(z|\sigma, D=1)}{f(z|\sigma)}.
$$

Since identification of both $\mu$ and $\bar{d}$ is non-parametric, they could in principle be estimated non-parametrically (e.g. using GMM, as discussed in the appendix of \citealt{publicationbias2019}). For moderate sample sizes (number of estimates in the meta-study), however, parametric estimation tends to be better behaved.

\subsection{Estimation and empirical examples}
\label{sec: selectivity empirical}

\subsubsection{Testing for selectivity}

We investigate the extent of selectivity across all of our running applications. The nature of selectivity depends upon researchers' priors and preferences about the size of the causal effect of interest. In all four of our applications, economic theory or nudge design predicts a \textit{positive} effect of the treatment (e.g. higher re-employment, longer unemployment, higher consumption, increase in outcome targeted by the nudge). The concern regarding selectivity is therefore that researchers and the publication process will select for positive significant effects (rather than positive and negative significant effects). 

We report results from three types of tests for selectivity discussed above.  For \cite{elliott2022detecting}, we test whether there is a discontinuity in the density of one-sided p-values and $Z$-statistics (a monotonic transformation of p-values) at $p = 0.025$ (corresponds to $Z = 1.96$) using a local polynomial density estimator \citep{cattaneo2020simple}. Estimates are shown in columns 4 and 5 of \autoref{tab:selectivity_detected}.\footnote{We show histograms of p-values and discuss sensitivity of our results to bandwidth selection in Appendix \ref{app:density_discont}.}
Second, for \cite{egger1997bias}, we estimate equation \ref{eqn: egger} and report $\beta$, where $\beta\neq0$ indicates the presence of selectivity. Third, we implement estimators using the model in \cite{publicationbias2019}, both assuming a latent $t$ distribution and making no assumption about the shape of the latent distribution. Given the discussion above about selectivity being specifically likely to generate positive significant effects, we report estimates for $P(\text{publish} | Z < 1.96) / P(\text{publish} | Z > 1.96)$.

\begin{table}[ht]
\centering
\caption{Testing for Selectivity in Four Empirical Applications}
\label{tab:selectivity_detected}
\small
\resizebox{\textwidth}{!}{%

\begin{tabular}{lccccc}
\toprule \toprule
\multicolumn{1}{c}{ } & \multicolumn{1}{c}{\makecell[c]{Regression\\Coefficient}} & \multicolumn{4}{c}{$P(\text{publish} | Z < 1.96) / P(\text{publish} | Z \geq 1.96)$} \\
\cmidrule(l{3pt}r{3pt}){2-2} \cmidrule(l{3pt}r{3pt}){3-6}
Study & Eggers & \makecell[c]{MLE\\Latent t} & \makecell[c]{MLE\\Nonparametric} & \makecell[c]{Density Discont.\\(p-value)} & \makecell[c]{Density Discont.\\(z-statistic)}\\
\midrule
Card, Kluve, \& Weber (2018) & 0.01 & 0.25 & 0.19 & 0.07 & 0.53\\
 & {}[-0.01, 0.04] & {}[0.00, 0.57] & {}[0.04, 0.34] & {}[0.00, 0.89] & {}[0.00, 1.38]\\
Cohen \& Ganong (2025) & 1.35 & 0.05 & 0.13 & 0.08 & 0.46\\
 & {}[0.92, 1.77] & {}[0.00, 0.11] & {}[0.01, 0.25] & {}[0.00, 1.27] & {}[0.00, 2.50]\\
Crosta et al. (2024)$^{1}$ & 0.47 & 0.28 & 0.12 & 0.27 & 0.8\\
 & {}[0.40, 0.54] & {}[0.01, 0.54] & {}[0.03, 0.21] & {}[0.00, 1.25] & {}[0.00, 24.03]\\
DellaVigna and Linos (2022)$^{2}$ & 2.71 & 0.12 &  & 0.63 & 0.73\\
 & {}[2.29, 3.13] & {}[0.00, 1.37] &  & {}[0.00, 1.93] & {}[0.00, 2.67]\\
\bottomrule
\end{tabular}

}

\begin{minipage}{\textwidth}
\footnotesize
\footnotesize
\textit{Notes:} This table re-analyzes four different meta-analyses: effect of active labor market policies, the effect of benefits on unemployment duration, the effect of nudges, the effect of cash transfers on consumption. Top row for each study is the estimate; the bottom row is a 95\% confidence interval. The first column implements the test proposed in \cite{egger1997bias}. The second and third columns show the relative publication probability using the test from \cite{publicationbias2019} for a $t$ distribution of latent estimates and a non-parametric distribution of latent estimates. The 4th column and 5th column show the ratio of density of $Z$-statistics (p-values) at the $z = 1.96$ ($p = 0.025$) cutoff; we implement suggested methods from \cite{cattaneo2022lpdensity} using quadratic fit.
$^1$For \cite{crosta2024unconditional}, MLE latent normal is used, not MLE latent $t$. We use a linear fit, not quadratic fit for density ratios.
$^2$For \cite{dellavigna2022rctstoscale}, we report relative publication probability for the academic journal subsample. We omit the MLE nonparametric estimate due to numerical issues.
\end{minipage}

\end{table}

We find widespread evidence of selectivity.  We find evidence of selectivity---in the sense that we can reject the null hypothesis of no selectivity---for three out of four applions using the tests from \cite{publicationbias2019} and the test from\cite{egger1997bias}.\footnote{\cite{card2018works} reports a finding of no selectivity based on the test in \cite{egger1997bias}; we replicate that finding,} The evidence using the density discontinuity approach from \cite{elliott2022detecting} is nuanced. In every case, the estimate implies that there is substantial selectivity at the cutoff but we are only once able to reject the null that the density ratio at the cutoff is 1.
The fact that we both estimate substantial selectivity and are unable to reject the null indicates the test for discontinuity is low-powered in our contexts.\footnote{\cite{elliott_power_2025}  show that in some contexts the power of discontinuity tests can be quite low. So a failure to reject the null of no selectivity should be interpreted as dispositive. We also find that testing for $f(p)$ increasing in $p$ is low-powered (another concern raised by \cite{elliott_power_2025} ).} Using the MLE method with a latent $t$  distribution, we find that the probability of reporting insignificant results relative to significant ones varies from 5\% to 28\% depending on the application. 

A finding of selectivity may be surprising in the context of RCTs (nudges and cash transfers) because trial pre-registration should in principle limit the selective reporting of findings. One possibility is that selectivity emerges not in terms of which studies report results, but which outcomes are reported or emphasized by study authors. \cite{dellavigna2022rctstoscale} find evidence of selective reporting in terms of the most significant $Z$-statistic (as opposed to all $Z$-statistics).
Although \cite{crosta2024unconditional} works to standardize outcomes as much as possible, there is still a range of consumption definitions which researchers can report. Sometimes food is excluded and sometimes it is included, sometimes durables are excluded and sometimes they are included, sometimes the time horizon is one month and sometimes it is one year.  Another possibility is that the identifying assumption of orthogonal estimates and standard errors is violated through power calculations, where there are some studies which are projected to have bigger effects for that subpopulation, but estimates for that subpopulation rely on smaller trials.\footnote{As one concrete example of this, many of the studies with the highest standard errors have framing which is tied to specific life circumstances (e.g. small business entrepreneurship or children). One might plausibly think that there are larger effects of cash in this subpopulation (indeed this is the motivation for targeting in the first place). } 

\subsubsection{Estimating the latent distribution of effects in presence of selectivity}
\label{sec: select correct}
We next ask whether selectivity affects the distribution of the latent estimates across our applications in  Table \ref{tab:corrected_mean_and_dispersion}. 

We use two approaches to correct for selectivity.\footnote{We do not use the correction approach from equation (\ref{eqn: egger}) because, as shown in Section \ref{sec: meta studies}, it is not identified; we do not use the correction approach from MLE nonparametric since we do not believe it is robust in our settings.} Our first approach analyzes highly powered studies using the approach proposed in  \cite{ioannidis2017power}. We estimate  $\hat{\alpha}$  from equation (\ref{eqn: egger}) and then take the subset of estimates with $\sigma_i<|\hat{\alpha}|/2.8$. \cite{ioannidis2017power} say that this approach filters to the studies which will capture the true latent mean (of $\hat{\alpha}$) with 80\% power. They then calculate a precision-weighted mean among this subset of ``highly-powered'' studies.  The highly-powered studies method (at least as implemented following the procedure in \citealt{ioannidis2017power}) only delivers a meaningful subset of studies to analyze in two out of four meta-analyses. The fact that there are very few studies which meet this criteria motivates alternative selection correction methods that use all the available data. In our second approach, we use the model from \citet{publicationbias2019}. 

Correcting for selectivity using this method reduces the mean of the latent distribution $\bar\theta$ in our four applications. More specifically, the ratio of the corrected mean to the simple mean ranges from 12\% to 21\% depending on the application.\footnote{Our estimates imply that the latent distribution for \cite{dellavigna2022rctstoscale} is Cauchy. The mean is undefined for a Cauchy distribution. In this case, $\bar\theta$ can be interpreted as the median.} 
In the context of selection oriented towards positive significant findings (which we believe is a good description of the four meta-analyses that we study), this is precisely what one would expect to find.  The distribution of published findings---which features many positive significant findings but has ``deleted'' some of the positive insignificant and negative insignifcant findings---will have too high of a mean. Selectivity can therefore lead an unsophisticated literature review to overstate the average causal effects of the policies studied.

Table \ref{tab:corrected_mean_and_dispersion} also reveals that the precision weighting and incorporating fat tails from Section \ref{sec:aggregation_no_x} provide a partial back door approach to selection correction. If the true data-generating process involves selectivity in favor of positive significant findings, then estimates with large standard errors will on average be more biased than estimates with small standard errors. Precision-weighting reduces the influence of these more-biased estimates. Our findings also reflect this pattern: the precision-weighted means (shown earlier in Table~\ref{tab:intercept_only_models} and repeated for convenience in Table \ref{tab:corrected_mean_and_dispersion}) are smaller than the simple means in all four applications.\footnote{This has the interesting implication that selection correction will \textit{not} always change the conclusion of a meta-analysis. As one concrete example,  \cite{crosta2024unconditional} focus on MLE estimates which precision weight each estimate (but do not correct for selectivity). In that case, the simple mean for the estimated average effect of \$1 on monthly consumption is \$0.12, the estimated effect when using MLE without correcting for selectivity is \$0.03, and the estimated effect of using the MLE with correcting for selectivity of \$0.02. So even without correcting for selectivity, precision weighting generates similar conclusions about the mean of the effect distribution to a selection correction.} This is what we would expect to see in the presence of selectivity; in the absence of selectivity, the precision-weighted means could plausibly be larger or smaller than the simple means. 

Last but not least, it is useful to note that the empirical estimates from the MLE latent $t$ model falsify an identifying assumption from the highly-powered approach. The problem, which we describe in the theory section, is that if there are some latent effects $\theta_i$ which are zero, then no standard error will be sufficiently small as to qualify as ``highly powered.'' Across all applications, the distributions implied by the estimates for $\bar\theta$ and  $\tau$ imply that there exists a meaningful distribution of latent effects in the neighborhood of zero. 

\begin{table}[p]
    \centering
    \caption{Estimating Mean Average Effects and Dispersion in Presence of Selectivity}
    \resizebox{\textwidth}{!}{%
    
\begin{tabular}{lccc|cc}
\toprule \toprule
\multicolumn{4}{c}{ } & \multicolumn{2}{c}{\makecell[c]{Model latent distribution\\with selectivity}} \\
\cmidrule(l{3pt}r{3pt}){5-6}
Hyper-parameter & \makecell[c]{Simple\\Mean} & \makecell[c]{Prec-wtd.\\Mean (eqn 2)} & \makecell[c]{MLE\\Latent t} & \makecell[c]{Highly\\Powered} & \makecell[c]{MLE\\Latent t}\\
\midrule
\addlinespace[0.3em]
\multicolumn{6}{l}{\textit{\makebox[0pt][l]{Card, Kluve, \& Weber (2018) -- Effect of Active Labor Market Program on Employment}}}\\

\midrule\hspace{1em}$\bar{\theta}$ (Mean) & \makecell{0.083\\(0.008)} & \makecell{0.040\\(0.007)} & \makecell{0.057\\(0.020)} & \makecell{0.040\\(0.001)} & \makecell{0.010\\(0.016)}\\
\hspace{1em}\% of Simple Mean &  & 48\% & 69\% & 48\% & 12\%\\
\hspace{1em}$\tau$ (Between-Study Dispersion) &  & \makecell{0.000\\(2.615)} & \makecell{0.064\\(0.023)} &  & \makecell{0.030\\(0.014)}\\
\hspace{1em}$\nu$ (Degrees of Freedom) &  &  & \makecell{4.11\\(4.13)} &  & \makecell{1.49\\(0.37)}\\
\addlinespace[0.3em]
\multicolumn{6}{l}{\textit{\makebox[0pt][l]{Cohen \& Ganong (2025) -- Elasticity of Unemp Duration w.r.t. Unemp Benefits}}}\\

\midrule\hspace{1em}$\bar{\theta}$ (Mean) & \makecell{0.513\\(0.050)} & \makecell{0.448\\(0.044)} & \makecell{0.382\\(0.046)} & \makecell{0.350\\(0.003)} & \makecell{0.107\\(0.089)}\\
\hspace{1em}\% of Simple Mean &  & 87\% & 74\% & 68\% & 21\%\\
\hspace{1em}$\tau$ (Between-Study Dispersion) &  & \makecell{0.345\\(0.059)} & \makecell{0.240\\(0.032)} &  & \makecell{0.244\\(0.053)}\\
\hspace{1em}$\nu$ (Degrees of Freedom) &  &  & \makecell{4.04\\(1.69)} &  & \makecell{2.89\\(1.14)}\\
\addlinespace[0.3em]
\multicolumn{6}{l}{\textit{\makebox[0pt][l]{Crosta et al. (2024)$^1$ -- Effect of \$1 Transfer on Monthly Consumption}}}\\

\midrule\hspace{1em}$\bar{\theta}$ (Mean) & \makecell{0.127\\(0.049)} & \makecell{0.011\\(0.004)} & \makecell{0.028\\(0.004)} &  & \makecell{0.018\\(0.005)}\\
\hspace{1em}\% of Simple Mean &  & 9\% & 22\% &  & 14\%\\
\hspace{1em}$\tau$ (Between-Study Dispersion) &  & \makecell{0.000\\(2.212)} & \makecell{0.020\\(0.003)} &  & \makecell{0.019\\(0.003)}\\
\addlinespace[0.3em]
\multicolumn{6}{l}{\textit{\makebox[0pt][l]{DellaVigna and Linos (2022) -- Effect of Nudge on Targeted Outcome}}}\\

\midrule\hspace{1em}$\bar{\theta}$ (Mean) & \makecell{0.031\\(0.004)} & \makecell{0.026\\(0.003)} & \makecell{0.004\\(0.001)} &  & \makecell{0.005\\(0.026)}\\
\hspace{1em}\% of Simple Mean &  & 84\% & 13\% &  & 16\%\\
\hspace{1em}$\tau$ (Between-Study Dispersion) &  & \makecell{0.070\\(0.002)} & \makecell{0.005\\(0.001)} &  & \makecell{0.010\\(0.122)}\\
\hspace{1em}$\nu$ (Degrees of Freedom) &  &  & \makecell{1.00\\(0.08)} &  & \makecell{1.28\\(6.84)}\\
\bottomrule
\end{tabular}
}
    \label{tab:corrected_mean_and_dispersion}
    \begin{minipage}{\textwidth}

\textit{Notes:}
\footnotesize
This table re-analyzes four different meta-analyses. The first column shows the simple mean. The second column shows estimates from equation \ref{eqn: prec wtd mean} (which weights observations by $\frac{1}{\tau^2 + \sigma_i^2}$). The third column assumes a latent $t$ distribution and uses Maximum Likelihood Estimation. The fourth column implements \citet{stanley2017finding} (which weights observations by $\frac{1}{\sigma_i^2}$). The number of highly-powered estimates is, in descending order of rows, 31/169, 45/71, 2/75, and 6/315. We omit the highly-powered estimates for \cite{crosta2024unconditional} and \cite{dellavigna2022rctstoscale} since there are too few studies. The fifth column implementing \citet{publicationbias2019} with a latent t distribution is a direct counterpart to the third column. \\
$^1$For \cite{crosta2024unconditional}, MLE latent normal is used, not MLE latent t. MLE latent $t$ estimates are similar with estimated $df$ very large.
\end{minipage}
\end{table}
\noindent

\section{A model with selectivity and covariates}
\label{sec:unified model} 
\subsection{Theory}
\label{sec:unified model theory}
Many meta-analyses want to combine covariates and selectivity -- that is, they want a predictive model for new problem instances characterized by covariates that also accounts for selectivity. 

Our discussion in \autoref{sec:selectivity theory} sidestepped study level covariates. Covariates can, however, be incorporated either by partitioning the observed studies and then separately implementing any of the selectivity methods\footnote{See for example \cite{brodeur2020methods}, who consider p-curves separately depending on the methods employed in different empirical studies.} or by directly building covariates into the models used. The former approach requires sample sizes larger than those present in our empirical examples so we will focus on the latter.

\paragraph{Likelihood}
Recall the assumptions stated in \autoref{sec:setup}, where we assumed a sampling distribution of
$\hat \theta_i |\theta_i, \sigma_i^2,X_{i} \sim N(\theta_i, \sigma_i^2),$ and selection based solely on $Z$-statistics $Z_i = \hat \theta_i/\sigma_i$ according to 
$P(D_i =1 | \hat \theta_i, \sigma_i, X_i) = \bar{d}(Z_{i}).$

Suppose now additionally, as in the meta-regression model (equation \ref{eq:metaregression}), that
$$
{\theta}_i| X_i =x \sim N(x\cdot \beta, \tau^2).
$$
Then, again as before, $\hat{\theta}_{i} |X_{i}, \sigma_{i} \sim N(x \cdot \beta, \tau^{2} + \sigma_{i}^{2}),$ and thus
$$Z_{i} |X_{i}, \sigma_{i} \sim N(x \cdot \beta / \sigma_i, \tau^{2}/\sigma_{i}^{2} + 1).$$
We get the conditional likelihood of published $Z$-statistics by Bayes rule (this implicitly also conditions on model parameters $\beta,\tau^{2}$ and $\bar{d}$), 
$$
f(Z_{i} |X_{i}, \sigma_{i}, D_i = 1)
= \underbrace{ \frac{1}{\sqrt{\frac{\tau^{2}}{\sigma_i^{2}}+1 }} \varphi\left(\frac{Z_{i}-X_{i}\cdot\beta}{\sqrt{\frac{\tau^{2}}{\sigma_i^{2}}+1 }} \right) }_{ f(Z_i|X_i, \sigma_i) }\cdot \underbrace{ \frac{\bar{d}(Z_{i})}{E[\bar{d}(Z)| X_{i}, \sigma_{i}]} }_{ \frac{P(D_i=1 |Z_i, \sigma_i,X_i)}{P(D_i=1|X_i, \sigma_i)} }.
$$
If we consider a parametric specification for $\bar{d}$ of the form $\bar d(z) = \mathbf{1}(z>1.96) + \gamma\cdot\mathbf{1}(z \leq 1.96)$ (selection for positive significant effects), then the denominator in the last fraction equals
$$E[\bar{d}(Z)| X_{i}, \sigma_{i}] = 1 + (\gamma-1) \cdot\Phi\left(\frac{1.96-X_{i}\cdot\beta}{\sqrt{\tfrac{\tau^{2}}{\sigma_i^{2}}+1 }} \right). $$

\paragraph{Prior and posterior}
We can combine this likelihood with a prior for the hyper-parameters $\beta,\tau^{2}$ and $\gamma$ to get a complete hierarchical Bayes specification. We can then use Hamiltonian Monte Carlo (as implemented in Stan) to sample from the posterior. Using the posterior mean $\bar{\beta}$ of $\beta$, we can furthermore make predictions for new instances, via $X_{0}\cdot\bar{\beta}$.

For our empirical application, we assume a normal prior for $\beta$, $\beta \sim N(0, \Sigma)$, as in the Ridge regression models discussed earlier, a half-normal prior for $\tau^2$ with large variance, and a lognormal prior for $\gamma$, $\log(\gamma) \sim N(0,1)$.

\subsection{Empirical example of a model with selectivity and covariates}
\label{sec: unified model empirical}

The meta-analyses that want to combine covariates and selectivity 
include three out of four of our applications: \cite{cohen2025disemployment}, \cite{dellavigna2022rctstoscale}, \cite{card2018works}.\footnote{The first article studies the combination using Bayesian Model Averaging, the second article studies the combination by using a complex simulation procedure (Table VI), and the third article is a priori interested in both questions, but did not find evidence of selectivity and therefore views covariate-only models as sufficient.} \cite{cohen2025disemployment} in particular is interested in the optimal replacement rate for unemployment insurance. It finds that higher baseline replacement rates ($X_i$) are associated with a higher elasticity of unemployment duration with respect to benefits ($\hat\theta_i$). In the model they study, solving for the optimal replacement rate requires predicting how the elasticity varies with baseline replacement rate.

\begin{figure}[h]
  \centering
  \caption{Correcting Covariate Coefficients for Selectivity \citep{cohen2025disemployment}}
  \includegraphics[width=\textwidth]{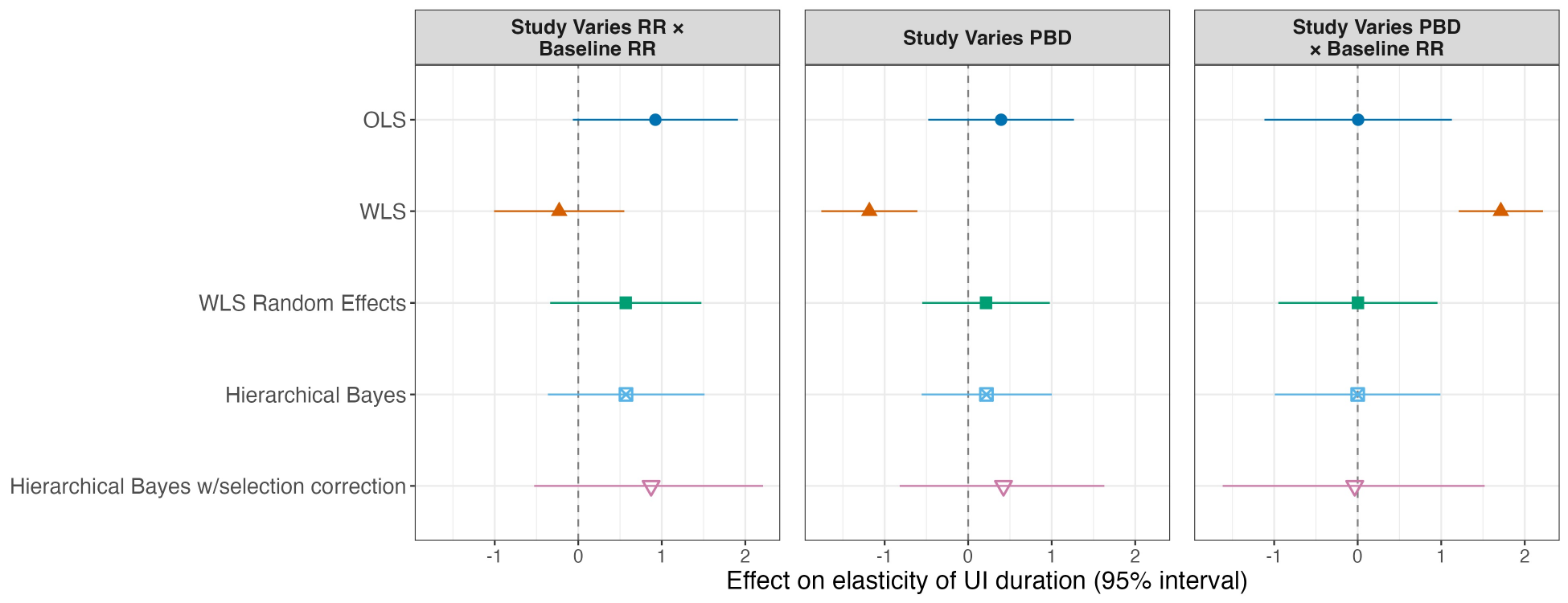}
  \label{fig: unified model forest}
  \begin{minipage}{\textwidth}
  \scriptsize
    \textit{Notes:} This figure shows the coefficients and the 90\% confidence interval of the coefficients for
    $\hat{\theta}_i = \alpha + \beta_1\,\text{Baseline RR}\times\mathbf{1}(\text{Study Varies RR}) + \beta_2\,\text{Baseline RR}\times\mathbf{1}(\text{Study Varies PBD}) + \beta_3\,\mathbf{1}(\text{Study Varies PBD}) + \epsilon_i$. RR stands for replacement rate; PBD stands for potential benefit duration. The data is from \cite{cohen2025disemployment} and the circles represent regression models; squares represent hierarchical Bayes models. From the top, the first model does an equal weight regression; the second model does WLS with $w_i = \frac{1}{\sigma_i^2}$; the third model does WLS with $w_i = \frac{1}{\sigma_i^2 + \tau^2}$ where $\tau^2$ is the posterior mean from the fourth model; the fourth model estimates a hierarchical Bayes model with diffuse priors and no selection correction; the fifth model estimates a hierarchical Bayes model with diffuse priors and a selection correction. When hierarchical Bayes is used, the posterior mean of the coefficients is used as the estimate.
    \end{minipage}
\end{figure}%

\autoref{fig: unified model forest} builds up to the model described in \autoref{sec:unified model theory}. We start with a multivariate OLS regression. \cite{cohen2025disemployment} find that estimates for the elasticity of unemployment duration with respect to  replacement rate are larger in contexts with a larger baseline replacement rate. We replicate this finding by running the minimal regression needed to investigate this interaction effect:
\begin{align*}
    \hat{\theta}_i = &\alpha + \beta_1\,\text{Baseline RR}\times \mathbf{1}(\text{Study Varies RR}) \\ &+ \beta_2\,\text{Baseline RR}\times \mathbf{1}(\text{Study Varies PBD})
    + \beta_3\,\mathbf{1}(\text{Study Varies PBD})  + \epsilon_i
\end{align*}
where ``RR'' is replacement rate and ``PBD'' is potential benefit duration.

Then, we account for differing precision by switching to weighted least squares with weights $w_i = \frac{1}{\sigma^2}$. This leads to large changes in the coefficients and in particular shows no difference in elasticity with respect to RR for changes in the baseline RR. Accounting further for heterogeneity through weighting by $w_i = \frac{1}{\sigma_i^2 + \tau^2}$ (with the $\tau^2$ obtained by estimating the hierarchical Bayes model without a selection correction) leads to more equal weights across studies; generally this will lead to coefficent estimates that fall between those of OLS and WLS with $w_i = \frac{1}{\sigma^2}$.\footnote{In a multivariate regression this is not a rule -- as demonstrated by this relationship failing with covariate ``PBD x baseline RR'' -- because changes in the values of other coefficients will change the coefficients of collinear covariates. The hierarchical Bayes model without a selection correction almost exactly replicates this WLS with $w_i = \frac{1}{\sigma_i^2 + \tau^2}$.} 
We get identical results by running the hierarchical Bayes model without a selection correction. 

The final model accounts for selection. We find that when the replacement rate increases by 10\%, the elasticity increases by 8.6\%. The hierarchical Bayes model without a selection correction would estimate an elasticity increase of 5.6\%. The selection-corrected model implies an optimal replacement rate of 32\% in the US context studied by \cite{cohen2025disemployment} if all other factors except the slope of the elasticity vs. baseline replacement rate are held constant. The uncorrected model implies an optimal replacement rate of 29\%.

\section{Cookbook}
\label{sec:cookbook}

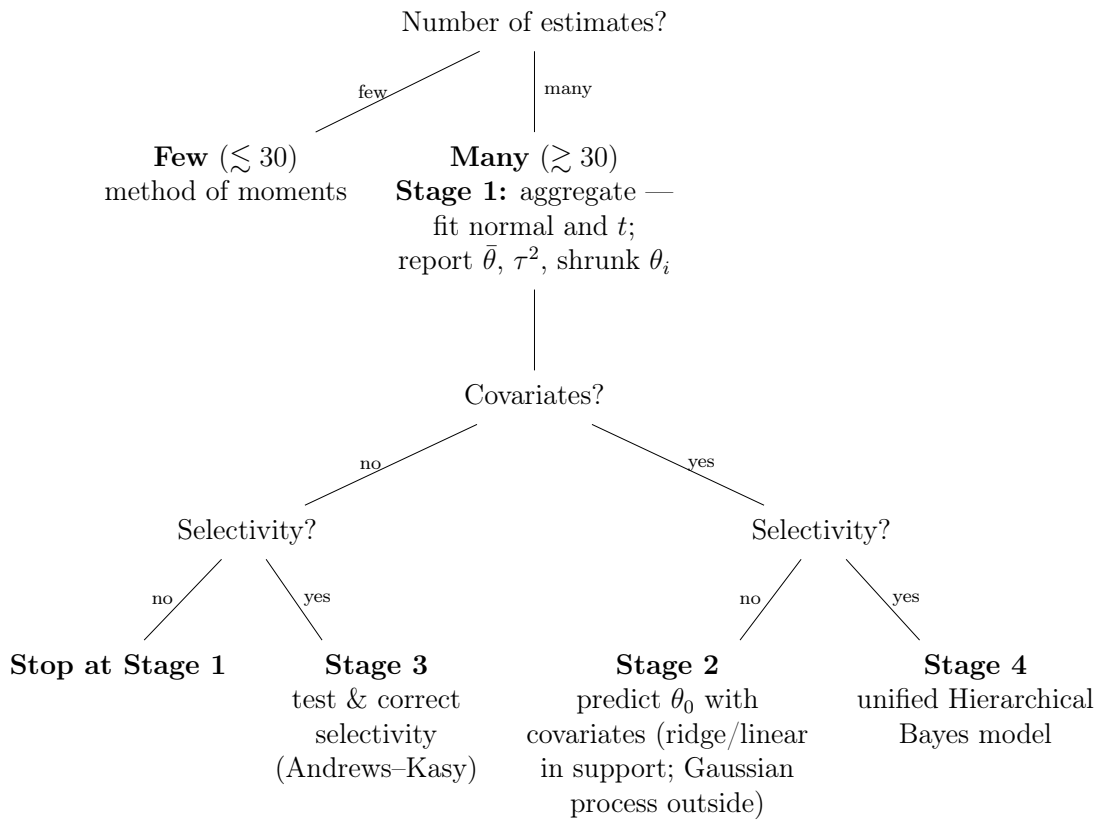
\begin{figure}[h!]
  \centering
  \resizebox{\textwidth}{!}{%
  \begin{forest}
  for tree={
    align=center,
    l sep=34pt,
    s sep=12pt,
    edge={-},
  }
  [{Number of estimates?}, calign=child, calign child=2
    [{\textbf{Few} ($\lesssim 30$)\\method of moments}, edge label={node[midway,left,font=\scriptsize]{few}}]
    [{\textbf{Many} ($\gtrsim 30$)\\\textbf{Stage 1:} aggregate ---\\fit normal and $t$;\\report $\bar\theta$, $\tau^2$, shrunk $\theta_i$}, edge label={node[midway,right,font=\scriptsize]{many}}
      [{Covariates?}
        [{Selectivity?}, edge label={node[midway,left,font=\scriptsize]{no}}
          [{\textbf{Stop at Stage 1}}, edge label={node[midway,left,font=\scriptsize]{no}}]
          [{\textbf{Stage 3}\\test \& correct\\selectivity\\(Andrews--Kasy)}, edge label={node[midway,right,font=\scriptsize]{yes}}]
        ]
        [{Selectivity?}, edge label={node[midway,right,font=\scriptsize]{yes}}
          [{\textbf{Stage 2}\\predict $\theta_0$ with\\covariates (ridge/linear\\in support; Gaussian\\process outside)}, edge label={node[midway,left,font=\scriptsize]{no}}]
          [{\textbf{Stage 4}\\unified Hierarchical\\Bayes model}, edge label={node[midway,right,font=\scriptsize]{yes}}]
        ]
      ]
    ]
  ]
  \end{forest}}
  \caption{Decision tree for the cookbook pipeline.}
  \label{fig:cookbook_alt_tree}
\end{figure}

This section collects our recommendations into a single pipeline. The pipeline runs in six stages, summarized in \autoref{fig:cookbook_alt_tree}: (0) scoping and data, (1) aggregation, (2) prediction with covariates,  (3) selectivity, (4) the combined model, and (5) decisions.

\paragraph{Stage 0 -- Scope, collect, assess credibility.} Fix the estimand $\theta_0$, the search procedure, and the coding rules and write down the proposed procedure. This matters because apparent selectivity can be an artifact of which \emph{outcome} was coded rather than which study was published, as in the cash-transfer application (\autoref{sec: selectivity empirical}). AI tools make extraction cheap; see \cite{cook2026aimeta} for pitfalls.  The methods in this paper make sense when there are at least three prior estimates, so you should stop if you cannot find three prior estimates. 

At this stage, also decide whether to collect covariates and which covariates to collect, since it determines the model in Stage 3. Do not discard lower-quality studies. Instead, collect measures of study quality as part of the covariates (and return to this question in step 3). This could be something relatively objective like the research design (e.g. DiD, RD, RCT), a subjective assessment of the credibility of the paper's identifying assumptions, or both.\footnote{Although we recommend this practice, we did not implement it as part of our empirical examples in this article. That is because the notion of study quality is difficult to treat in a consistent way across very different applications. We advocate for using the meta-analyst's assessment of quality (rather than say journal quality, which could be related to selectivity  \cite{forking2021}.} 

\paragraph{Stage 1 -- Aggregate without covariates or selection.}

Ask whether the dispersion of the $\hat\theta_i$ exceeds what their standard errors imply (\autoref{eqn: tau mom}). Report a precision-weighted mean (\autoref{eqn: prec wtd mean}), not a simple mean, and form shrunken study-level estimates (\autoref{eqn: study level correction}).  Whether you proceed to the subsequent steps depends on the number of estimates you have collected. The subsequent steps need a soft floor of roughly thirty estimates (\citealt{irsova2024meta}).\footnote{We take the upper bound from \cite{irsova2024meta} who suggest at least 10 primary studies with 30 total estimates.} 

Next, if you have enough studies, fit and compare three latent distributions (\autoref{tab:intercept_only_models}): normal , $t$ for heavy tails, and nonparametric via nonparametric maximum with Tweedie posterior means (\autoref{eqn:tweedie}). Read off whether tails are heavy (small $\nu$) and whether $\mu$ is skewed -- these choices materially move the estimated mean and shrink $\tau^2$. 

\paragraph{Stage 2 -- Predict with covariates.}
Standardize covariates by their empirical standard deviation. For a context \emph{inside} the support of the covariates, use a meta-regression estimated as random effects WLS (\autoref{eq:metaregression}), equivalently the Gaussian-process posterior mean (\autoref{eq:GPposterior}); the equivalence makes explicit that the prediction is a transparent weighted average of the $\hat\theta_i$. Although we emphasize the importance of distributions that allow for fat tails in stage 1, this becomes less important once covariates are available; this is because it is likely that the distribution is assumed to be normal \textit{after conditioning on the prediction $X_i\beta$ or $\bar{\theta}(X_i)$}, where the covariates $x$ can explain why a specific estimate $\hat\theta_i$ is an outlier.

For extrapolation \emph{outside} the support, use a Gaussian process prior with a squared-exponential kernel (equation \ref{eqn: se kernel}). Make an active decision about whether to use a long or short length scale. The short length scale will lead to a much larger posterior standard deviation. This choice should be informed by your economic model of how outcomes are determined (see discussion in Section \ref{sec: predictive models with x} as an example). 

A second use of covariates is to incorporate study quality into the prediction problem. If you are interested in the likely effects of a policy, do the prediction exercise where $\theta_0$ is defined as the highest quality strata of study.\footnote{As one example, \cite{cohen2025disemployment} report predictions for a regression discontinuity design when doing policy analysis.} The researcher's choice of prior will decide how much consideration to give to the studies from the lower-quality strata. Another option is to do the pipeline for just the high-quality studies; this is implicitly what is already done in the context of the small-sample meta-analysis (Tables \ref{tab:pollution_elasticities}, \ref{tab:disclosure_laws}, and \ref{tab: paid leave}). 

For any such prediction model for $\theta_i$ given $X_i$, it is important to keep in mind that (i) the resulting coefficients are not causal and (ii) any coefficients in a multivariate regression capture variation \emph{holding all other regressors constant}.

\paragraph{Stage 3 -- Test and correct for selectivity.}
Assess whether you believe that equations (\ref{eqn: theta se indep}) and (\ref{eqn: select only on z}) are satisfied.\footnote{As an example of where they may not be satisfied, \cite{dube2024minwage} find in the context of minimum wage own-wage elasticities, ``credible designs may use less minimum wage variation, resulting in lower precision, but they may also have less bias''. In randomized control trials, researchers have priors about the size of the causal effect they use in power calculations to choose sample sizes, leading to larger expected causal effects being systematically less precise. See \cite{chen2025precision} for an example of a model with co-dependence (but is not about meta-analysis)} If they are not satisfied, then you are limited to the a $p$-curve density-discontinuity test \citep{elliott2022detecting}. However, it is important to know that discontinuity tests can have low power \citep{elliott_power_2025}, so non-rejection may not imply the absence of selectivity. If equations (\ref{eqn: theta se indep}) and (\ref{eqn: select only on z})  are satisfied, you can also use the \cite{egger1997bias} meta-regression, and the relative publication probability from \cite{publicationbias2019}. 

For \emph{correction}, we emphasize that two commonly used methods are invalid. The Egger intercept is a valid test but not a valid bias correction -- under selection, $E[Z_i\mid\sigma_i]$ is nonlinear in $1/\sigma_i$ and linear extrapolation to $\sigma_i=0$ can return almost any value (\autoref{fig:nonlinear_metaregression}). The highly-powered-studies estimator \citep{stanley2017finding, ioannidis2017power} is valid only when such studies exist; when $\mu$ has mass near zero they may not, as in two of our four applications (\autoref{tab:corrected_mean_and_dispersion}). Our workhorse is the \cite{publicationbias2019} model, which jointly estimates $\mu$ and the selection function from the joint distribution of $\hat\theta_i$ and $\sigma_i$ (\autoref{sec: ak theory}).

One note of caution regarding the correction method of \cite{publicationbias2019} is that it can deliver results which are fragile to assumptions about the distribution of latent effects as well as the chosen thresholds for selective reporting. Although we report only one selectivity-corrected MLE model here for brevity, we recommend that authors assess sensitivity to both assumptions about the distribution of latent effects as well as chosen thresholds. See \cite{cohen2025disemployment} for an example of such sensitivity tests.

\paragraph{Stage 4 -- Combine covariates and selectivity.}
Build covariates into the selection model: the linear meta-regression $\theta_i\mid X_i\sim N(X_i\cdot\beta,\tau^2)$ combined with a step-function $\bar d$, estimated by Hamiltonian Monte Carlo in Stan. We recommend doing this with the \texttt{\href{https://github.com/wwiecek/baggr/tree/master}{baggr}} package.
Build it up in visible steps -- OLS, WLS, random effects WLS, hierarchical Bayes without and then with selection (\autoref{fig: unified model forest}) -- so the reader sees what each ingredient does.  

\paragraph{Optional Stage 5 -- From estimates to decisions.}
For a binary choice, carry the prediction through to the rule rather than stopping at a coefficient. Under welfare $U(A_i,\theta_i)=A_i\theta_i$, choose $A_i=\mathbf 1\!\left(E[\theta_i\mid \hat\theta_i,\sigma_i,X_i]\ge 0\right)$ for an evaluated site and $A_0=\mathbf 1\!\left(E[\theta_0\mid X_0,\hat\theta_{1:n}]\ge 0\right)$ for a new context (equation \autoref{eqn: action}).

\bibliographystyle{apalike}
\bibliography{library, library_appendix}

\newpage
\appendix

\clearpage
\phantomsection
\begin{center}
  {\large\bfseries Appendixes -- for online publication only}
\end{center}
\addcontentsline{toc}{section}{Appendixes -- for online publication only}
\vspace{1em}
\beginappendixtoc
\clearpage

\section{Derivation of Tweedie's formula}
\label{app:heavy_tails}
This appendix gives the derivation of Tweedie's formula, equation~\eqref{eqn:tweedie}, summarized in Section~\ref{ssec:nonparametric_eb_summary}. To describe the argument without excessive notation, let us consider the normalized parameters $\omega_{i} = {\theta_{i}} / {\sigma_{i}}$. Denote the distribution of the normalized parameters $\omega_i$ by $\nu$. Then the marginal density of the $Z$-statistics $Z_i = \hat \theta_i / \sigma_i$ (absent p-hacking or publication bias) is given as follows:
\begin{align}
Z_i|\omega_i & \sim N(\omega_i, 1)\nonumber \\
\omega_{i} & \sim \nu\nonumber\\
F(z) &= P(Z_i \leq z)  =\int \Phi(z - \omega) d \nu(\omega)\nonumber \\
f(z) &= \frac{ \partial P(Z_{i} \leq z)}{ \partial z}  = \int \varphi(z - \omega) d \nu(\omega)
\label{eq:convolutiondensity}
\end{align}
Mathematically, we say that the distribution of $Z$ is given by the convolution of $\nu$ with the standard normal density $\varphi$. Note that the distribution of $Z$ is directly identified, and we can estimate the density $f(z)$ using any of a number of methods. We can correspondingly also estimate the distribution $\nu$ of $\omega$. This is known as deconvolution, cf. \citeapp{meister2009}. Perhaps surprisingly, non-parametric maximum likelihood for $\nu$ also yields a well behaved estimator in this setting \citepapp{koenker2014convex}.

An interesting property of the normal distribution allows us to relate the density $f(z)$ to the empirical Bayes estimate $E[\omega_{i}|Z_{i}=z]$ of $\omega_{i}$. Since $\varphi(z) = (2\pi)^{-1/2} \cdot \exp\left( -\frac{z^{2}}{2} \right)$, we have that $\varphi'(z) = -z \cdot \varphi(z)$. This property allows us to derive Tweedie's formula \citepapp{efron2011tweedie}:
$$
\frac{ \partial \log(f(z)) }{ \partial z }
=\frac{\int \varphi'(z - \omega) d \nu(\omega)}{\int \varphi(z - \omega) d \nu(\omega)}
=-\frac{\int (z-\omega)\varphi(z - \omega) d \nu(\omega)}{\int \varphi(z - \omega) d \nu(\omega)}
=-E[Z_i-\omega_i | Z_i=z],
$$
and thus
\begin{equation}
   \label{eqn:tweedie_normalized}
E[\omega_{i}|Z_{i}=z] = z +\frac{ \partial \log(f(z)) }{ \partial z }.
\end{equation}
We thus obtain an expression for the empirical Bayes estimator which applies for any distribution of $\omega$.

Generalizing this argument to the setting with covariates and to the un-normalized estimates $\hat{\theta}_{i}$, we recover equation~\eqref{eqn:tweedie}, and more generally
$$
E\left[ \theta_{i}| \hat{\theta}_{i}, \sigma_{i}, X \right] = \hat{\theta}_{i} + \sigma_{i}^{2} \cdot \frac{ \partial }{ \partial \hat \theta } \log(f(\hat{\theta}_{i} | \sigma_{i}, X_{i})).
$$
Here $f(\hat{\theta} | \sigma, X)$ is the conditional density. Both the density $f(z)$ and the conditional density $f(\hat{\theta} | \sigma, X)$ can, in principle, be estimated parametrically or non-parametrically, from the observed empirical distribution of $(\hat{\theta}_{i}, \sigma_{i},X_{i})$. In practice, however, estimation in the tails and estimation of conditional densities might be poorly behaved.

\clearpage

\section{``Small'' meta-analysis}

\subsection{Elasticity of Housing Prices with Respect to Air Pollution}
\label{sec: appendix housing prices}
Panel A: estimates use nonattainment designation as an instrumental variable for pollution, with change in log house prices as the outcome. \citetapp{chay2005daq} regressor is Total Suspended Particles (PM$_{100}$): flexible-controls specification, Table~5 Panel~A col.~3. \citetapp{bento2015} regressor is PM$_{10}$: elasticity stated in Section~V.C, derived from the overidentified IV coefficient of 1.33 (SE\,$=\,$0.50) on $\Delta$PM$_{10}$ in Table~3. \citetapp{grainger2012} regressor is PM$_{10}$: elasticity derived from the IV coefficient of -0.0145 (SE\, $=\,$ (0.0058) in Table~5 Panel~A col.~1. Converted to elasticity using the 1990 PM$_{10}$ mean reported in Table~1.

Panel D reports \citetapp{sager2025} regressor is PM$_{2.5}$: preferred M1DiD-IV specification, Table~3 Panel~A. SE of implied elasticity computed as $\widehat{\mathrm{SE}}_{\varepsilon}=\widehat{\mathrm{SE}}_{\hat{\theta}}
\times\bar{c}$, where $\bar{c}$ is the mean pollutant concentration
implied by the reported elasticity.

\subsection{Effect of Pay Transparency Laws on the Gender Pay Gap}
\label{sec: appendix pay transparency}
Panel A: estimates are percent effect relative to conditional pre-policy gender pay gap. \citetapp{baker2023} average over four specification with gender pay gap as dependent variable: Table~4 col.~1-4. Converted to percent effect using the conditional gender pay gap of 6 percent at the time of the reform (Figure~2). \citetapp{bennedsen2022} underlying coefficient comes from Table~4 col.~2. Converted to percent effect from their statement that the estimate represented a 13\% reduction in the conditional gender pay gap. \citetapp{gulyas2023} estimate derived from their statement that they can rule of 0.4pp effect at the 95\% percent confidence level.

Panel D reports \citetapp{blundell2025} treated times post regressor: Table~3 col.~1.

\subsection{Effect of Paid Family Leave on Mother's Earnings}
\label{sec: appendix paid leave}
All estimates are the effect of paid-leave \emph{availability} (the intent-to-treat or reduced-form effect of the policy, not of individual take-up of leave), expressed as a percent of the control-group mean. Studies that report an availability or difference-in-differences coefficient are used directly; studies that report a take-up effect (two-stage least squares or a local average treatment effect) are converted to the reduced form by multiplying by the paper's own first-stage take-up.

Panel A: \citetapp{rossinslater2013} is the intent-to-treat effect on log annual wage income for mothers of one-year-olds, Table~7. \citetapp{campbell2017} is the two-stage least squares effect on annual wages in quarters 4--7 for the claimants sample, Table~3, converted to a reduced form by multiplying by the first-stage take-up of 0.817 (crossing the eligibility threshold raises the probability of receiving paid leave by 81.7 percentage points). \citetapp{baumruhm2016} is the difference-in-differences effect on log hourly wages one year after birth for mothers employed at all during pregnancy, Table~5 of the published version. \citetapp{timpe2024} is the difference-in-differences effect on log hourly wages over the first five years for women aged 18--45, Table~3 column~1. The annual-earnings studies (Rossin-Slater et al., Campbell et al., and Bailey et al.) and the hourly-wage studies (Baum and Ruhm, and Timpe) share the percent-effect-on-pay interpretation but are not identical objects.

Panel E reports \citetapp{bailey2025}, Table~3, Panel~B (cumulative real wage earnings in the short run, years 1--3). To match the all-mothers populations of the prior studies, we pool Bailey's first-birth and higher-order-birth estimates, weighting each by its sample size: the first-birth take-up effect is $-12.6\%$ (SE $5.7$) on $n=283{,}594$ and the higher-order-birth take-up effect is $-0.5\%$ (SE $7.3$) on $n=441{,}589$, giving a pooled take-up effect of $-5.2\%$. The standard error of the pool follows from the delta method. The pooled take-up effect is then converted to the availability scale via Bailey's first stage --- women who could take California's paid family leave consecutively after disability leave were 16 percentage points more likely to take it up --- giving a reduced-form effect of $-5.2\% \times 0.16 = -0.8\%$ (SE $0.8$).

\subsection{Additional settings from AEJ 2025 where small sample meta-analysis is feasible}
\label{app:aej_review}
\begin{landscape}
\captionsetup{font=normalsize}
\scriptsize
\setlength{\LTleft}{0pt}
\setlength{\LTright}{0pt}
\setlength{\tabcolsep}{3pt}
\setlength{\LTcapwidth}{\dimexpr3.2cm+4.8cm+4.3cm+1.1cm+7.6cm}
\renewcommand{\arraystretch}{1.2}
\begin{longtable}{@{}>{\raggedright\arraybackslash\hspace{0pt}}p{3.2cm}>{\raggedright\arraybackslash\hspace{0pt}}p{4.8cm}>{\raggedright\arraybackslash\hspace{0pt}}p{4.3cm}>{\centering\arraybackslash\hspace{0pt}}p{1.1cm}>{\raggedright\arraybackslash\hspace{0pt}}p{7.6cm}@{}}
\caption{Meta-analysis Candidates in the 2025 American Economic Journal: Economic Policy}
\label{tab:aej_candidates}\\
\toprule
\rowcolor{gray!35}
\textbf{Authors and year} & \textbf{Paper title} & \textbf{Quantity of interest} & \textbf{No. of other papers} & \textbf{Other estimates of quantity of interest among already-cited papers (first four shown)} \\
\midrule
\endfirsthead
\toprule
\rowcolor{gray!35}
\textbf{Authors and year} & \textbf{Paper title} & \textbf{Quantity of interest} & \textbf{No. of other papers} & \textbf{Other estimates of quantity of interest among already-cited papers (first four shown)} \\
\midrule
\endhead
\midrule
\multicolumn{5}{r}{Continued on next page}\\
\midrule
\endfoot
\bottomrule
\multicolumn{5}{@{}p{\dimexpr3.2cm+4.8cm+4.3cm+1.1cm+7.6cm}@{}}{\footnotesize \textit{Notes:} The ``Other estimates of quantity of interest among already-cited papers'' column lists the first four cited studies when a paper cites more than four (the remainder are indicated by ``\ldots''). An asterisk ($^{*}$) in the ``No.\ of other papers'' column denotes that the count is the number of studies in a cited meta-analysis rather than individually listed papers.}\\
\endlastfoot
Bailey et al. (2025a) & The Long-Run Effects of California's Paid Family Leave Act on Women's Careers and Childbearing: New Evidence from a Regression Discontinuity Design and US Tax Data & Effects of CA paid leave act on labor market outcomes (employment, childbearing) & 4 & Rossin-Slater et al. (2013); Campbell et al. (2018); Baum \& Ruhm (2016); Timpe (2024) \\
\rowcolor{gray!12}
Bailey et al. (2025b) & Show Me the Money! A Field Experiment on Electric Vehicle Charge Timing & Price elasticity of electric vehicle charging time & 3 & Burkhardt, Gillingham, Kopalle (2023); Qiu et al (2022); Ito, Ida, Tanaka (2018) \\
Bhuller et al. (2025) & Mental Health Consequences of Correctional Sentencing & Effect of rehabilitative sentencing on inmate mental health & 5 & Binswanger et al (2007); Weidner, Schultz (2019); Haglund et al (2014); Sailas et al (2006); \ldots \\
\rowcolor{gray!12}
Bilicka and Güçeri (2025) & Dividend Taxation and Firm Performance with Heterogeneous Payout Responses & Dividend reduction after introduction of a tax & 4 & Alstadsæter, Fjærli (2009); Bach, et al. (2023) - led to retraction of BM 2022; Alstadsæter, Kopczuk, Telle (2014); Hanlon, Hoopes (2014) \\
Bilicka and Güçeri (2025) & Dividend Taxation and Firm Performance with Heterogeneous Payout Responses & Elasticity of dividend response & 3 & Yagan (2015); Chetty, Saez (2006); Poterba (1987) \\
\rowcolor{gray!12}
Blundell et al. (2025) & Pay Transparency and Gender Equality & Decrease in pay gap after transparency act & 3 & Baker et al. (2023); Bennedsen et al. (2022); Gulyas et al. (2023) \\
Brough et al. (2025) & Eliminating Fares to Expand Opportunities: Experimental Evidence on the Impacts of Free Public Transportation on Economic and Social Disparities & Employment effects of free fares & 3 & Phillips (2014); Franklin (2017); Abebe et al (2021) \\
\rowcolor{gray!12}
Brülhart et al. (2025) & Who Bears the Burden of Local Taxes? & Incidence of local taxes on high-income population & 3 & Suárez-Serrato, Zidar (2016); Fuest, Peichl, Siegloch (2018); Löffler, Siegloch (2021) \\
Bukovina et al. (2025) & Corporate Minimum Tax and the Elasticity of Taxable Income: Evidence from Administrative Tax Records & Elasticity of taxable income & 3 & Pomeranz (2015); Almunia, Lopez-Rodriguez (2018); Naritomi (2019) \\
\rowcolor{gray!12}
Cabral et al. (2025) & The Impact of Provider Payments on Health Care Utilization of Low-Income Individuals: Evidence from Medicare and Medicaid & Demand elasticity for Medicare & 5 & Roberts, Desai (2021); Brown, Kowalski, Lurie (2019); Fung et al (2021); Saez (2010); \ldots \\
Castell et al. (2025) & Take-up of Social Benefits: Experimental Evidence from France & Effects of various treatments on program take-up & 11 & Duflo, Saez (2003); Kling et al (2012); Bhargava, Manoli (2015); Bettinger et al (2012); \ldots \\
\rowcolor{gray!12}
Daysal et al. (2025) & Do Medical Treatments Work for Work? Evidence from Breast Cancer Patients & Mortality effects of radiotherapy & 4 & Overgaard et al (1997); Ragaz et al (1997); Clarke et al (2005); Darby et al (2011) \\
Daysal et al. (2025) & Do Medical Treatments Work for Work? Evidence from Breast Cancer Patients & Labor market outcomes of breast cancer survivors & 3 & Bradley, Bednarek, Neumark (2002); Bradley et al (2005); Heinesen, Kolodziejczyk (2013) \\
\rowcolor{gray!12}
Daysal et al. (2025) & Do Medical Treatments Work for Work? Evidence from Breast Cancer Patients & \% of breast cancer survivors that return to work & 6 & Drolet et al (2005); Balak et al (2008); Johnsson et al (2009); Damkjær et al (2011); \ldots \\
Dupas et al. (2025) & Informing Mothers about the Benefits of Conversing with Infants: Experimental Evidence from Ghana & Effect of program on \# of mothers conversing with infants & 102$^{*}$ & Meta-analysis by Jeong et al. (2021) \\
\rowcolor{gray!12}
Edwards et al. (2025) & HBCU Enrollment and Longer-Term Outcomes & Effects of attending an HBCU on degree completion & 6 & Fryer and Greenstone (2010); Price, Spriggs, Swinton (2011); Elu et al (2019); Ehrenberg, Rothstein (1999); \ldots \\
Getik and Meier (2025) & The Long-Run Effects of Peer Gender on Occupational Sorting and the Wage Gap & Effect of occupational sorting on the female wage & 3 & Black, Devereux, Salvanes (2010); Annelli, Peri (2019); Brenøe, Zölitz (2020) \\
\rowcolor{gray!12}
Guttman-Kenney et al. (2025) & The Semblance of Success in Nudging Consumers to Pay Down Credit Card Debt & Effects of various nudges on credit card debt levels & 4 & Agarwal et al (2015); Seira, Elizondo, Laguna-Müggenberg (2017); Adams et al (2022); Medina (2021) \\
Harris and Mills (2025) & Should College Be "Free"? Evidence on Free College, Early Commitment, and Merit Aid from an Eight-Year Randomized Trial & Effect of financial aid programs on college attainment \& college outcomes conditional on attainment & 5 & Angrist, Autor, Pallais (2021); Barrow et al (2014); Goldrick-Rab et al (2016); Carlson et al (2019); \ldots \\
\rowcolor{gray!12}
Ito and Zhang (2025) & Do Consumers Distinguish Fixed Cost from Variable Cost? "Schmeduling" in Two-Part Tariffs in Energy & Percentage decrease in energy use & 6 & Wolak (2011); Ito (2014); Ito (2015); Ito, Ida, Tanaka (2018); \ldots \\
Kessler and Roth (2025) & Increasing Organ Donor Registration as a Means to Increase Transplantation: An Experiment with Actual Organ Donor Registrations & Effect of "yes-no" framing vs "opt-in" framing on organ donor registrations & 2 & Johnson, Goldstein (2004); van Dalen, Henkens (2014) \\
\rowcolor{gray!12}
Li et al. (2025) & The Effects of a Multifaceted Poverty Alleviation Program on Rural Income and Household Behavior in China & \% increase in rural/poor incomes X years after program implementation & 2 & Meng (2013); Park, Wang (2010) \\
Lichter et al. (2025) & Profit Taxation, R\&D Spending, and Innovation & Elasticity of R\&D spending & 2 & Agrawal, Rosell, Simcoe (2020); Dechezleprêtre et al (2023) \\
\rowcolor{gray!12}
Mehmood et al. (2025) & Transmitting Rights: Effective Cooperation, Inter-gender Contact, and Student Achievement & Effects of curricular reforms on teacher/student attitudes & 2 & Alan et al (2021); Dhar et al (2022) \\
Moscona (2025) & The Management of Aid and Conflict in Africa & Effects of foreign aid on rate of violent conflict & 7 & Crost, Felter, Johnston (2014); Nunn, Qian (2014); Dube, Naidu (2015); Collier, Hoeffler (2010); \ldots \\
\rowcolor{gray!12}
Sabet et al. (2025) & Terrorism and Voting: The Rise of Right-Wing Populism in Germany & Effect of terror shocks on right-wing party vote shares & 9 & Gould, Klor (2010); Berrebi, Klor (2011); Getmansky, Zeitzoff (2014); Rehman, Vanin (2017); \ldots \\
Sager and Singer (2025) & Clean Identification? The Effects of the Clean Air Act on Air Pollution, Exposure Disparities, and House Prices & Elasticity of house prices with respect to pollution levels & 3 & Chay and Greenstone (2005); Bento et al. (2015); Grainger (2012) \\
\rowcolor{gray!12}
Shastry and Tortorice (2025) & Effective Health Aid: Evidence from Gavi's Vaccine Program & Effects of Gavi vaccine program on child mortality & 5 & Lu et al (2006); Jaupart, Dipple, Dercon (2019); Dykstra et al (2019); Ikilezi et al (2020); \ldots \\
Siegloch et al. (2025) & Spillover, Efficiency, and Equity Effects of Regional Firm Subsidies & Employment effects of GRW & 7 & Brachert, Dettmann, Titze (2019); Kline, Moretti (2014); Alder, Shao, Zilibotti (2016); Criscuolo et al (2019); \ldots \\
\rowcolor{gray!12}
Sigurdsson (2025) & Labor Supply Responses and Adjustment Frictions: A Tax-Free Year in Iceland & Frisch elasticity & 5 & Martínez, Saez, Siegenthaler (2021); Fehr, Goette (2007); Farber (2014); Chetty, Friedman, Saez (2013); \ldots \\
\end{longtable}
\normalsize
\end{landscape}

\clearpage
\section{Binary implementation decisions}
\label{app:binary_decisions}

Binary decisions $A_i$ might correspond to hypothesis tests, to decisions about implementing different treatments, or decisions about the same treatment for different sub-populations, etc. For binary decisions, we take $\theta_{i}$ to be the return to a positive decisions ($A_i=1$) in problem-instance $i$.

Suppose that each of the studies $i$ corresponds to an (experimental or quasi-experimental) evaluation of a different treatment. Suppose further that for each of these treatments, we wish to make a decision $A_{i} \in \{0,1\}$ whether to implement them on a larger scale, with social return $\theta_{i}$, so that $U(A_{i}, \theta_{i}) = A_{i} \cdot \theta_{i}$, $A_{i}\in \{0,1\}$. For this interpretation, $\theta _i$ should be defined such that it captures all relevant costs and benefits of the treatment.

Ideally, if we could observe $\theta_{i}$, we would choose $A_{i} = \mathbf{1}(\theta_{i} \geq 0)$. In practice, we only observe $(\hat{\theta}_{i}, \sigma_{i}, X_{i})$. Denote the posterior expected return to $A_i$ by $t(\hat{\theta}, \sigma) = E\left[ \theta_{}| \hat{\theta}, \sigma \right]$. We derived expressions for this posterior expected return, under various assumptions, in Section~\ref{sec:empiricalbayes} above.
The optimal decision for a Bayesian decision-maker who maximizes expected welfare equals
$$
A_{i} = \mathbf{1}(t(\hat{\theta}_{i}, \sigma_{i},X_i) \geq 0),
$$
To estimate the posterior expected return, we recall Tweedie's formula from Equation~\eqref{eqn:tweedie}, $t(\hat{\theta}, \sigma, x) = \hat{\theta} + \sigma^2 \cdot \frac{ \partial }{ \partial \hat \theta } \log(f(\hat{\theta} | \sigma,x))$, where $f(\hat{\theta} | \sigma,x)$ is the conditional density. Under the parametric model where $\theta_i |X_i = x \sim N(x \cdot \beta, \tau^2)$,  we can specialize this to $E[\theta_{i} | \hat{\theta}_{i}, \sigma_{i}, X_{i}] = (1-\kappa_{i}) \cdot (X_{i} \cdot \beta) + \kappa_{i} \cdot\hat{\theta}_{i}$, where $\kappa_{i} = \frac{\tau^{2}}{\tau^{2} + \sigma^{2}_{i}}$. Note that the sign of $t(\hat{\theta}_{i}, \sigma_{i},X_i)$ might differ from the sign of $\hat{\theta}_i$, depending on the distribution of $\hat{\theta}_i$ across $i$, and depending on the value of the covariates $X_i$.\\

A closely related decision problem concerns the implementation of a binary decision $A_{0}$ for a new problem instance, for which no $\hat{\theta}_{0}$ or $\sigma_{0}$ has been observed, but covariates $X_0$ are available. In this setting, the optimal decision is given by
$$
A_{0} = \mathbf{1}(E[\theta_{0} | X_{0}]\geq 0).
$$
Under the parametric linear model, $E[\theta_{0}|X_{0}] = X_{0}\cdot\beta$; we discussed estimation of $\beta$ using (appropriately weighted) Ridge regression in Section~\ref{sec:empiricalbayes_with_x}. The posterior expected welfare for the optimal treatment decision in instance $i$, when an estimate $\hat{\theta}_i$ is available, equals
$$
E[U(A_{i}, \theta_{i}) |\hat{\theta}_{i}, \sigma_{i}, X_i] = \max(t(\hat{\theta}_i, \sigma _i, X_i), 0).
$$
This expression for posterior expected welfare will be important below, when we consider the allocation of limited research resources, where the implied goal is to maximize the (ex-ante) \emph{value of information} of observing $\hat{\theta}_i$,  which can be written as
\begin{equation}
	\label{eq:valueofinformation}
E\left[\max\left(t(\hat{\theta}_i, \sigma _i, X_i), 0\right)\big|X_i\right] - \max \left( E\left[t(\hat{\theta}_i, \sigma _i, X_i)\big|X_i\right], 0\right).
\end{equation}

\clearpage

\section{Covariates}\label{app:covariates}

Table~\ref{tab:ckw_tr_long_mixed_ui_vs_ltu_panel_b} reports the per-study estimates, standard errors, distances to each test point, citations, and sample sizes for every study that appears in Figure~\ref{fig: ckw gp weights} of Section~\ref{sec: predictive models with x} --- the union of the ten highest-$|w|$ studies for each of the two test contexts.

\begin{table}[htbp]
\centering
\caption{Posterior distribution of employment effects of training: Panel B}
\label{tab:ckw_tr_long_mixed_ui_vs_ltu_panel_b}
\scriptsize
\renewcommand{\arraystretch}{0.9}
\setlength{\tabcolsep}{4pt}
\begin{tabular}{lrcccccc}
\toprule
  & & \multicolumn{2}{c}{Study estimate} & \multicolumn{2}{c}{UI recipients} & \multicolumn{2}{c}{Long-term unemployed} \\
\cmidrule(lr){3-4}\cmidrule(lr){5-6}\cmidrule(lr){7-8}
 & $n$& $\hat{\theta}_i$ & SE & $k$(short $\ell$) & $k$(long $\ell$) & $k$(short $\ell$) & $k$(long $\ell$) \\
\midrule
 1. UI/Tr/L & 4,664 & +0.12 & 0.009 & 0.0056 & 0.0056 & 0.0017 & 0.0052 \\
 2. UI/Tr/L & 7,934 & +0.15 & 0.009 & 0.0056 & 0.0056 & 0.0017 & 0.0052 \\
 3. UI/Tr/L & 95,000 & +0.04 & 0.01 & 0.0056 & 0.0056 & 0.0017 & 0.0052 \\
 4. UI/Tr/L & 92,500 & +0.05 & 0.01 & 0.0056 & 0.0056 & 0.0017 & 0.0052 \\
 5. UI/Tr/L & 85,400 & +0.25 & 0.01 & 0.0056 & 0.0056 & 0.0017 & 0.0052 \\
 6. UI/Tr/L & 86,000 & +0.27 & 0.01 & 0.0056 & 0.0056 & 0.0017 & 0.0052 \\
 7. UI/Tr/L & 88,400 & +0.24 & 0.01 & 0.0056 & 0.0056 & 0.0017 & 0.0052 \\
 8. UI/Tr/S & 28,246 & -0.019 & 0.005 & 0.0018 & 0.0052 & 0.0006 & 0.0048 \\
 9. LTU/Tr/S & 23,182 & +0.023 & 0.006 & 0.0006 & 0.0048 & 0.0018 & 0.0052 \\
10. LTU/Tr/S & 15,532 & +0.021 & 0.007 & 0.0006 & 0.0048 & 0.0018 & 0.0052 \\
11. LTU/Tr/M & 23,182 & +0.056 & 0.006 & 0.0003 & 0.0046 & 0.0008 & 0.0049 \\
12. LTU/Tr/M & 15,532 & +0.018 & 0.008 & 0.0003 & 0.0046 & 0.0008 & 0.0049 \\
13. LTU/Ot/L & 2,268 & -0.037 & 0.017 & 0.0001 & 0.0043 & 0.0003 & 0.0046 \\
14. LTU/Ot/S & 3,705 & +0.041 & 0.001 & 0.0000 & 0.0040 & 0.0001 & 0.0043 \\
15. LTU/Ot/M & 3,705 & +0.05 & 0.0014 & 0.0000 & 0.0038 & 0.0000 & 0.0041 \\
16. Dis/Tr/S & 83,145 & -0.000295 & 0.0016 & 0.0007 & 0.0049 & 0.0002 & 0.0046 \\
\bottomrule
\end{tabular}
\begin{minipage}{0.95\textwidth}
\vspace{0.3em}\footnotesize\textit{Notes:} Per-study estimates and per-context kernel covariances for the union of studies displayed in either panel of the GP-weights figure. $\hat{\theta}_i$ and SE are the study estimate and standard error; $n$ is the study sample size. $k(x_0, x_i) = \rho^2 \exp(-d^2 / 2\ell^2)$ is the squared-exponential kernel covariance between study $i$ and the test point $x_0$, where $d$ is their Euclidean distance in the standardised covariate space, evaluated at the short and long length scales $\ell$ ($\rho^2 = 0.0056$, $\ell_{\mathrm{short}} = 1.58$, $\ell_{\mathrm{long}} = 6.32$). A larger covariance means study $i$ is more similar to the test point and receives more posterior weight. $\hat{\theta}_i$ and SE are shown to the number of decimals reported in the original study; $n$ to 0 and $k$ to 4 decimals.
\end{minipage}
\end{table}

\clearpage

\section{Smoothness and monotonicity of the p-curve}
\label{app:pcurve_derivation}
What properties should the p-curve exhibit in the absence of selectivity? For the following derivation, assume that p-values are based on $Z$-statistics $Z_i$ for one-sided hypothesis tests of the null that $\theta_i = 0$ , $P_{i} = \Phi(-Z_i)$. Let $z = \Phi^{-1}(1-p)$, so that $\frac{ \partial z }{ \partial p } = -\frac{1}{\varphi(z)}$.
In Equation~\ref{eq:convolutiondensity} in the preceding section, we showed that the density of $Z$ is given by a convolution of the distribution $\nu$ of $\omega$ with the standard normal density $\varphi$, $f(z) = \int \varphi(z - \omega) d \nu(\omega)$.
From this, we can derive the corresponding distribution of p-values as follows, taking into account the change of variables from $Z$ to $P$. Here $g(p)$ is the density of p-values, or p-curve:
$$
\begin{aligned}
P(P_{i} \leq p) & = 1- F(z) \\
g(p) & = \frac{ \partial P(P \leq p) }{ \partial p } =
\frac{f(z)}{\varphi(z)}, \\
g'(p) &= \frac{-1}{\varphi(z)^{3}} \cdot \left(f'(z) \varphi(z) - f(z) \varphi'(z) \right)  \\
&=\frac{-1}{\varphi(z)^{3}} \cdot \left(\int \left( \varphi'(z-\omega)\varphi(z) - \varphi(z-\omega) \varphi'(z) \right) d \nu(\omega) \right)
\end{aligned}
$$
Smoothness of the density $g(p)$ is immediate from this derivation. In fact, $g$ is analytic (infinitely differentiable) since $\varphi$ is analytic, $f$ is a convolution of $\nu$ with $\varphi$, and analyticity is preserved under convolution. The p-curve should thus have no jumps or kinks.

In addition to smoothness, this derivation also implies restrictions on the shape of $g(\cdot)$.
Since the standard normal density is given by $\varphi(z) = (2\pi)^{-1/2} \cdot \exp\left( -\frac{z^{2}}{2} \right)$, it follows that $\frac{\varphi'(z)}{\varphi(z)} = \frac{ \partial \log(\varphi(z)) }{ \partial z } =-z$. Consider the integrand $\varphi'(z-\omega)\varphi(z) - \varphi(z-\omega) \varphi'(z)$ in the expression for $g'(p)$ above. This integrand is positive iff
$$
\omega-z = \frac{\varphi'(z-\omega)}{\varphi(z-\omega)}\geq \frac{\varphi'(z)}{\varphi(z)} = -z.
$$
The integrand is therefore positive for $\omega\geq 0$. Therefore $g'(z) \leq 0$ if $\nu$ is supported on $\mathbb{R}^{+}$, and monotonicity of $g$ follows. A similar calculation yields monotonicity of $g$ for 2-sided hypothesis tests, where $P_{i} = 2\cdot\Phi(-|Z_{i}|)$.

\clearpage
\section{Density ratio in \cite{cohen2025disemployment}}
\label{app: cg density ratio}

\begin{figure}[H]
  \centering
  \caption{Cohen \& Ganong data: $Z$-statistic densities by SE group and density ratio}
  \includegraphics[width=\textwidth]{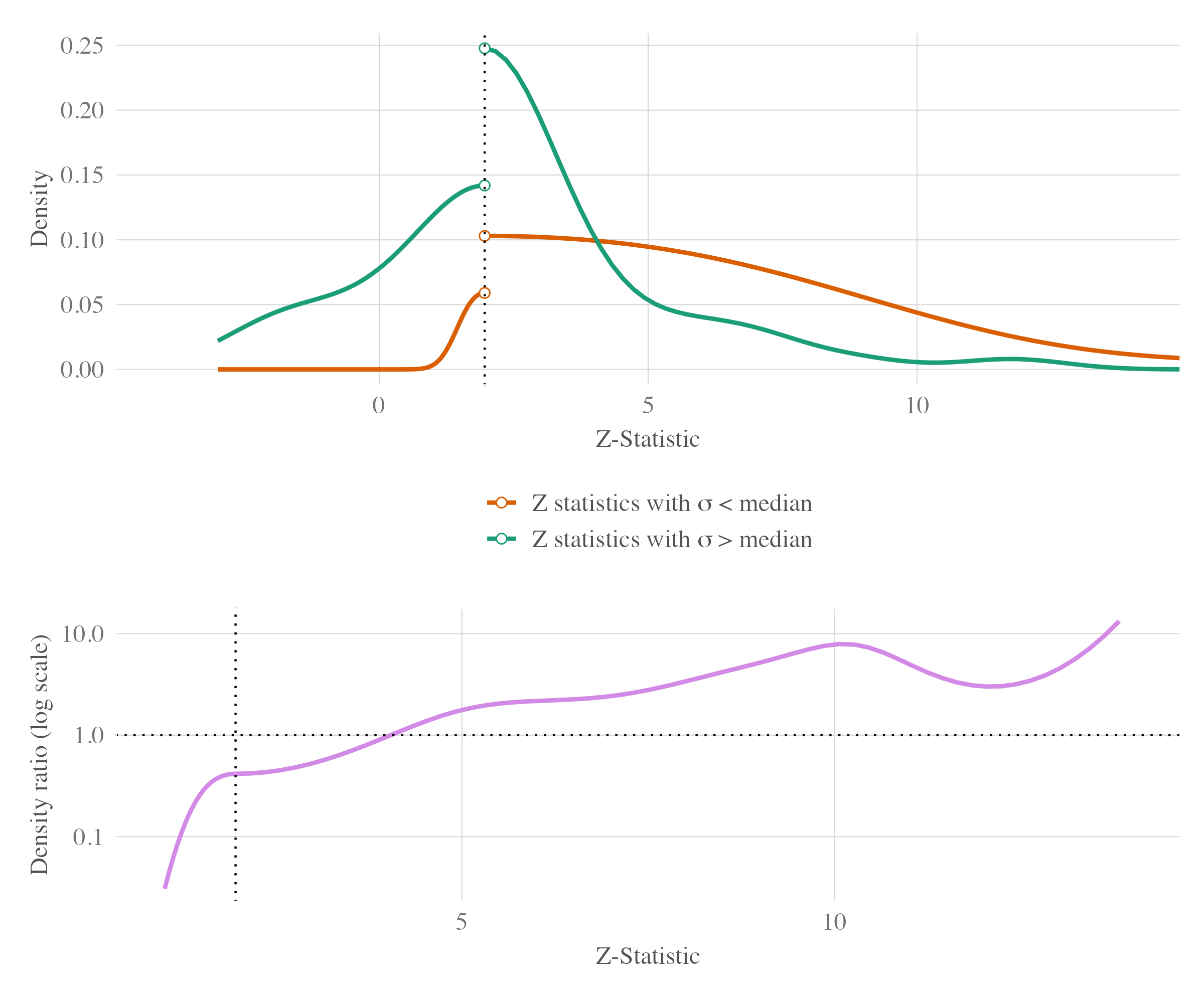}
  \label{fig: cg density ratio}
\end{figure}

To show how reduced-form evidence of a nonzero latent mean can appear in a real-life empirical setting with selectivity, we apply the plotting framework from the illustrative example to a real application: \citeapp{cohen2025disemployment} in Figure \ref{fig: cg density ratio}. We divide the sample into two groups based on the median standard error. The top panel of the figure shows the density of $Z$-statistics. The panel shows many of the same features as the simulated environment in Figure \ref{fig: z pub}: the $Z$-statistics for the more precise estimates in red are much more dispersed than the imprecise estimates in green and there is a jump in the densities at $Z = 1.96$. The key qualitative pattern to focus on is that the precise estimates in red are not simply a mean-preserving spread of the green estimates (as is the case in the left panel of Figure  \ref{fig: z pub}). Instead, the density of the red estimates is shifted to the right relative to the density of the green estimates (as is the case in the right panel of Figure  \ref{fig: z pub}). This is the ``signature'' of a positive value for the latent mean of the distribution. 

The bottom panel of Figure  \ref{fig: cg density ratio} makes this ``signature'' legible. The density ratio (of the imprecise estimates to the precise estimates) is less than one everywhere below 1.96 and is rising monotonically in the $Z$-statistic through $Z = 10$. This pattern is quite similar to the right panel of Figure  \ref{fig: density ratio} and is noticeably different from the U-shape in the density ratio in the left panel of Figure  \ref{fig: density ratio}. We emphasize that one should not use this procedure of splitting the sample in half to actually estimate a latent mean. Maximum likelihood, which uses all available information in $\sigma_i$, is optimal. Nevertheless, the patterns here are useful as a visual diagnostic.

\clearpage

\section{Density Estimation and Bandwidth Selection}\label{app:density_discont}

In the main text, Section \ref{sec: selectivity empirical}, we test for selectivity by examining whether the p-curve is discontinuous at 0.05 \citepapp{elliott2022detecting, elliott_power_2025}. Our method of local polynomial density estimation requires choosing a bandwidth \citepapp{cattaneo2020simple}. With a small set of studies, bandwidth choice becomes challenging and could meaningfully affect whether one rejects the null of no discontinuity in the p-curve at 0.05. To assess whether our results are robust to different bandwidth choices, Table \ref{tab: bandwidth choices} shows the p-values for three bandwidth choices: MSE optimal, 1/2 MSE optional, and 2x MSE optimal.

\begin{table}[h!]
    \centering
    \caption{Bandwidth Sensitivity in Discontinuity Tests.}
    \footnotesize
    
\begin{tabular}{llrrrr}
\toprule
\toprule
\multicolumn{2}{c}{ } & \multicolumn{2}{c}{Bandwidths} & \multicolumn{2}{c}{ } \\
\cmidrule(l{3pt}r{3pt}){3-4}
Study &  & $h_{p<0.025}$ & $h_{p \geq 0.025}$ & Density Ratio & p-value\\
\midrule
 & 0.5x MSE-optimal & 0.012 & 0.098 & 0.191 & 0.287\\
\cmidrule{2-6}
 & MSE-optimal & 0.025 & 0.196 & 0.634 & 0.221\\
\cmidrule{2-6}
\multirow{-3}{*}{\raggedright\arraybackslash \shortstack[l]{Effect of Nudge on\\Targeted Outcome\\ DellaVigna and Linos (2022)}} & 2.0x MSE-optimal & 0.025 & 0.392 & 0.642 & 0.208\\
\cmidrule{1-6}
 & 0.5x MSE-optimal & 0.011 & 0.126 & 0.039 & 0.075\\
\cmidrule{2-6}
 & MSE-optimal & 0.022 & 0.252 & 0.068 & 0.000\\
\cmidrule{2-6}
\multirow{-3}{*}{\raggedright\arraybackslash \shortstack[l]{Effect of Active Labor Market \\ Program on Employment \\ Card et al. (2018)}} & 2.0x MSE-optimal & 0.025 & 0.505 & 0.062 & 0.001\\
\cmidrule{1-6}
 & 0.5x MSE-optimal & 0.012 & 0.151 & 0.269 & 0.198\\
\cmidrule{2-6}
 & MSE-optimal & 0.025 & 0.301 & -0.814 & 0.095\\
\cmidrule{2-6}
\multirow{-3}{*}{\raggedright\arraybackslash \shortstack[l]{Effect of \$1 Transfer \\ on Monthly Consumption \\ Crosta et al. (2024)}} & 2.0x MSE-optimal & 0.025 & 0.603 & -0.539 & 0.091\\
\cmidrule{1-6}
 & 0.5x MSE-optimal & 0.012 & 0.263 & 0.068 & 0.374\\
\cmidrule{2-6}
 & MSE-optimal & 0.025 & 0.526 & 0.077 & 0.021\\
\cmidrule{2-6}
\multirow{-3}{*}{\raggedright\arraybackslash \shortstack[l]{Elasticity of Unemp Duration\\w.r.t. Unemp Benefits\\Cohen and Ganong (2026)}} & 2.0x MSE-optimal & 0.025 & 0.975 & 0.070 & 0.015\\
\bottomrule
\end{tabular}

    \label{tab: bandwidth choices}
    \vspace{2mm}
     \begin{minipage}{\textwidth}
    \textit{Notes:} This table presents p-values from tests of selectivity with different choices of bandwidths and with de-rounded data. It reports the p-value using MSE optimal bandwidths, 1/2 MSE optimal bandwidth, and 2x MSE optimal bandwidth. If the bandwidth below the cutoff is greater than 0.025, we use 0.025. If the bandwidth above the cutoff is greater than 0.975, we use 0.975. %
    \end{minipage}
\end{table}
In general, the results are not robust to different bandwidth choices. However, we consistently estimate a density ratio less than 1 but find insufficient evidence to reject the null hypothesis of no selection.

\begin{landscape}
    \begin{figure}
        \centering
        \caption{Density Discontinuity in the P-curve at 0.05.}
        \includegraphics[width=\linewidth]{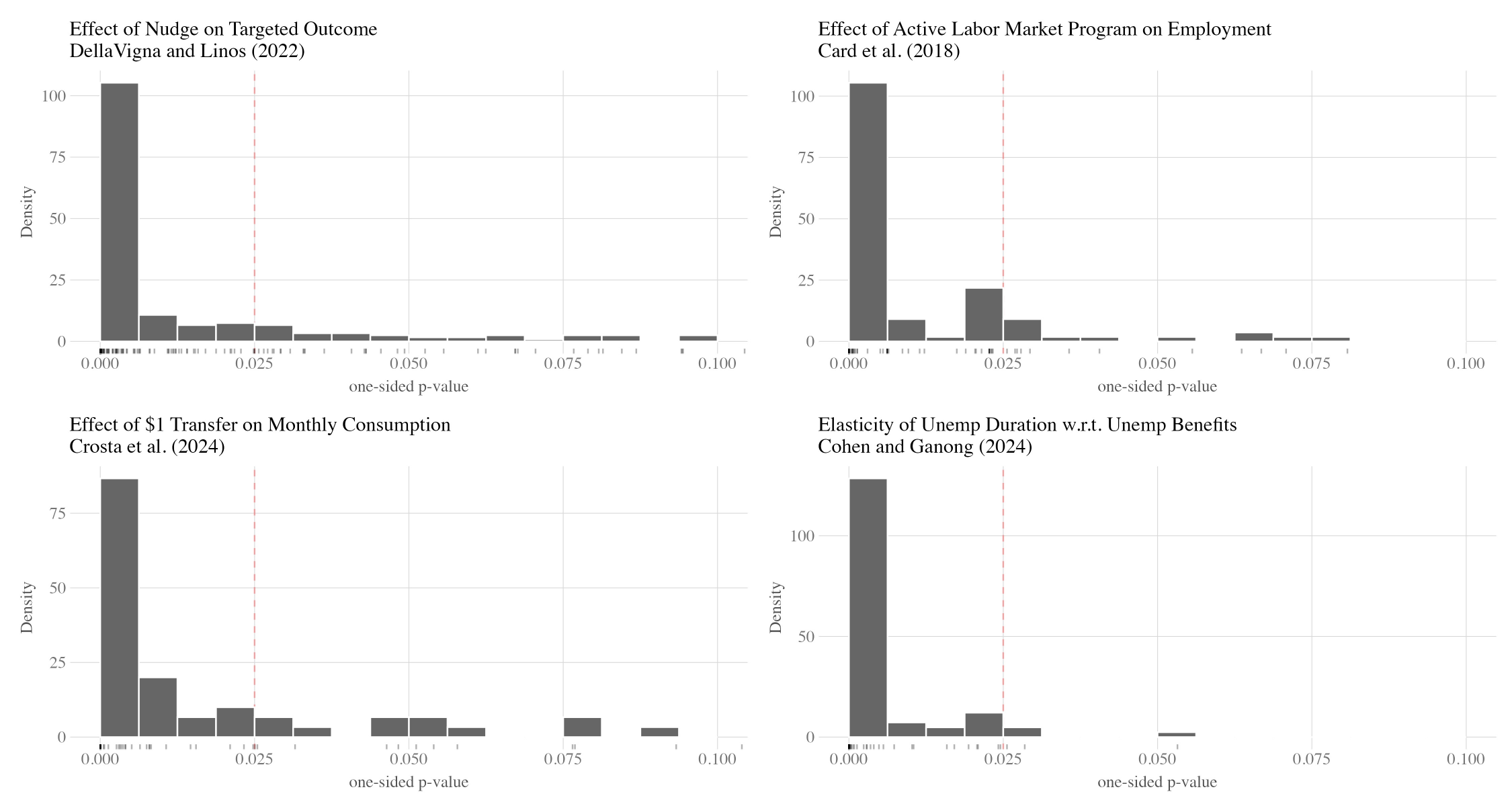}
        \label{fig: elliot density}
        \begin{minipage}{\textwidth}
    \textit{Notes:} This figure shows the histogram of p-values for estimates of studies in four meta-analyses. The y-axis is the number of estimates. Plot annotations are MSE-optimal bandwidth $h$ on each side of the cutoff, the estimated density (number of studies), the difference of the densities, and the p-value of the difference of these densities. For \cite{dellavigna2022rctstoscale}, we include only estimates from the academic journal subsample since we assume nudge unit estimates are always reported.
    \end{minipage}
    \end{figure}
\end{landscape}

\clearpage

\bibliographystyleapp{apalike}
\bibliographyapp{library_appendix}

\end{document}